\RequirePackage{tikz}
\PassOptionsToPackage{hidelinks}{hyperref}

\documentclass[sn-mathphys,hidelinks]{sn-jnl}% Math and Physical Sciences Reference Style
\usepackage{graphicx}

\usepackage[english]{babel}
\usepackage{bm}
\usepackage{latexsym}
\usepackage{epsfig}
\usepackage{psfrag}
\usepackage[normalem]{ulem}
\usepackage{textcomp}
\usepackage{color}
\usepackage[utf8]{inputenc}
\usepackage{comment}
\usepackage{xfrac}   % for \sfrac macro
\usepackage{bbm}

\usepackage{blindtext}
\usepackage{float}
\usepackage{microtype}
\usepackage{bm}
\usepackage{latexsym}
\usepackage{epsfig}
\usepackage{psfrag}
\usepackage{color}
\usepackage{xcolor}
\usepackage{paralist}
\usepackage{lmodern}
\usepackage{placeins}

\usepackage{epstopdf}
\usepackage{mathrsfs}
\usepackage{varioref}
\usepackage{nicefrac}

\usepackage[hidelinks]{hyperref}
\jyear{2021}%

\theoremstyle{thmstyleone}%
\theoremstyle{thmstyletwo}%

\theoremstyle{thmstylethree}%

\def\nn{\nonumber} 
\def\f{\frac}

\def\l{\left}
\def\r{\right}
\def\d{{\mathrm{d}}}
\def\pa{\partial}
\def\Mpl{M_{_{\mathrm{Pl}}}}
\def\cN{{\mathcal{N}}}

\def\ps{\mathcal{P}_{_{\mathcal{R}}}}

\def\ke{k_{\mathrm{e}}}

\def\ei{\eta_{\mathrm{i}}}

\def\e1i{\epsilon_{1\mathrm {i}}}

\def\phii{\phi_{\mathrm {i}}}

\allowdisplaybreaks[1]

\def\nn{\nonumber} 
\def\pa{{\partial}}
\def\f{\frac}
\def\l{\left}
\def\r{\right}
\def\d{{\rm d}}
\def\beq{\begin{equation}}
\def\eeq{\end{equation}} 
\def\beqa{\begin{eqnarray}}
\def\eeqa{\end{eqnarray}} 

\def\Omegar{\Omega^{\mathrm{R}}}
\def\Omegai{\Omega^{\mathrm{I}}}

\def\Omegasisir{\Omega_{\s\s}^{\mathrm{R}}}
\def\Omegasisii{\Omega_{\s\s}^{\mathrm{I}}}
\def\Omegassr{\Omega_{ss}^{\mathrm{R}}}
\def\Omegassi{\Omega_{ss}^{\mathrm{I}}}
\def\Omegasisr{\Omega_{\s s}^{\mathrm{R}}}
\def\Omegasisi{\Omega_{\s s}^{\mathrm{I}}}

\def\cA{{\mathcal{A}}}
\def\cB{{\mathcal{B}}}
\def\cC{{\mathcal{C}}}

\def\cR{{\mathcal{R}}}

\def\Z{{\textrm{Z}}}

\def\s{\sigma}

\newcommand{\boldmathsymbol}[1]{{\ensuremath{\boldsymbol{#1}}}}
\newcommand{\bmk}{\boldmathsymbol{k}}
\newcommand{\bmp}{\boldmathsymbol{p}}

\begin{document}
%%%%%%%%%%%%%%%%%%%%%%%%%%%%%%%%%%%%%%%%%%%%%%%%%%%%%%%%%%%%%%%%%%%%%%%%%%%%%%%

\title[Evolution of quantum state of perturbations during inflation]{Enhanced power 
on small scales and evolution of quantum state of perturbations in single and 
two field inflationary models}
\author*[1]{\fnm{Rathul Nath} \sur{Raveendran}}\email{rathulnath.r@gmail.com}
\affil*[1]{\orgdiv{School of Physical Sciences}, \orgname{Indian Association 
for the Cultivation of Science}, \orgaddress{\city{Kolkata~700032}, \country{India}}}
\author[2]{\fnm{Krishnamohan} \sur{Parattu}}\email{mailofkrishnamohan@gmail.com}
\affil[2]{\orgname{Chennai Mathematical Institute}, \orgaddress{\city{Kelambakkam, 
Tamil Nadu~603103}, \country{India}}}
\equalcont{These authors contributed equally to this work.}
\author[3]{\fnm{L.}~\sur{Sriramkumar}}\email{sriram@physics.iitm.ac.in} 
\affil[3]{\orgdiv{Centre for Strings, Gravitation and Cosmology, 
Department of Physics}, \orgname{Indian Institute of Technology Madras},
\orgaddress{\city{Chennai~600036}, \country{India}}}
\equalcont{These authors contributed equally to this work.}
%%%%%%%%%%%%%%%%%%%%%%%%%%%%%%%%%%%%%%%%%%%%%%%%%%%%%%%%%%%%%%%%%%%%%%%%%%%%%%%
\abstract{With the detection of gravitational waves from merging binary black 
holes, over the last few years, there has been a considerable interest in the 
literature to understand if these black holes could have originated in the
early universe.
If the primordial scalar power over small scales is boosted considerably when 
compared to the COBE normalized amplitude over large scales, then, such an
increased power can lead to a copious production of primordial black holes
that can constitute a significant fraction of the cold dark matter density
today.
Recently, many models of inflation involving single or two scalar fields 
have been constructed which lead to enhanced power on small scales.
In this work, we examine the evolution of the quantum state of the curvature 
perturbations in such single and two field models of inflation using measures
of squeezing, entanglement entropy or quantum discord.
We find that, in the single as well as the two field models, the extent of 
squeezing of the modes is enhanced to a certain extent (when compared to 
the scenarios involving only slow roll) over modes which exhibit increased 
scalar power.
Moreover, we show that, in the two field models, the entanglement entropy 
associated with the curvature perturbation, arrived at when the isocurvature 
perturbation has been traced out, is also enhanced over the small scales.
We also take the opportunity to discuss the relation between entanglement
entropy and quantum discord in the case of the two field model. 
We conclude with a brief discussion on the wider implications of the results.}
%%%%%%%%%%%%%%%%%%%%%%%%%%%%%%%%%%%%%%%%%%%%%%%%%%%%%%%%%%%%%%%%%%%%%%%%%%%%%%%

\maketitle
	
%%%%%%%%%%%%%%%%%%%%%%%%%%%%%%%%%%%%%%%%%%%%%%%%%%%%%%%%%%%%%%%%%%%%%%%%%%%%%%%

\section{Introduction}\label{sec:intro}
	
The cosmic microwave background~(CMB) is a vestige of the radiation dominated 
epoch of the early universe.
It is almost perfectly thermal in nature, and it carries pristine information 
regarding the state of the universe during the early stages.
The temperature of the CMB, while it is isotropic to a great degree, also 
contains small anisotropies (of the order of one part in~$10^5$) superimposed
upon it (for recent observations, see Refs.~\cite{Planck:2015mrs,Planck:2018nkj,
BICEPKeck:2021bsr}).
It is these anisotropies that are supposed to have evolved into the inhomogeneities 
in the large scale structure through gravitational instability, which we observe
today as galaxies and clusters of galaxies. 
However, in the conventional hot big bang model, it is only a rather small part 
of the CMB sky (about $1^\circ$) that could have been causally connected at 
the time of decoupling, when the CMB started streaming freely towards us.
Therefore, explaining the extent of isotropy of the CMB poses a challenge within
the model, an issue that is often referred to as the horizon problem.
Among the different paradigms that have been proposed to overcome these challenges,
without doubt, it is the inflationary scenario which provides the most simple 
and attractive mechanism to overcome the horizon problem as well as generate the
primordial perturbations.

Inflation refers to a brief period of accelerated expansion during the early 
stages of the radiation dominated epoch.
The accelerated expansion is usually driven with the aid of one or more scalar
fields that are often encountered in high energy physics and string theory.
While the classical component of the scalar fields is supposed to drive the rapid
expansion, it is the quantum fluctuations associated with the scalar fields that
are supposed to be responsible for the primordial perturbations (see, for example,
the reviews~\cite{Mukhanov:1990me,Martin:2003bt,Martin:2004um,Bassett:2005xm,
Sriramkumar:2009kg,Baumann:2008bn,Baumann:2009ds,Sriramkumar:2012mik,Linde:2014nna,
Martin:2015dha}).
The quantum fluctuations are expected to grow and turn in to classical perturbations
during the later stages of inflation (see, for instance, Refs.~\cite{Albrecht:1992kf,
Polarski:1995jg,Kiefer:1998qe,Kiefer:1998jk,Kiefer:2006je,Martin:2007bw,Kiefer:2008ku,
Martin:2012pea,Martin:2012ua,Martin:2015qta}).
One of the challenges in cosmology today is to gain a clear understanding as 
well as identify possibly observable signatures of the quantum-to-classical 
transition of the primordial perturbations.
	
There exist many models of inflation which are consistent with the CMB data
on large scales, say, over $10^{-5}<k<1\;\mathrm{Mpc}^{-1}$ (for the constraints
from the Planck data, see Refs.~\cite{Planck:2015sxf,Planck:2018jri}; for a long
list of inflationary models involving a single canonical scalar field that are 
consistent with the Planck data, see Refs.~\cite{Martin:2013tda,Martin:2013nzq}).
But, the constraints on the primordial power spectra over smaller scales, say,
over $k > 1\,\mathrm{Mpc}^{-1}$, are considerably weaker.  
The recent discovery of gravitational waves form merging binary black holes 
has led to a strong interest in investigating whether these black holes could 
have formed in the early universe~\cite{Bird:2016dcv,DeLuca:2020qqa,
Jedamzik:2020ypm,Jedamzik:2020omx}.
The simpler models of slow roll inflation produce an insignificant number of 
primordial black holes (PBHs)~\cite{Chongchitnan:2006wx}.
Hence, if a significant amount of PBHs have to be produced, say, comparable
to the cold dark matter density today, then the strength of the inflationary 
scalar power spectrum has to be considerably enhanced on small scales, in
contrast to the COBE normalized amplitude on the CMB scales.
During the past few years, a variety of inflationary models involving one or 
two scalar fields have been examined in this context.
In the case of single field models of inflation, it has been found that, a 
point of inflection in the potential leads to an epoch of ultra slow roll,
which can give rise to the required boost in the scalar amplitude over 
small scales (for a small list of such efforts, see
Refs.~\cite{Garcia-Bellido:2017mdw,Ballesteros:2017fsr,Germani:2017bcs,
Kannike:2017bxn,Dalianis:2018frf,Ragavendra:2020sop}).
However, often, in these models, there seems to arise a tension with the 
constraints on the scalar spectral index from the recent Planck CMB data. 
The two field models of inflation, in contrast, offer a richer dynamics to
achieve the boost in power on small scales.
For example, one can either consider hybrid models of inflation or, apart 
from the canonical scalar field, invoke a non-canonical scalar field in 
order to arrive at the required background evolution and the desired power 
spectra (in this context, see, for instance, Refs.~\cite{Palma:2020ejf,
Fumagalli:2020adf,Braglia:2020eai,Choi:2021yxz}).

Our aim in this work is to examine the evolution of the quantum state of the 
perturbations in single and two field models of inflation.
We shall specifically focus on the evolution of the quantum state of the 
curvature perturbation in models of inflation that have been considered in 
the context of enhanced formation of PBHs (for related discussions, see 
Ref.~\cite{dePutter:2019xxv,Figueroa:2021zah}). 
We shall utilize measures such as squeezing, entanglement entropy or quantum 
discord to track the evolution of the quantum state and compare their behavior
in these models with those that occur in simpler models that permit only slow 
roll inflation.

This paper is organized as follows. 
In the following section, we shall discuss the evolution of the quantum
state of the curvature perturbations in single field models of inflation.
We shall consider a specific inflationary model that permits a period of
ultra slow roll inflation which leads to an enhanced power on small scales.
We shall compare the behavior of the squeezing amplitude in such models 
with the corresponding behavior in slow roll inflationary models.
In Sec.~\ref{sec:tfm}, we shall extend these arguments to the case of 
inflationary models involving two fields.
We shall trace out the degree of freedom associated with the isocurvature
perturbations to arrive at the reduced density matrix describing the curvature
perturbations.
We shall show that the entanglement entropy corresponding to the reduced
density matrix is the same as the quantum discord associated with the 
system.
Apart from the squeezing amplitude, we shall study the behavior of the 
entanglement entropy in an inflationary model leading to enhanced power
on small scales and compare the behavior in a situation which only 
involves slow roll inflation.
We shall conclude in Sec.~\ref{sec:d} with a brief discussion.

At this stage of our discussion, let us make a few clarifying remarks concerning 
the conventions and notations that we shall adopt in this work.
We shall be working in $(3+1)$~spacetime dimensions with the metric signature 
of $(-,+,+,+)$. 
We shall adopt the natural units wherein $\hbar=c=1$, and set the Planck mass 
to be $\Mpl=(8\, \pi\, G)^{-1/2}$. 
We shall assume that the homogeneous and isotropic background is described by 
the spatially flat, Friedmann-Lema\^itre-Robertson-Walker (FLRW) line element 
characterized by the scale factor~$a$.
As is customary, an overdot shall denote differentiation with respect to the 
cosmic time~$t$, while an overprime shall denote differentiation with respect 
to the conformal time~$\eta$. 
Also, the Hubble parameter $H$ is defined as~$H=\dot{a}/a=\mathcal{H}/a$, where
$\mathcal{H}$ is usually referred to as the conformal Hubble parameter.
Moreover, we shall use e-folds, denoted by~$N$, as another convenient time variable,
particularly when illustrating the evolution of the background and the perturbed 
quantities.

%%%%%%%%%%%%%%%%%%%%%%%%%%%%%%%%%%%%%%%%%%%%%%%%%%%%%%%%%%%%%%%%%%%%%%%%%%%%%%%

\section{Evolution in single field models of inflation}\label{sec:sfm}

In this section, we shall discuss the evolution of the quantum state describing
the curvature perturbation in models of inflation driven by a single scalar field.

%%%%%%%%%%%%%%%%%%%%%%%%%%%%%%%%%%%%%%%%%%%%%%%%%%%%%%%%%%%%%%%%%%%%%%%%%%%%%%%

\subsection{Quantization of the perturbations} \label{sec:quant-Sch}
 
In the case of inflationary models driven by a single, canonical scalar 
field, the scalar perturbations are governed by a single quantity, which
we shall assume to be the Mukhanov-Sasaki variable that is often denoted 
as~$v$~\cite{Mukhanov:1990me,Martin:2003bt,Martin:2004um,Bassett:2005xm,
Sriramkumar:2009kg,Baumann:2008bn,Baumann:2009ds,Sriramkumar:2012mik,
Linde:2014nna,Martin:2015dha}.
At the linear order in perturbation theory, the variable~$v$ is governed
by the following action~\cite{Albrecht:1992kf,Mukhanov:1990me,Polarski:1995jg}: 
\begin{eqnarray}
S[v(\eta,\bm{x})] 
= \f{1}{2}\,\int \d \eta \int\d^3 {\bm x}\, \biggl[v^{\prime 2}- (\pa_i v)^2
-\, 2\,\f{z'}{z}\, v'\, v+\l(\f{z'}{z}\r)^2\, v^2\biggr],\label{eq:soa}
\end{eqnarray}
where $z= a\,\dot{\phi}/H$, with $\phi$ being the inflaton.
In the spatially flat, FLRW background of our interest, we can decompose 
the Mukhanov-Sasaki variable~$v$ in terms of the Fourier modes $v_{\bm k}(\eta)$
as
\begin{equation}
v(\eta,{\bm x})
=\int\f{\d^3{\bm k}}{(2\,\pi)^{3/2}}\,
v_{\bm k}(\eta)\,{\rm e}^{i\,{\bm k}\cdot{\bm x}}.\label{eq:v-fd}
\end{equation}
The condition that the Mukhanov-Sasaki variable~$v$ is a real quantity leads
to the constraint $v_{-\bm{k}}=v_{\bm{k}}^\ast$.
We find that the action~\eqref{eq:soa} can be expressed in
terms of the quantities $v_{\bm k}$ and $v_{\bm k}^\ast$ as follows:
\begin{eqnarray}
S[v_{\bm k}(\eta)]
&=& \int \d \eta\,  \int_{{\mathbb{R}^3}/2}\d^3 {\bm k}\, 
\biggl[v_{\bm k}'\,{v_{\bm k}^{\ast}}'
- \f{z'}{z}\,\l(v_{\bm k}'\, v_{\bm k}^{\ast}
+ {v_{\bm k}^{\ast}}'\, v_{\bm k}\r)\nn\\ 
& &	-\,\l(k^2-\f{z'^2}{z^2}\r)\,v_{\bm k}\,v_{\bm k}^{\ast}\biggr],
\label{eq:a-sfi}
\end{eqnarray}
where $k=\vert{\bm k}\vert$. 
In this action, to deal with only the independent degrees of freedom, we 
have used the constraint $v_{-\bm{k}}=v_{\bm{k}}^\ast$ and have restricted
the integration over ${\bm k}$ to be over half of the Fourier space, i.e. 
$\bm k \in {\mathbb{R}^3}/2$~\cite{Martin:2015qta}. 
(Note that the division of $\mathbb{R}^3$ into two may be done using any plane 
through the origin, which ensures that any ${\bm k}$ and $-{\bm k}$ occur on 
either side of the plane. 
But then, the $\bmk$'s on the plane have to be divided into two using any line 
through the origin, and further the line itself has to be divided into two 
using the origin.)  

Let us now write~\cite{Mukhanov-Winitzky,Martin:2007bw,Martin:2012pea,Battarra:2013cha}
\begin{equation}
v_{\bm k}=\frac{1}{\sqrt{2}}\,\l(v_{\bm k}^{\mathrm{R}}
+i\,v_{\bm k}^{\mathrm{I}}\r),\label{eq:v-ri}
\end{equation}
where, evidently, $v_{\bm k}^{\mathrm{R}}$ and $v_{\bm k}^{\mathrm{I}}$ are 
the real and imaginary parts of $v_{\bm k}$. 
Note that the constraint $v_{-\bm{k}}=v_{\bm{k}}^\ast$ translates to 
the conditions $v_{-\bm k}^{\mathrm{R}}=v_{\bm k}^{\mathrm{R}}$ and 
$v_{-\bm k}^{\mathrm{I}}=-v_{\bm k}^{\mathrm{I}}$.
If we focus on a single Fourier mode, then the action governing
either $v_{\bm k}^{\mathrm{R}}$ or $v_{\bm k}^{\mathrm{I}}$ is given 
by~\cite{Martin:2007bw,Martin:2012pea,Battarra:2013cha}
\begin{equation}
S[v(\eta)]=\f{1}{2}\, \int {\rm d} \eta\, 
\l[{v'}^2- 2\,\f{z'}{z}\,v'\,v- \l(k^2-\f{z'^2}{z^2}\r)\,v^2\r],
\label{eq:a-vf}
\end{equation}
where we have dropped the subscript ${\bm k}$ and the superscripts $\mathrm{R}$
and $\mathrm{I}$ for convenience. 
Upon varying the above action, we obtain the equation of motion describing the 
variable~$v$ to be~\cite{Kiefer:1991xy,Polarski:1995jg,Kiefer:1998qe,Kiefer:2006je,
Martin:2007bw,Kiefer:2008ku,Martin:2012ua,Martin:2012pea}
\begin{equation}
v''+\l(k^2-\f{z''}{z}\r)v=0,\label{eq:de-v}
\end{equation}
which is essentially the equation describing a harmonic oscillator having a 
time-dependent frequency.

We shall work in the Schrodinger picture to understand the nature and evolution 
of the quantum state describing the variable~$v$.
The conjugate momentum associated with the variable~$v$ can be obtained from the 
action~\eqref{eq:a-vf} to be
\begin{equation}
p=v'-\f{z'}{z}\, v,\label{eq:p-def}
\end{equation}
and the corresponding Hamiltonian, say, $H_1$, can be constructed to be
\begin{equation} 
H_1= \f{p^2}{2}+\f{z'}{z}\,p\,v+\f{k^2}{2}\,v^2.\label{eq:H-sfm}
\end{equation}
It is useful to note that the classical, first order, Hamilton's equations of
motion for the system are given by
\begin{equation}
v'=p+\f{z'}{z}\, v,\quad
p'=-\f{z'}{z}\,p-k^2v.\label{eq:cl-H-eqs-sfm}
\end{equation}
We should also point out that these two equations can be combined to arrive at 
the second order equation of motion~\eqref{eq:de-v} governing the variable~$v$.
If we denote the wave function associated with $v$ as $\Psi(v,\eta)$, then
the Schrodinger equation takes the form
\begin{equation}
i\,\f{\partial \Psi}{\pa \eta}
= -\f{1}{2}\,\f{\pa^2 \Psi}{\pa v^2}
- \f{i}{2}\,\f{z'}{z}\,\l(\Psi+2\,v\, \f{\pa \Psi}{\pa v}\r)
+ \f{k^2}{2}\, v^2\,\Psi,\label{eq:se-sfm}
\end{equation}
where the quantity $p\,v$ in the Hamiltonian~\eqref{eq:H-sfm} has been replaced 
by the symmetrized operator $(1/2)\l(\,\hat{p}\,\hat{v}+\,\hat{v}\,\hat{p} \r)$. 
We shall assume the following Gaussian ansatz for the wave 
function~\cite{Polarski:1995jg,Kiefer:1991xy,Kiefer:1998qe,Kiefer:2006je,
Martin:2007bw,Kiefer:2008ku,Martin:2012ua,Martin:2012pea}:
\begin{equation}
\Psi(v,\eta)=\cN(\eta)\,\exp\l[-\f{\Omega(\eta)\,v^2}{2}\r],\label{eq:ga-sfm}
\end{equation}
where $\cN$ and $\Omega$ are functions that depend on time, which need to be 
determined by solving the Schrodinger equation~\eqref{eq:se-sfm}. 
In fact, upon normalizing the wave function, we obtain the relation between 
the functions~$\cN$ and~$\Omega$ to be
\begin{equation}\label{eq:N-or}
\vert \cN\vert=\l(\f{\Omega+\Omega^\ast}{2\,\pi}\r)^{1/4}
=\l(\f{\Omegar}{\pi}\r)^{1/4},
\end{equation}
where $\Omegar$ denotes the real part of~$\Omega$.
Clearly, $\Omegar$ has to be positive for the wave function to be normalizable.
Also, if we can determine $\Omega$, we can obtain $\cN$ using the above relation,
barring an overall phase factor which does not play an important role in our
discussions.
On substituting the ansatz~\eqref{eq:ga-sfm} in the Schrodinger equation~\eqref{eq:se-sfm}, 
we find that the function $\Omega$ satisfies the differential equation~\cite{Martin:2007bw}
\begin{equation}
\Omega'=-\,i\, \Omega^2-2\,\f{z'}{z}\,\Omega+i\, k^2.\label{eq:omega-e-sfm}
\end{equation}
Let us now write the quantity $\Omega$ as 
\begin{equation}
\Omega = -\f{i\,\pi^{f\ast}}{f^\ast}, \label{eq:omega-sfm}
\end{equation} 
where $\pi^f$ can be considered to be the momentum conjugate to~$f$,
and is given by 
\begin{equation}
\pi^f={f}'-\f{z'}{z}\,f.\label{Bog_mom}
\end{equation}
When we do so, we find that Eq.~\eqref{eq:omega-e-sfm} leads to a second order 
differential equation for~$f^\ast$, which is the same as the one satisfied 
by the Mukhanov-Sasaki variable~$v$ [cf. Eq.~\eqref{eq:de-v}].
Note that, since the coefficients of the equation governing~$v$ are real, 
$f$ too satisfies the same equation.
In other words, we can construct the wave function describing the quantum 
system in terms of the classical solutions.
Usually, the so-called Bunch-Davies initial conditions are imposed on the 
Mukhanov-Sasaki variable~$v$ when the modes are well inside the Hubble 
radius, i.e. when $k \gg (a\,H)$ or, more precisely, when $k \gg
\sqrt{z''/z}$~\cite{Mukhanov:1990me,Martin:2003bt,Martin:2004um,Bassett:2005xm,
Sriramkumar:2009kg,Baumann:2008bn,Baumann:2009ds,Sriramkumar:2012mik,Linde:2014nna,
Martin:2015dha}.
If we choose the same initial conditions for the quantity~$f$ and its 
conjugate momentum~$\pi^f$, viz. that
\begin{equation}
f(\ei) = \f{1}{\sqrt{2\,k}}\,\mathrm{e}^{-i\,k\,\ei},\quad
\pi^f(\ei) = -\f{1}{\sqrt{2\,k}}\,\l[i\,k+\f{z'(\ei)}{z(\ei)}\r]\,
\mathrm{e}^{-i\,k\,\ei},\label{eq:bd-ic-sfm}
\end{equation}
where $\ei$ is the time when the modes are in the sub-Hubble domain, then 
these conditions correspond to evolving the wave function $\Psi$ from the 
Bunch-Davies vacuum.
It is useful to note that the Wronskian $\mathcal{W}=\l(f\,\pi^{f\ast}
-f^{\ast}\,\pi^f\r)$ associated with the differential equation~\eqref{eq:de-v} 
is a constant, and the above Bunch-Davies initial conditions lead to $\mathcal{W}=i$.

%%%%%%%%%%%%%%%%%%%%%%%%%%%%%%%%%%%%%%%%%%%%%%%%%%%%%%%%%%%%%%%%%%%%%%%%%%%%%%%

\subsection{Wigner function and squeezing parameters}

In the case of single field inflationary models, we shall utilize two 
standard measures to examine the evolution of the quantum state of the
system, viz. the Wigner function \cite{Polarski:1995jg,Kiefer:1998jk,
Martin:2007bw,Kiefer:2008ku,Martin:2012pea,Martin:2015qta,Hollowood:2017bil} 
and the squeezing parameters~\cite{Grishchuk:1993ds,Albrecht:1992kf,
Polarski:1995jg,Kiefer:1998jk,Martin:2007bw,Kiefer:2008ku,Martin:2012pea,
Martin:2015qta,Hollowood:2017bil}.
The Wigner function is a quasi-probability distribution that helps us visualize 
the dynamics of a quantum system in phase space (in this context, see, for  
instance, Refs.~\cite{Hillery:1983ms,case2008wigner}).
For a system with a single degree of freedom, say, $x$, given a wave function 
$\Psi(x,t)$, the corresponding Wigner function $W(x,p,t)$ is defined by the 
following integral:
\begin{equation}
W(x,p,t) = \f{1}{\pi}\, \int_{-\infty}^\infty \d z\, \Psi(x-z,t)\, 
\Psi^\ast(x+z,t)\, {\mathrm{e}}^{2\, i\, p\, z},\label{eq:wf-sfm-d}
\end{equation}
where $p$ is the momentum conjugate to the variable~$x$.
While it is straightforward to show that the Wigner function $W(x,p,t)$
is always real, one finds that it does {\it not}\/ prove to be positive 
definite for all states of a system.
It is for this reason that the Wigner function is often referred to as a 
`quasi'-probability distribution.
However, for the Gaussian wave function of our interest here, it turns out
to be positive.

We can arrive at the Wigner function in the phase space $v$-$p$ corresponding 
to the wave function~\eqref{eq:ga-sfm} by substituting it in the 
expression~\eqref{eq:wf-sfm-d} and evaluating the Gaussian integral involved.
We obtain that 
\begin{equation}
W(v,p,\eta)= \frac{1}{\pi}\, 
{\mathrm{exp}} \l[-\Omegar\, v^2 
- \f{1}{\Omegar}\, \l(p+\Omegai \, v\r)^2\r],\label{eq:wf-sfm}
\end{equation}  
where $\Omegai$ represents the imaginary part of~$\Omega$.
Evidently, the evolution of the Wigner function is determined by the behavior 
of the argument in the exponential.
Often, the evolution of the Wigner function is tracked by the behavior of the
ellipse in the phase space $v$-$p$ defined through the relation~\cite{Battarra:2013cha,Hollowood:2017bil}
\begin{equation}
\Omegar\, v^2 + \f{1}{\Omegar}\, \l(p+\Omegai \, v\r)^2 =1.\label{eq:we-sfm}  
\end{equation}
Note that the time evolution of the Wigner ellipse is determined by the 
behavior of the quantities $\Omegar$ and $\Omegai$. 
If we now make use of the definition~\eqref{eq:omega-sfm} of $\Omega$, we 
obtain that 
\begin{subequations}
\begin{eqnarray}
\Omegar=\f{1}{2}\,\l(\Omega+ \Omega^{\ast}\r)
&=&-i\,\f{\l(f\,\pi^{f\ast}-f^\ast\,\pi^f\r)}{2\,\vert f\vert^2}
=\f{1}{2\,\vert f\vert^2},\label{eq:or-fv}\\
\Omegai=\f{1}{2\,i}\, \l(\Omega- \Omega^{\ast}\r)
&=& -\f{\l(f\,\pi^{f\ast} + f^\ast\,\pi^f\r)}{2\,\vert f\vert^2},
\end{eqnarray}
\end{subequations}
where, in the last equality of the first equation, we have made use of the 
fact that the Wronskian is given by $\mathcal{W}=i$.
In other words, the behavior of the Wigner ellipse can be followed using
the solution to the classical equation of motion.

As is well known, the time evolution essentially leads to rotation and squeezing 
of the quantum state, which will be reflected in the behavior of the Wigner 
function we described above.
In order to relate to the squeezing parameters describing the state of
the system, let us rewrite the Wigner function~\eqref{eq:wf-sfm} in terms 
of the so-called covariance matrix.
To do so, let us first introduce canonical variables of the same dimension, viz. 
\begin{equation}
\tilde{v}= \sqrt{k}\,v,\quad \tilde{p}= \f{1}{\sqrt{k}}\, p.\label{eq:dcv}
\end{equation}	
Upon defining the vector $\Z$ such that its transpose is $\Z^T= (\tilde{v},\tilde{p})$,
we find that the Wigner function can be expressed as~\cite{weedbrook2012gaussian}
\begin{equation}
W(\tilde{v},\tilde{p},\eta)
= \f{1}{2\,\pi\,\sqrt{\mathrm{det}~V}}\;
\mathrm{exp}\l(-\f{\Z^T\, V^{-1}\,\Z}{2}\r),\label{eq:wf-sfm-cm}
\end{equation}
where $V$ is the covariance matrix. 
In terms of the operator vector $\hat{\Z}$ defined by $\hat{\Z}^T= (\hat{\tilde{v}},
\hat{\tilde{p}})$, the covariance matrix has the elements 
\begin{eqnarray}
V_{ij}&=& \f{1}{2}\,\l[\l\langle \l(\hat{\Z}_i -\l\langle \hat{\Z}_i \r\rangle\r)\,
\l(\hat{\Z}_j -\l\langle \hat{\Z}_j \r\rangle\r)\r\rangle
+\{i\Leftrightarrow j\}\r]\nn\\ 
&=& \f{1}{2}\,\langle\hat{\Z}_i\,  \hat{\Z}_j+\hat{\Z}_j\,  \hat{\Z}_i \rangle
-\langle \hat{\Z}_i\, \rangle\, \langle \hat{\Z}_j \rangle
= \f{1}{2}\,\langle \hat{\Z}_i\,  \hat{\Z}_j+\hat{\Z}_j\, \hat{\Z}_i \rangle
\label{eq:cm-sfm-def}
\end{eqnarray} 
and, in the last equality, we have specialized to our case wherein $\langle \hat{\Z}_i 
\rangle=0$.

The evolution of the Wigner function in phase space can be succinctly captured 
by introducing the parameters $\l(r,\vartheta,\varphi\r)$ which capture the 
geometry of the Wigner ellipse, specifically its shape since it can be proved 
that its area remains constant for a pure state (in this regard, see 
Refs.~\cite{narcowich1990geometry,Koksma:2010zi,Martin:2021znx}; also see
App.~\ref{app:gwe}). 
The parameters $\l(r,\vartheta,\varphi\r)$ are called the squeezing amplitude, 
rotation angle and squeezing angle, respectively.
In terms of these shape parameters, the covariance matrix has the 
form~\cite{cariolaro2015quantum,weedbrook2012gaussian,Martin:2021znx}
\begin{equation}
V = \frac{1}{2}\,\begin{bmatrix}
\cosh\, (2\,r) +  \sinh\, (2\,r)\,\cos\, (2\, \varphi)  
& \sinh\, (2\,r)\,\sin\, (2\, \varphi)\\
\sinh\, (2\,r)\,\sin\, (2\, \varphi)   
& \cosh\, (2\,r) - \sinh\, (2\,r) \,\cos\, (2\, \varphi)
\end{bmatrix}.\label{eq:cm-sfm}
\end{equation}
(We should mention that, in App.~\ref{app:conventions-sq}, we have briefly
commented on the conventions we have adopted to write the above covariance 
matrix.)
Evidently, in the case of the single field model of our interest, the components
of the covariance matrix are given by 
\begin{equation}
V = \begin{bmatrix}
\langle \hat{\tilde{v}}^2\rangle & \f{1}{2}\,\langle
\hat{\tilde{v}}\,\hat{\tilde{p}}+\hat{\tilde{p}}\,\hat{\tilde{v}}\rangle\\
\f{1}{2}\,\langle \hat{\tilde{v}}\,\hat{\tilde{p}}
+\hat{\tilde{p}}\,\hat{\tilde{v}}\rangle & \langle \hat{\tilde{p}}^2\rangle 
\end{bmatrix}.
\end{equation}
Upon using the wave function~\eqref{eq:ga-sfm}, the definition~\eqref{eq:omega-sfm}
of $\Omega$ and the expression~\eqref{eq:cm-sfm} for the covariance matrix in terms 
of the squeezing parameters, we obtain that
\begin{subequations}\label{eq:corr}
\begin{eqnarray}
\langle\hat{v}^2\rangle 
&=& \vert f\vert^2=\f{1}{2 \, k}\,\l[\cosh\,(2\, r) 
+\sinh\,(2 \, r)\, \cos\,(2\, \varphi)\r],\label{eq:v2-sfm}\\
\langle\hat{p}^2\rangle 
&=&\vert \pi^f\vert^2=\f{k}{2}\,\l[\cosh\,(2\,r)
-\sinh\,(2\, r)\, \cos\,(2\,\varphi)\r],\label{psq_pure}\\
\f{1}{2}\,\langle \hat{v}\, \hat{p}+ \hat{p}\, \hat{v}\rangle
&=&\f{1}{2}\,\l(f\,\pi^{f^\ast} + f^\ast\,\pi^f\r)\, 
=\frac{1}{2}\,\sinh\,(2\, r)\, \sin\,(2\,\varphi).\label{vp_pure}
\end{eqnarray}
\end{subequations}
To be precise, instead of the above variances, we will have relations 
such as $\langle\hat{v}_\bmk\,\hat{v}_\bmp\rangle= \vert f\vert^2\,
\delta^{(3)}(\bmk-\bmp)$ (see 
Refs.~\cite{Martin:2012pea,Martin:2015qta}).
For convenience, we have dropped the delta functions, and we should
caution that this will modify the dimensions of the quantities 
involved.

We can invert the above relations to express the squeezing parameters as
\begin{equation}
\cosh\, (2\,r) =k\, \vert f\vert^2 +\f{\vert \pi^f\vert^2}{k},\quad
\cos\, (2\, \varphi) =\f{1}{\sinh\,(2 \, r)}\,
\l(k\,\vert f\vert^2-\f{\vert \pi^f\vert^2}{k}\r).\qquad\label{eq:rphi-g}
\end{equation}
Thus, if we know the solution to the classical equation governing $v$ [see
Eq.~\eqref{eq:de-v}], we can determine the squeezing amplitude~$r$ and the 
squeezing angle $\varphi$ from the amplitudes of the mode function~$f$ and its 
conjugate momentum~$\pi^f$.
In the following subsection, we shall discuss the behavior of the Wigner 
function and the squeezing parameters in specific examples of single field 
models permitting slow roll and ultra slow roll inflation.
		
%%%%%%%%%%%%%%%%%%%%%%%%%%%%%%%%%%%%%%%%%%%%%%%%%%%%%%%%%%%%%%%%%%%%%%%%%%%%%%%

\subsection{Behavior in slow roll and ultra slow roll inflation}

One of the primary quantities of observational interest in the inflationary
scenario is the scalar power spectrum $\ps(k)$, where the subscript $\cR$ 
refers to the curvature perturbation.
It is defined in terms of the Mukhanov-Sasaki variable as 
follows~\cite{Mukhanov:1990me,Martin:2003bt,Martin:2004um,Bassett:2005xm,
Sriramkumar:2009kg,Baumann:2008bn,Baumann:2009ds,Sriramkumar:2012mik,Linde:2014nna,
Martin:2015dha}:
\begin{equation}
\ps(k) =\f{k^3}{2\,\pi^2}\,\f{\langle \hat{v}^2\rangle}{z^2},
\label{eq:ps-d}
\end{equation}
with the expectation value to be evaluated in the Bunch-Davies vacuum.
On utilizing the result~\eqref{eq:v2-sfm} for $\langle \hat{v}^2\rangle$,
we find that we can express the power spectrum in terms of the solution~$f$ 
and the squeezing parameters $r$ and $\varphi$ as follows:
\begin{equation}
\ps(k)=\f{k^2}{4\,\pi^2\, z^2}\,\l[\cosh\,(2\, r)
+\sinh\,(2 \, r)\, \cos\,(2\, \varphi)\r]
=\f{k^3}{2\,\pi^2}\, \f{\vert f\vert^2}{z^2}.
\end{equation}
Note that the quantity $z$ is determined by the evolution of the background 
in a given model of inflation.
With~$z$ at hand, we can solve Eq.~\eqref{eq:de-v} with suitable initial 
conditions [see Eq.~\eqref{eq:bd-ic-sfm}] to determine the quantity~$f$ and 
arrive at the power spectrum.
In this subsection, we shall compare the behavior of the squeezing 
amplitude~$r$ in inflationary models leading to slow roll and ultra slow
roll inflation.
We shall also briefly comment on the behavior of the Wigner function in these 
cases.

The slow roll model we shall consider is the popular Starobinsky model described 
by the potential~\cite{Starobinsky:1979ty,Starobinsky:1980te}
\begin{equation}
V(\phi)=V_0\,\l[1-\mathrm{exp}\l(-\sqrt{\f{2}{3}}\,\f{\phi}{\Mpl}\r)\r]^2.
\label{eq:sm1}
\end{equation}
The COBE normalization of the scalar power spectrum on the CMB scales fixes the 
overall amplitude of the potential to be $V_0 = 1.43 \times 10^{-10}\, \Mpl^4$.
We shall evaluate the background and the perturbations numerically using 
well established procedures (in this context, see, for instance, 
Refs.~\cite{Hazra:2012yn,Ragavendra:2020old,Ragavendra:2020sop}).
As far as the background is concerned, we have chosen the initial values of 
the field and the first slow roll parameter to be $\phi_{\mathrm{i}} =5.6\,
\Mpl$ and $\e1i = 1.453\times10^{-4}$, respectively.
For the above-mentioned value of $V_0$, these initial conditions lead to 
about $69.5$ e-folds before inflation ends.  

As we had mentioned earlier, many models of inflation which lead to an epoch of 
ultra slow roll have been considered in the literature.
In contrast to models that lead to only slow roll inflation, potentials that 
permit departures from slow roll (such as an epoch of ultra slow roll) require 
features in the potential.  
It is the features in the potential that are responsible for the features in 
the scalar power spectrum.
Therefore, to a certain extent, potentials have to be designed in order to 
generate specific background dynamics and hence the desired features in the 
scalar power spectrum (for a discussion on, say, reverse engineering required 
potentials, see Refs.~\cite{Ragavendra:2020sop,Balaji:2022zur,Franciolini:2022pav}).
In particular, the potentials can require a certain level of fine tuning if they 
have to enhance scalar power on small scales and simultaneously generate scalar 
and tensor power spectra that are consistent with the constraints from the CMB 
over large scales (in this context, see, for instance, Ref.~\cite{Iacconi:2021ltm}).
Interestingly, as we had pointed out, it has been found that potentials that 
contain a point of inflection permit an epoch of ultra slow roll, which 
inevitably generates higher power on smaller scales.
Among the various models that have been examined, we shall consider the model 
described by the potential~\cite{Dalianis:2018frf,Ragavendra:2020sop}
\begin{equation}
V(\phi) = V_0\,\biggl\{\mathrm{tanh}\l(\f{\phi}{\sqrt{6}\,\Mpl}\r)
+ A\,\sin\l[\f{1}{f_\phi}\,
\mathrm{tanh}\l(\f{\phi}{\sqrt{6}\,\Mpl}\r)\r]\biggr\}^2.\label{eq:usr}
\end{equation}
The potential contains a point of inflection, and we shall choose to 
work with the following values of the parameters involved:~$V_0 = 2
\times10^{-10}\,\Mpl^4$, $A = 0.130383$ and $f_\phi = 0.129576$.
For these values of the parameters, the point of inflection in the 
potential is located at $\phi_0 = 1.05\,\Mpl$~\cite{Ragavendra:2020sop}.
Also, if we choose the initial value of the field to be $\phii=6.1\,\Mpl$, 
with $\e1i=10^{-4}$, we obtain about $66$~e-folds of inflation in the model.
Moreover, we shall assume that the pivot scale exits the Hubble radius about
$56.2$~e-folds prior to the termination of inflation.

\begin{figure}[!t]
\centering
\includegraphics[width=0.74\linewidth]{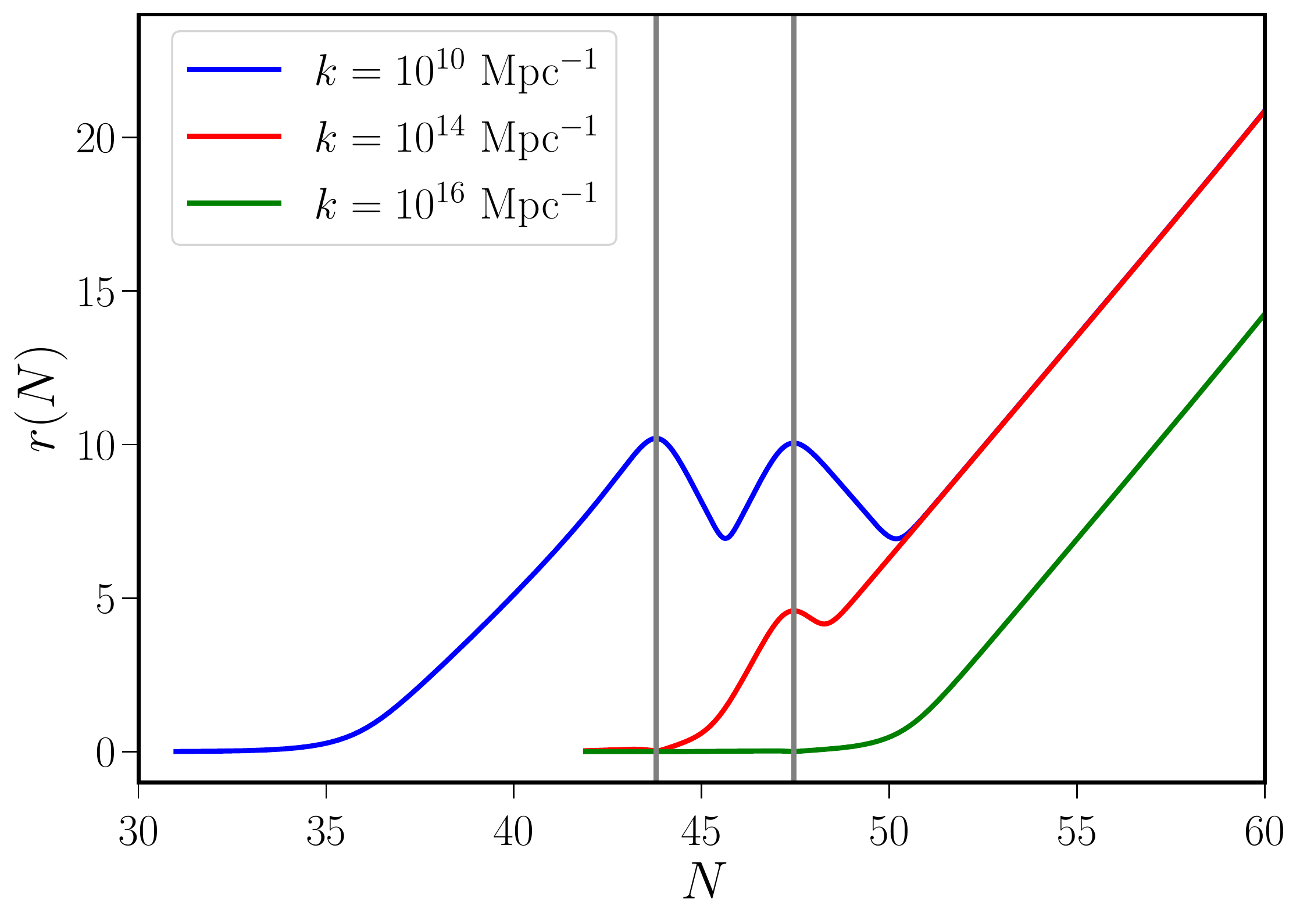}
\vskip 5pt
\includegraphics[width=0.74\linewidth]{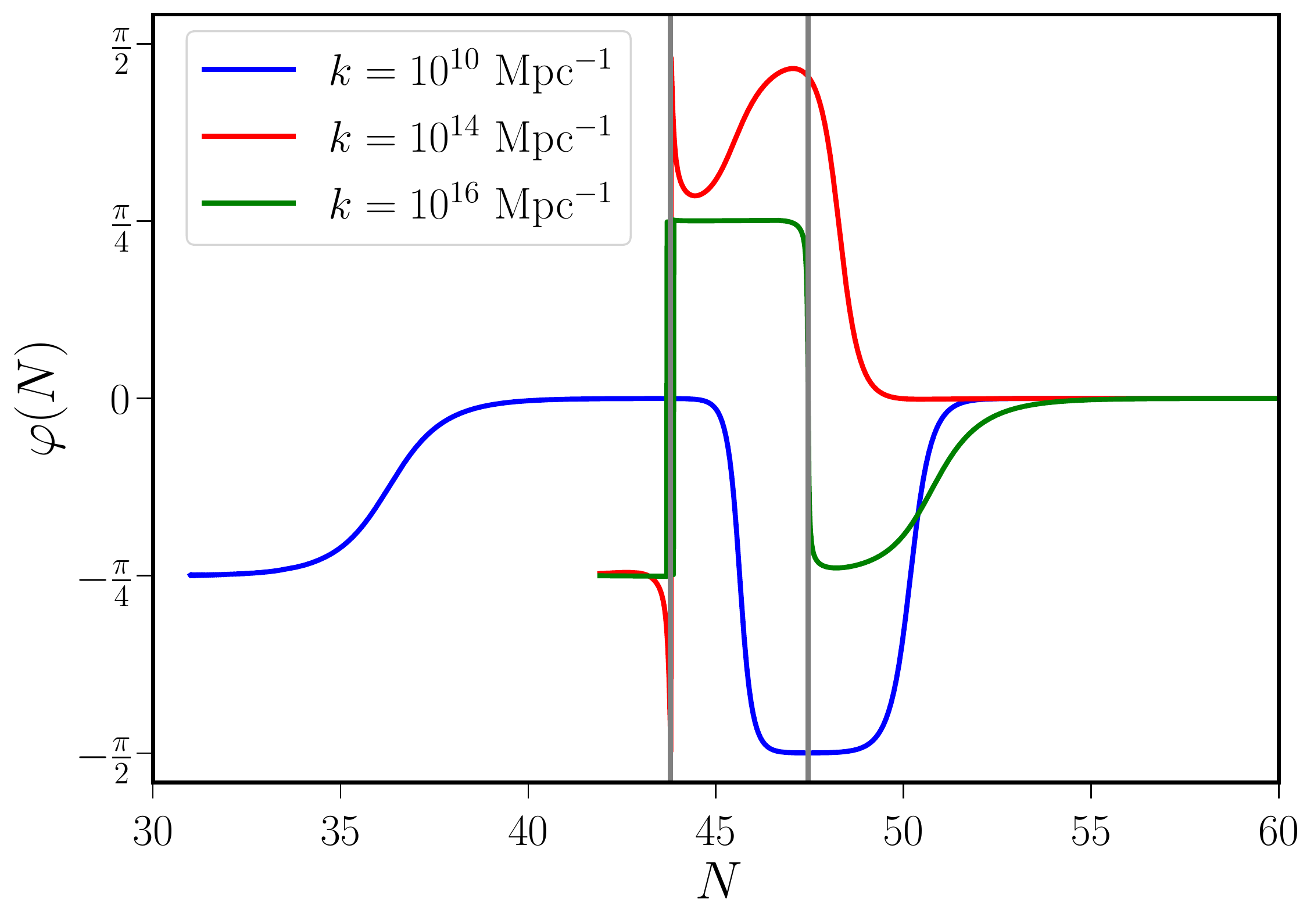}
\caption{The evolution of the squeezing amplitude $r$ and the squeezing 
angle $\varphi$ have been plotted (on top and at the bottom, respectively) 
against the number of e-folds $N$ in the ultra slow roll model of interest. 
In the figures, we have demarcated the period of ultra slow roll (by gray 
vertical lines).
We have plotted the evolution of the squeezing parameters for modes with the 
wave numbers $k=(10^{10}, 10^{14}, 10^{16})\,\mathrm{Mpc}^{-1}$ (in blue, red
and green).
While the first and the last of these modes exit the Hubble radius before
and after the ultra slow roll phase, the second mode leaves the Hubble radius
during the period of ultra slow roll.
As we have pointed out, in slow roll inflation, the squeezing parameter $r$ 
behaves as $r\propto N$ after the modes leave the Hubble radius (in a manner
similar to the case of the third wave number, plotted in green).
We find that the epoch of ultra slow roll modifies this behavior for modes
that leave the Hubble radius immediately prior to or during the ultra slow
roll phase (as in the cases for the wave numbers plotted in blue and red).
Similarly, in a slow roll scenario, the squeezing parameter $\varphi$ 
approaches zero monotonically once the modes are on super-Hubble scales.
In contrast, as should be evident from the above figure, the epoch of ultra 
slow roll induces non-trivial behavior.
However, note that the parameter $\varphi$ eventually approaches zero for
all the wave numbers (barring those that leave the Hubble radius very close
to the end of inflation), indicating that the Wigner ellipse orients itself 
along the $v$-axis towards the end of inflation.}\label{fig:r-phi-usr}
\end{figure}

Let us first discuss the time evolution of the squeezing parameters~$r$ 
and~$\varphi$ in these models.
In slow roll inflation, using the solutions for the modes~$f$ in the de Sitter
approximation, it is easy to establish that, while on sub-Hubble scales $r(N) 
\simeq 0$, on super-Hubble scales, $r(N)\propto N$.
Also, it can be shown that the squeezing parameter $\varphi$ vanishes at 
late times for modes with wave numbers $k\ll \ke$, where $\ke$ denotes the 
wave number that leaves the Hubble radius at the end of inflation.
One finds that the parameter $\varphi$ indicates the angle between the 
$v$-axis and the major axis of the Wigner ellipse in phase space 
[see Eq.~\eqref{eq:we-sfm}; in this regard, also see App.~\ref{app:gwe}].
Hence, the fact that the parameter $\varphi$ goes to zero towards the end of
inflation implies that the Wigner ellipse eventually orients itself along the 
$v$-axis for modes with wave numbers such that $k\ll \ke$.
In models leading to an epoch of ultra slow roll, we find that the squeezing 
parameters evolve non-trivially during the phase of ultra slow roll.
In Fig.~\ref{fig:r-phi-usr}, we have plotted the time evolution of the 
squeezing parameters~$r$ and~$\varphi$ obtained numerically in the model 
permitting ultra slow roll inflation.
%%%%%%%%%%%%%%%%%%%%%%%%%%%%%%%%%%%%%%%%%%%%%%%%%%%%%%%%%%%%%%%%%%%%%%%%%%%%%%%
%%%%%%%%%%%%%%%%%%%%%%%%%%%%%%%%%%%%%%%%%%%%%%%%%%%%%%%%%%%%%%%%%%%%%%%%%%%%%%%
We have plotted the evolution for modes with three different wave numbers that 
leave the Hubble radius just prior to, during and after the period of ultra 
slow roll.
Note that, despite the non-trivial evolution during the intermediate period, 
the parameter $\varphi$ eventually approaches zero at adequately late times.  
We shall soon discuss the impact of the ultra slow roll phase on the squeezing
amplitude~$r$.

\begin{figure}[!t]
\centering
\includegraphics[width=0.75\linewidth]{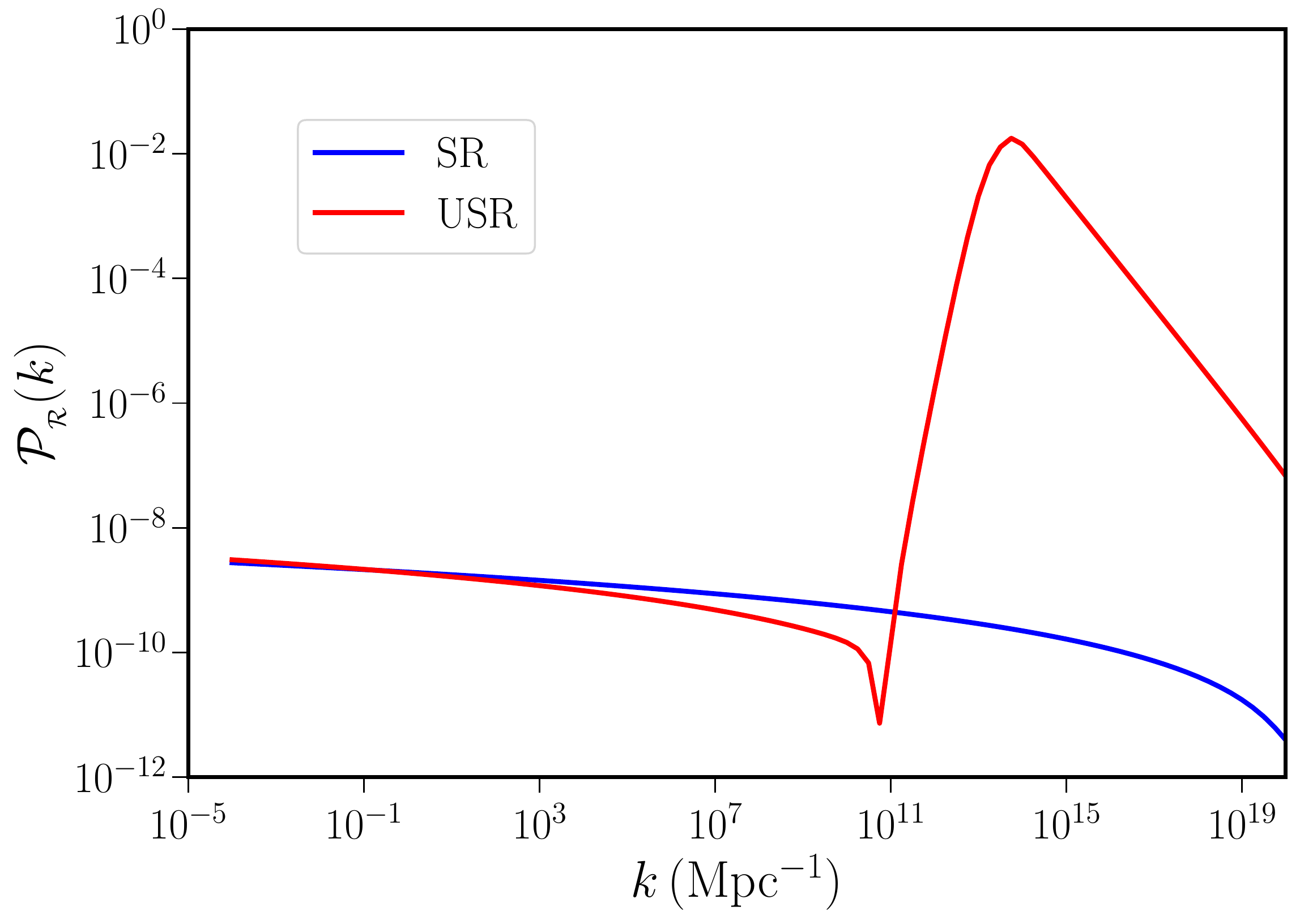}
\vskip 5pt
\includegraphics[width=0.75\linewidth]{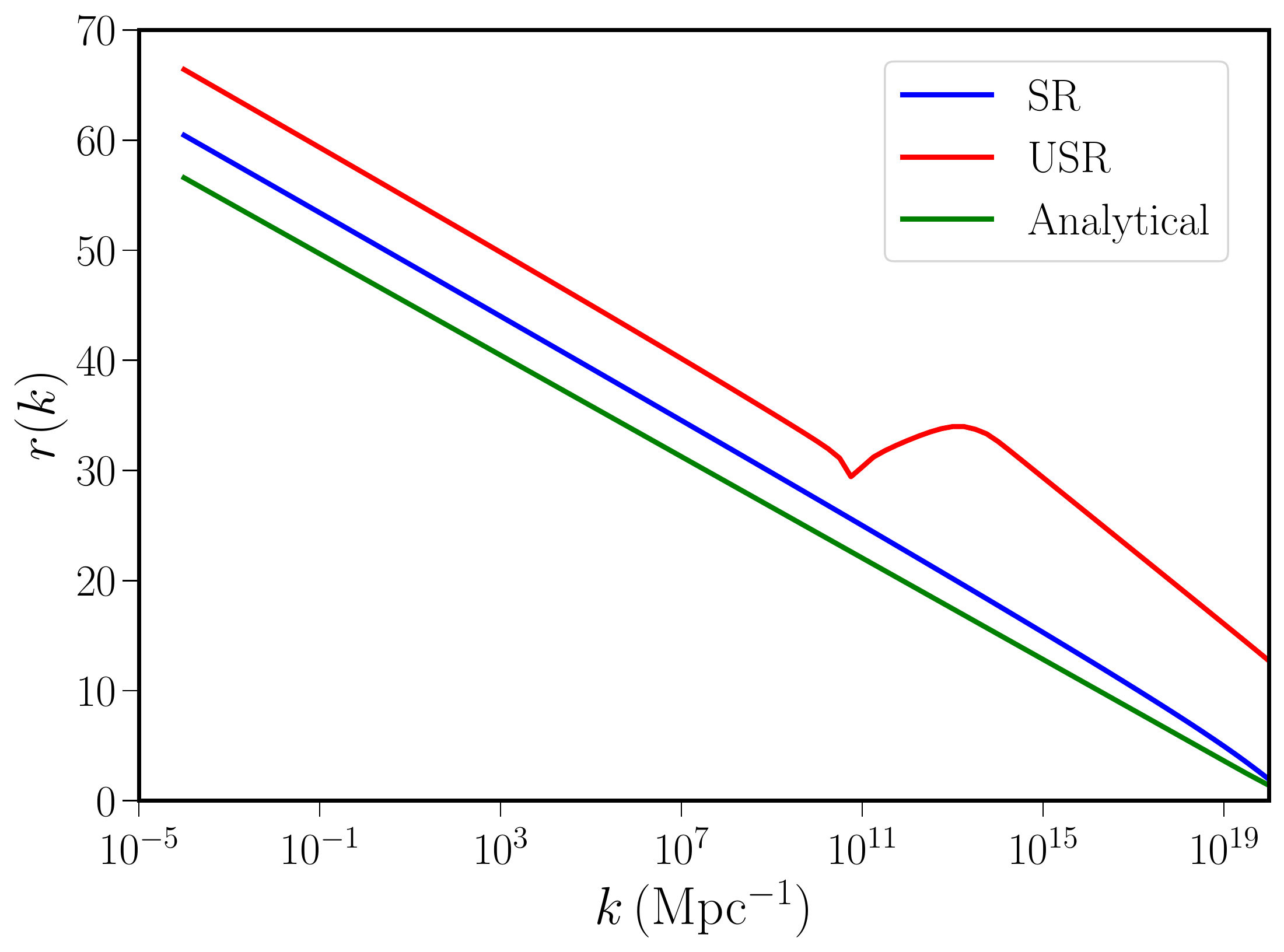}
\caption{The scalar power spectrum $\ps(k)$ (on top) and the squeezing 
amplitude~$r(k)$ (at the bottom) evaluated towards the end of inflation have
been plotted as a function of the wave number for the two single field inflationary 
models of our interest, viz. the popular model due to Starobinsky that admits 
only slow roll inflation (SR, in blue) and the other model that leads to a brief 
period of ultra slow roll (USR, in red).
We have evaluated these quantities numerically and, in the case of slow roll
inflation, we have also plotted the analytical results for the squeezing amplitude,
arrived at in the de Sitter approximation (in green, at the bottom).
In slow roll inflation, since $r(N)\propto N$ on super-Hubble scales, the value 
of the squeezing parameter $r(k)$ associated with a given wave number roughly 
corresponds to the number of e-folds between the time the mode leaves the Hubble 
radius and the end of inflation.
Note that, in the case of ultra slow roll inflation, the squeezing amplitude 
corresponding to the small scale modes which exhibit higher power is enhanced 
to a certain extent when compared to the slow roll case.}\label{fig:ps-r-sfm}
\end{figure}

In Fig.~\ref{fig:ps-r-sfm}, we have plotted the scalar power spectra~$\ps(k)$
and the squeezing amplitude~$r(k)$, obtained numerically, in the above two 
models for a wide range of wave numbers.
%%%%%%%%%%%%%%%%%%%%%%%%%%%%%%%%%%%%%%%%%%%%%%%%%%%%%%%%%%%%%%%%%%%%%%%%%%%%%%%
%%%%%%%%%%%%%%%%%%%%%%%%%%%%%%%%%%%%%%%%%%%%%%%%%%%%%%%%%%%%%%%%%%%%%%%%%%%%%%%
As expected, the power spectrum is nearly scale invariant in the slow roll 
model, while it exhibits a peak with higher power at small scales in the 
model permitting ultra slow roll inflation.
It is the peak with the enhanced power that proves to be responsible for
producing a significant number of PBHs in such models~\cite{Garcia-Bellido:2017mdw,
Ballesteros:2017fsr,Germani:2017bcs,Kannike:2017bxn,Dalianis:2018frf,Ragavendra:2020sop}.
Since $r(N)\simeq 0$ on sub-Hubble scales and $r(N) \propto N$ on super-Hubble 
scales in the slow roll model, towards the end of inflation, we can expect that
$r(k)\simeq 60$ for the largest scales, which is clearly reflected in the results 
plotted in the figure. 
In fact, using the standard modes in the de Sitter approximation, one can show
that, in slow roll, 
\begin{equation}
\mathrm{cosh}\,(2\,r) \simeq 1+\f{\ke^2}{2\,k^2},\quad
\mathrm{tan}\,(2\,\varphi) \simeq -\f{2\,k}{\ke},
\end{equation}
where, as we mentioned, $\ke$ denotes the wave number that leaves the Hubble 
radius at the end of inflation.
In the figure, we have also plotted the above analytical result for $r(k)$.
Clearly, the analytical estimate captures the more accurate numerical results 
fairly well in the slow roll case.
Interestingly, in the model involving ultra slow roll inflation, we find that
$r(k)$ is boosted to a certain extent over wave numbers which exhibit enhanced
scalar power.
This can evidently be attributed to the non-trivial evolution of the modes 
during the epoch of ultra slow roll.

%%%%%%%%%%%%%%%%%%%%%%%%%%%%%%%%%%%%%%%%%%%%%%%%%%%%%%%%%%%%%%%%%%%%%%%%%%%%%%%

\section{Evolution in two field inflationary models}\label{sec:tfm} 

Having discussed the evolution of the quantum state of the curvature perturbations
in single field models of inflation, let us now turn to discuss the behavior in
two field models.
In this section, we shall examine the evolution of the Wigner function, the 
squeezing parameters and the entanglement entropy describing the curvature 
perturbation in a model permitting slow roll inflation and a model that leads 
to enhanced power on small scales, achieved through a turn in the field space.
We shall also show that the entanglement entropy for the system of our interest
is equivalent to the quantum discord.

%%%%%%%%%%%%%%%%%%%%%%%%%%%%%%%%%%%%%%%%%%%%%%%%%%%%%%%%%%%%%%%%%%%%%%%%%%%%%%%

\subsection{Action and equations governing the background}\label{sec:models}

The two field models of our interest consist of a canonical scalar field~$\phi$ 
and a non-canonical scalar field~$\chi$.
The system is governed by the following action (see, for instance,
Refs.~\cite{DiMarco:2002eb,Lalak:2007vi,Raveendran:2018yyh,Braglia:2020fms,
Braglia:2020eai,Palma:2020ejf}):
\begin{equation} 
S[\phi,\chi]=\int\d^4 x\sqrt{-g}\,\biggl[-\f{1}{2}\,\pa_\mu\phi\,\pa^\mu\phi
-\,\f{\mathrm{e}^{2\, b(\phi)}}{2}\,\pa_\mu\chi\,\pa^\mu\chi
-V(\phi,\chi)\biggr],\label{eq:a-tfm-f}
\end{equation}
where, evidently, it is the exponential term involving the function~$b(\phi)$
that makes the scalar field $\chi$ non-canonical.
The equations of motion governing the homogeneous scalar fields $\phi$ and
$\chi$ can be obtained to be
\begin{subequations}
\begin{eqnarray}
\ddot{\phi}+3\,H\,\dot{\phi}
+ V_{\phi}&=&b_{\phi}\,\mathrm{e}^{2\,b}\,\dot{\chi}^2,\label{eq:bg-phi}\\
\ddot{\chi}+(3\,H+2\,b_{\phi}\,\dot{\phi})\,\dot{\chi}
+\mathrm{e}^{-2\,b}\,V_{\chi}&=&0,\label{eq:bg-chi}
\end{eqnarray}
\end{subequations}
where the subscripts $\phi$ and $\chi$ denote differentiation of the potential 
$V(\phi,\chi)$ and the function $b(\phi)$ with respect to the fields. 
We should also mention that the Hubble parameter and its time derivative 
are described by the equations
\begin{eqnarray}
H^2 &=& \f{1}{3\,\Mpl^2}\, \l(\f{\dot{\phi}^2}{2}+\mathrm{e}^{2\,b}\,
\f{\dot{\chi}^2}{2}+V\r),\\
\dot{H}&=&-\f{1}{2\,\Mpl^2}\,\l(\dot{\phi}^2+\mathrm{e}^{2\,b}\,\dot{\chi}^2\r).
\end{eqnarray}

%%%%%%%%%%%%%%%%%%%%%%%%%%%%%%%%%%%%%%%%%%%%%%%%%%%%%%%%%%%%%%%%%%%%%%%%%%%%%%%

\subsection{Action describing the perturbations and quantization}

Recall that, in the case of inflation driven by a single scalar field, the 
perturbations are characterized by the Mukhanov-Sasaki variable associated
with the curvature perturbation.
In inflationary models involving two fields, it is well known that, apart 
from the curvature perturbations, isocurvature perturbations also arise (in 
this context, see the reviews~\cite{Bassett:2005xm,Malik:2008im}).
Let $v_\sigma$ and~$v_s$ be the Mukhanov-Sasaki variables associated with 
the curvature and the isocurvature perturbations (or, equivalently, the
adiabatic and entropic perturbations), respectively.
In order to arrive at the action describing the variables~$v_{\sigma}$ 
and~$v_s$, it is convenient to introduce the quantities~\cite{DiMarco:2002eb,
Lalak:2007vi,Raveendran:2018yyh,Braglia:2020fms,Braglia:2020eai,Palma:2020ejf}
\begin{equation}
\mathrm{cos}\,\theta=\f{\dot{\phi}}{\dot{\sigma}},\quad
\mathrm{sin}\,\theta=\mathrm{e}^{b}\,\f{\dot{\chi}}{\dot{\sigma}},\quad
\dot{\sigma}^2= \dot{\phi}^2+\mathrm{e}^{2\,b}\,\dot{\chi}^2
\end{equation}
and define the quantities $V_{\s}$ and $V_s$ to be
\begin{equation}
V_\sigma = \mathrm{cos}\,\theta\,V_\phi
+\mathrm{sin}\,\theta\,{\mathrm{e}}^{-b}\,V_\chi,\quad
V_s =-\mathrm{sin}\,\theta\,V_\phi
+\mathrm{cos}\,\theta\,{\mathrm{e}}^{-b}\,V_\chi.\label{eq:Vsigma-Vs}
\end{equation}
It can be shown that, at the linear order in perturbation theory, the action
governing the complete system is given by~\cite{DiMarco:2002eb,Lalak:2007vi,
Raveendran:2018yyh,Braglia:2020fms,Braglia:2020eai,Palma:2020ejf}
\begin{eqnarray}
S[v_{\s}(\eta,\bm{x}),v_{s}(\eta,\bm{x})] 
&=& \f{1}{2}\,\int \d \eta \int\d^3 {\bm x}\, \Biggl\{v_\sigma^{\prime 2}
- (\pa_i v_\sigma)^2+ v_s^{\prime 2}- (\pa_i v_s)^2\nn\\
& &-\, 2\,\f{z'}{z}\, v_\sigma'\, v_\sigma
- 2\,\f{a'}{a}\,v_s'\, v_s-  2\, \xi\, v_{\sigma}'\, v_{s}
+ 2\, \xi\, \f{z'}{z}\, v_{\sigma}\, v_{s}\nn\\ 
& &+\, \l(\f{z'}{z}\r)^2\, v_{\sigma}^2
+ \l[\l(\f{a'}{a}\r)^2-\mu_s^2\,a^2\r]\, v_{s}^2
\Biggr\},
\label{eq:soa-tfm}
\end{eqnarray}
where $z = a\,\dot{\sigma}/H$ and $\xi= -2\,a\, V_s/\dot{\sigma}$. 
Moreover, the quantity $\mu_s^2$ has been defined to be
\begin{equation}
\mu _{s} ^2  
=  V_{ss} -\l(\frac{V_s}{\dot{\sigma}}\r)^2
+\mathrm{cos}\,\theta\,\l(1+\mathrm{sin}^2\theta\r)\, b_\phi\,V_\sigma
+ \mathrm{cos}^2\,\theta\,\mathrm{sin}\,\theta\,b_\phi\, V_s 
- \dot{\sigma}^2\, \l(b_\phi^2+b_{\phi\phi}\r),\label{eq:mus2}
\end{equation}
where $V_{ss}$ is given by
\begin{equation}
V_{ss} = \mathrm{sin}^2\theta\,V_{\phi\phi} 
-2\,\mathrm{sin}\,\theta\,\mathrm{cos}\,\theta\,{\mathrm{e}}^{-b}\, V_{\phi\chi}
+\mathrm{cos}^2\theta\, {\mathrm{e}}^{-2\, b}\, V_{\chi\chi}.
\end{equation}

In the spatially flat, FLRW background, we can decompose both the Mukhanov-Sasaki 
variables~$v_\sigma$ and $v_s$ in terms of the corresponding Fourier modes, say, 
$v_\sigma^{\bm{k}}$ and $v_s^{\bm{k}}$, as we had done earlier in the case of the 
single field models [see Eq.~\eqref{eq:v-fd}].
Moreover, the fact that $v_\s(\eta,{\bm x})$ and $v_s(\eta,{\bm x})$ are real 
quantities leads to the conditions $v_{\sigma}^{-\bm{k}}=v_{\sigma}^{\bm{k}\,\ast}$
and $v_{s}^{-{\bm{k}}}=v_{s}^{\bm{k}\,\ast}$ on the Fourier modes.
Further, we can express the Mukhanov-Sasaki variables in Fourier space, viz. 
$v_{\sigma}^{\bm{k}}$ and $v_{s}^{\bm{k}}$, in terms of their real and imaginary
parts as we had done before in Eq.~\eqref{eq:v-ri}.
When we do so, we find that the real and the imaginary parts of the variables 
$v_{\sigma}^{\bm{k}}$ and $v_s^{{\bm k}}$ are governed by the action~\cite{Battarra:2013cha}
\begin{eqnarray}
S[v_{\s}(\eta), v_s(\eta)]
&=& \f{1}{2}\, \int \d\eta\, 
\biggl(v_{\sigma}^{\prime 2} + v_{s}^{\prime 2} 
- 2\,\f{z'}{z}v_{\sigma}'\,v_\sigma - 2\,\f{a'}{a}\,v_s'\, v_s\nn\\ 
& &-\,  2\, \xi\, v_{\sigma}'\, v_{s} + 2\, \xi\, \f{z'}{z}\, v _{\sigma}\, v _{s}
- m _{\sigma}^2\, v_{\sigma}^2 - m_{s}^2\, v_{s}^2\biggr), \label{eq:a-tfm-fv}
\end{eqnarray}
where, for convenience, we have dropped the superscript~${\bm k}$, while
the quantities $m_{\s}^2$ and $m_s^2$ are given by
\begin{equation}
m_\s^2 = k ^2 - \f{z^{\prime 2}}{z^2}, \quad
m_{s}^2 = k^2 - \f{a^{\prime 2}}{a^2} + \mu_{s}^2\,a^2.
\end{equation}
From above action, the equations of motion governing the variables $v_\s$ and 
$v_s$ can be obtained to be
\begin{subequations}\label{eq:de-vsi-vs}
\begin{eqnarray}
v_{\sigma}'' +\l(k ^2 - \f{z''}{z}\r)\, v_{\sigma}
&= &\f{1}{z}\, \l(z\, \xi\, v_s\r)',\\
v_s'' + \l(k^2- \f{a''}{a}+\mu_{s}^2\,a^2\r)\, v_{s}
&=&-z\, \xi\, \l(\f{v_\sigma }{z}\r)'.
\end{eqnarray}
\end{subequations}	
	
The momenta, say, $p_\s$ and $p_s$, that are conjugate to the variables~$v_\s$ 
and~$v_s$ can be determined from the action~\eqref{eq:a-tfm-fv} to be
\begin{equation}
p_\s=v_{\s}'-\f{z'}{z}\, v_{\s}-\xi\, v_s,
\quad p_s= v_{s}'- \f{a'}{a}\, v_s.
\end{equation}
The Hamiltonian, say, $H_2$, of the complete system is then given by
\begin{eqnarray} 
H_2  &=&  \f{1}{2}\, \l(p_\s + \f{z'}{z}\, v_\s + \xi\,  v_s\r) ^2 
+ \f{1}{2}\, \l(p_s + \f{a'}{a}\, v_s \r)^2
- \xi\, \frac{z'}{z}\, v_\s\, v_s\nn\\
& &+\, \f{1}{2}\, m_{\sigma} ^2\, v_\s^2 + \f{1}{2}\, m _{s}^2\, v_s^2,\label{eq:H-tfm}
\end{eqnarray}
and the corresponding, first order, Hamilton's equations of motion can be
obtained to be
\begin{subequations}\label{eq:cl-H-eqs-tfm}
\begin{eqnarray}
v_{\s}'&=&p_\s+\f{z'}{z}\, v_{\s}+\xi\, v_s,\quad
p_{\s}'= - \f{z'}{z}\, p_{\s}-k^2\, v_{\s},\\
v_s' &=& p_s+ \f{a'}{a}\, v_s,\qquad\qquad
p_{s}'= - \f{a'}{a}\, p_{s}-\xi\, p_{\s}
-\l(k^2+\mu_{s}^2\,a^2 +\xi^2\r)\, v_{s}.\qquad\quad
\end{eqnarray}
\end{subequations}
Evidently, these equations can be combined to arrive at the corresponding second
order equations~\eqref{eq:de-vsi-vs} describing the variables $v_\s$ and $v_s$. 
Moreover, note that, if we set $v_s$ to zero, the above equations reduce to the 
Hamilton's equations~\eqref{eq:cl-H-eqs-sfm} in the single field case, as required.

If we denote the wave function of the complete system 
as~$\Psi(v_{\s},v_s,\eta)$, then the wave function satisfies the 
Schrodinger equation~\cite{Battarra:2013cha}
\begin{eqnarray}
i\,\f{\pa \Psi}{\pa \eta} 
&=& - \f{1}{2}\,\biggl[\f{\pa^2\Psi}{\pa v_{\sigma}^2} 
+ i\, \f{z'}{z}\, \l(\Psi+2\, v_{\s}\, \f{\pa \Psi}{\pa v_{\sigma}}\r)
+ \f{\pa^2\Psi}{\pa v_{s}^2} 
+ i\, \f{a'}{a}\, \l(\Psi+ 2\, v_s\,\f{\pa\Psi}{\pa v_{s}}\r)\nn\\
& &+\,2\, i\, \xi\, v_{s}\,\f{\pa\Psi}{\pa v_{\s}}
- k^2\, v_{\sigma}^2\,\Psi -\l(k^2+\xi^2+ \mu_s^2\,a^2\,\r)\,
v_{s}^2\,\Psi\biggr].\label{eq:se-tfm}
\end{eqnarray}
As in the single field case, we shall look for a solution to this Schrodinger 
equation which has a Gaussian form.
Let us start with the following ansatz~\cite{Battarra:2013cha,Vachaspati:2018hcu}:
\begin{equation}
\Psi(v_{\s},v_s,\eta)  
= \cN(\eta)\;
\mathrm{exp} \l[-\f{1}{2}\,\Omega_{\sigma\sigma}(\eta)\, v_{\sigma}^2 
- \f{1}{2}\, \Omega_{ss}(\eta)\, v_s^2
- \Omega_{\sigma s}(\eta)\, v_\sigma\, v_ s\r]\label{eq:ga-tfm}
\end{equation}
so that the normalization of the wave function leads to the condition
\begin{equation}
\vert \cN\vert 
=\l[\f{\Omegasisir\,\Omegassr-\l(\Omegasisr\r)^2}{\pi}\r]^{1/4},
\end{equation}
where $(\Omegasisir, \Omegassr,\Omegasisr)$ denote the real parts of the 
quantities $(\Omega_{\s\s},\Omega_{ss},\Omega_{\s s})$.
Clearly, if we can determine the quantities $(\Omega_{\s\s},\Omega_{ss},
\Omega_{\s s})$, we can obtain $\cN$ modulo an overall phase factor, as 
in the case of single field models.
On substituting the Gaussian ansatz~\eqref{eq:ga-tfm} in the Schrodinger
equation~\eqref{eq:se-tfm}, we find that the quantities $(\Omega_{\s\s},
\Omega_{ss}, \Omega_{\s s})$ satisfy the differential equations
\begin{subequations}\label{eq:de-Omega-tfm}
\begin{eqnarray}
i\, \Omega_{\s \s}'
&=& \l(\Omega_{\s\s} - i\, \f{z'}{z}\r)^2  + \Omega_{\s s}^2- m_{\s}^2,\\
i\, \Omega_{ss}'
&=& \l(\Omega_{ss} - i\, \f{a'}{a}\r)^2  + \l(\Omega_{\sigma s}- i \,\xi \r)^2
- m_{s}^2,\\
i\, \Omega_{\s s}'
&=& \l[\Omega_{\s \s}+\Omega_{ss}- i \l(\f{z'}{z}+\frac{a'}{a}\r) \r]\,
\Omega_{\s s} - i\, \xi\,\Omega_{\s \s}.
\end{eqnarray}
\end{subequations}

In the case of two field models, two sets of uncorrelated initial conditions 
are imposed on the curvature and isocurvature perturbations when the modes 
are well inside the sub-Hubble radius (for discussions on this point, see, 
for instance, Refs.~\cite{Lalak:2007vi,Raveendran:2018yyh,Braglia:2020fms,
Braglia:2020eai}).
To account for the two sets of initial conditions, let us now write the quantities
$(\Omega_{\s\s},\Omega_{ss}, \Omega_{\s s})$ as follows~\cite{Battarra:2013cha}:
\begin{subequations}\label{eq:omega-tfm}
\begin{eqnarray}
\Omega_{\s\s} &=&  -i\,\f{g_{s}^{\ast}\, \pi_{\sigma}^{f\ast} 
-f_s^{\ast}\,\pi_{\sigma}^{g\ast}}{g_{s}^{\ast}\, f_{\sigma}^{\ast}
- f_{s}^{\ast}\, g_{\sigma}^{\ast}},\label{eq:solCorr2}\\
\Omega_{ss} &=&  -i\,\f{f_{\sigma}^{\ast}\,\pi_{s}^{g\ast}  
- g_{\sigma}^{\ast}\, \pi_{s}^{f\ast} }{f_{\sigma}^{\ast}\, g_{s}^{\ast} 
- g_{\sigma}^{\ast}\,f_{s}^{\ast}},\label{eq:solCorr3}\\
\Omega_{\sigma s} &=& -i\,\f{f_{\sigma}^{\ast}\,\pi_{\sigma}^{g\ast} 
- g_{\sigma}^{\ast}\,\pi_{\sigma}^{f\ast}}{f_{\sigma}^{\ast}\, g_{s}^{\ast} 
- g_{\sigma}^{\ast}\, f_{s}^{\ast}}
= -i\,\f{g_{s}^{\ast}\, \pi_{s}^{f\ast}-f_s^{\ast}\, \pi_{s}^{g\ast}}{g_{s}^{\ast}\,
f_{\sigma}^{\ast} - f_{s}^{\ast}\, g_{\sigma}^{\ast}},
\end{eqnarray}
\end{subequations}
where the quantities $(\pi^f_\s, \pi^f_s)$ and $(\pi_\s^g,\pi_s^g)$ are defined as
\begin{subequations}\label{eq:pi-tfm}
\begin{eqnarray}
\pi^f_\s &=& f_{\s}'-\f{z'}{z}\, f_{\s}-\xi\, f_s,
\quad \pi^f_s= f_{s}'- \f{a'}{a}\, f_s,\\
\pi^g_\s&=&g_{\s}'-\f{z'}{z}\, g_{\s}-\xi\, g_s,
\quad \pi^g_s= g_{s}'- \f{a'}{a}\, g_s.
\end{eqnarray}
\end{subequations}
If we now substitute the above expressions for $(\Omega_{\s\s},\Omega_{ss}, 
\Omega_{\s s})$ in Eqs.~\eqref{eq:de-Omega-tfm}, then we find that the 
quantities $(f_\s,f_s)$ as well as $(g_\s,g_s)$ satisfy the differential 
equations~\eqref{eq:de-vsi-vs} governing the Mukhanov-Sasaki variables~$(v_\s,v_s)$.
Clearly, the quantities $(f_\s,f_s)$ and $(g_\s,g_s)$ correspond to the 
solutions to the Mukhanov-Sasaki variables for the two sets of initial 
conditions, while the quantities $(\pi^f_\s, \pi^f_s)$ and $(\pi_\s^g,\pi_s^g)$ 
can be considered as describing momenta that are conjugate to these solutions.
As in the case of single field models, the initial conditions can be imposed 
when the modes are deep inside the Hubble radius, i.e. when $k\gg \sqrt{z''/z}$.
In such a domain, one finds that the equations of motion governing the 
variables $(v_\s,v_s)$ decouple.
Hence, the two sets of initial conditions corresponding to the Bunch-Davies
vacuum are given by~\cite{Lalak:2007vi,Raveendran:2018yyh,Braglia:2020fms,
Braglia:2020eai}
\begin{subequations}
\begin{eqnarray}
f_\s(\ei) &=& \f{1}{\sqrt{2\,k}}\,\mathrm{e}^{-i\,k\,\ei},\quad
\pi_\s^f(\ei) = -\f{1}{\sqrt{2\,k}}\,\l[i\,k+\f{z'(\ei)}{z(\ei)}\r]\,
\mathrm{e}^{-i\,k\,\ei},\nn\\
f_s(\ei) &=& 0,\quad
\pi_s^f(\ei)=-\xi(\ei)\,\f{1}{\sqrt{2\,k}}\,\mathrm{e}^{-i\,k\,\ei},\\
g_\s(\ei) &=& 0,\quad
\pi_\s^g(\ei) = -\xi(\ei)\,\f{1}{\sqrt{2\,k}}\,\mathrm{e}^{-i\,k\,\ei},\nn\\
g_s(\ei) &=& \f{1}{\sqrt{2\,k}}\,\mathrm{e}^{-i\,k\,\ei},\quad
\pi_s^g(\ei) = -\f{1}{\sqrt{2\,k}}\,\l[i\,k+\f{a'(\ei)}{a(\ei)}\r]\,
\mathrm{e}^{-i\,k\,\ei}.
\end{eqnarray}
\end{subequations}
In the two field case, there arise four Wronskians~\cite{Battarra:2013cha}. 
By choosing the above initial conditions, we have set them to the following 
canonical values:
\begin{subequations}\label{eq:Ws}
\begin{eqnarray}
f_\sigma\,\pi_{\sigma}^{f\ast}+f_s\,\pi_{s}^{f\ast}
-\l(f_\sigma\,\pi_{\sigma}^{f\ast}+f_s\,\pi_{s}^{f\ast}\r)^{\ast} &=& i,\\
g_\sigma\,\pi_{\sigma}^{g\ast}+g_s\,\pi_{s}^{g\ast} 
-\l(g_\sigma\,\pi_{\sigma}^{g\ast}+g_s\,\pi_{s}^{g\ast}\r)^{\ast} &=& i,\\
f_\sigma\,\pi_{\sigma}^{g\ast}+f_s\,\pi_{s}^{g\ast}
-\l(g_\sigma\,\pi_{\sigma}^{f\ast} + g_s\,\pi_{s}^{f\ast}\r)&=& 0,\\
f_\sigma\,\pi_{\sigma}^{g} + f_s\,\pi_{s}^{g}
-\l(g_\sigma\,\pi_{\sigma}^{f} + g_s\,\pi_{s}^{f}\r) &=& 0.
\end{eqnarray}
\end{subequations}
Note that, for better numerical accuracy, we shall solve the first order
equations~\eqref{eq:cl-H-eqs-tfm} to determine the different measures 
describing the evolution of the quantum state of the system.

%%%%%%%%%%%%%%%%%%%%%%%%%%%%%%%%%%%%%%%%%%%%%%%%%%%%%%%%%%%%%%%%%%%%%%%%%%%%%%%

\subsection{Wigner function and squeezing parameters}
		
Let $x$ and $y$ denote the variables associated with a system with two degrees
of freedom.
If $p_x$ and $p_y$ represent the corresponding conjugate momenta, the Wigner 
function associated with the wave function $\Psi(x,y,t)$ describing the system
is defined as (see, for instance, Ref.~\cite{Hillery:1983ms})
\begin{eqnarray}
W(x,p_x,y,p_y,t) 
&=& \f{1}{\pi^2}\, 
\int_{-\infty}^\infty d z_1 \int_{-\infty}^\infty d z_2\; \Psi(x-z_1,y-z_2,t)\nn\\
& &\times\,\Psi^\ast(x+z_1,y+z_2,t)\,\mathrm{e}^{2\,i\,(p_x\,z_1+p_y\,z_2)}.
\end{eqnarray}
Since  the density matrix associated with the wave function $\Psi(x,y,t)$ is given by
\begin{equation}
\rho(x,y,\bar{x},\bar{y},t) = \Psi(x,y,t)\, \Psi^{\ast}(\bar{x},\bar{y},t),
\end{equation} 
the above Wigner function can be expressed in terms of the density matrix as follows:
\begin{eqnarray}
W(x,p_x,y,p_y,t) 
&=& \f{1}{\pi^2}\, 
\int_{-\infty}^\infty d z_1 \int_{-\infty}^\infty d z_2\; 
\rho(x-z_1,y-z_2,x+z_1,y+z_2,t)\nn\\ 
& &\times\,\mathrm{e}^{2\,i\,(p_x\,z_1+p_y\,z_2)}.
\label{eq:wf-tfm-dm}
\end{eqnarray}
In fact, such an expression for the Wigner function allows us to extend the
definition to even mixed states~\cite{Hillery:1983ms,case2008wigner}.

In principle, we can evaluate the complete Wigner function associated with the 
Gaussian wave function $\Psi(v_\s,v_s,\eta)$, which we had assumed to describe the 
system of our interest [see Eq.~\eqref{eq:ga-tfm}].
However, in this work, we shall focus only on the reduced dynamics associated with 
the curvature perturbation or, more precisely, the corresponding Mukhanov-Sasaki 
variable~$v_\s$.
Further, in due course, we shall be interested in evaluating the entanglement 
entropy (or, equivalently, the quantum discord) associated with the variable~$v_\s$.
For these reasons, we shall first evaluate the reduced density matrix describing
$v_\s$ and utilize it to arrive at the corresponding Wigner function~$W(v_\s,p_\s,\eta)$.

Motivated by the expression~\eqref{eq:wf-tfm-dm} for the  Wigner function in terms
of the density matrix, we can define the reduced Wigner function associated with the
variable $v_\s$ to be~\cite{Prokopec:2006fc,Battarra:2013cha}
\begin{equation}
W(v_\s,p_\s,\eta) 
= \f{1}{\pi}\,	\int_{-\infty}^\infty d z_1\, 
\rho_{r}(v_\s-z_1,v_\s+z_1,\eta)\, \mathrm{e}^{2\,i\,p_\s\,z_1}, \label{def:wf-r}
\end{equation}
where the reduced density matrix $\rho_{r}(v_\s,\bar{v}_\s,\eta)$ is given by
\begin{equation}
\rho_{r}(v_\s,\bar{v}_\s,\eta)  
= \int \d v_s\, \Psi(v_\s,v_s,\eta)\,\Psi^\ast(\bar{v}_\s,v_s,\eta). \label{def:rdm}
\end{equation}
Upon substituting the wave function $\Psi(v_\s,v_s,\eta)$ from Eq.~\eqref{eq:ga-tfm} 
in this expression and carrying out the integral over $v_s$, we find that the reduced 
density matrix can be written as
\begin{eqnarray}
\rho_{r}(v_{\sigma},\bar{v}_{\sigma},\eta)
&=& \bar{\cN}\, \mathrm{exp}\biggl[-\l(\f{C_{SS}}{4}+C_{DD}\r)\,
\f{(v_{\sigma}^2+\bar{v}_{\sigma}^2)}{2}\nn\\
& &-\, i\, C_{SD}\,\f{(v_{\sigma}^2-\bar{v}_{\sigma}^2)}{2}
+\l(C_{DD}-\f{C_{SS}}{4}\r)\, v_{\sigma}\,\bar{v}_{\sigma}\biggr],\label{eq:rdm}
\end{eqnarray}
where $\bar{\cN}$ is a normalization factor which ensures that the trace of the 
density matrix is unity.
The quantities $(C_{SS}, C_{DD},C_{SD})$ are given by
\begin{subequations} \label{the_Cs}
\begin{eqnarray}
C_{SS} & = & 2\, \Omegasisir\, 
\l(1 -\f{(\Omegasisr)^2}{\Omegasisir\,\Omegassr}\r),\label{eq:css}\\
C_{DD} & = & \f{\Omegasisir}{2}\,\l(1 + \f{(\Omegasisi)^2}{\Omegasisir\,\Omegassr}\r),\\
C_{SD} & = & \Omegasisii\, 
\l(1 - \f{\Omegasisi\,\Omegasisr}{\Omegasisii\,\Omegassr}\r),
\end{eqnarray}
\end{subequations}
where $(\Omegasisii,\Omegassi,\Omegasisi)$ denote the imaginary parts of the 
quantities $(\Omega_{\s\s},\Omega_{ss},\Omega_{\s s})$.
Upon using the expressions~\eqref{eq:omega-tfm} for $(\Omega_{\s\s},\Omega_{ss}, 
\Omega_{\s s})$, we find that the quantities $(C_{SS}, C_{DD},C_{SD})$ can be 
written as
\begin{subequations}\label{eq:c's}
\begin{eqnarray}
C_{SS} &=& \l(\vert f_{\s}\vert^2+\vert g_{\s}\vert^2\r)^{-1},\\
C_{DD} &=& \f{C_{SS}}{4}\,
\l(1+ 4\, \vert g_{\s}\,\pi_{\s}^f-f_{\s}\, \pi_{\s}^g\vert^2\r),\\
C_{SD} &=& \frac{C_{SS}}{2}\,\l[\l(f_{\s}\,\pi^{f\ast}_{\s}
+g_{\s}\,\pi^{g\ast}_{\sigma}\r)
+\l(f_{\s}\,\pi^{f\ast}_{\s}+g_{\s}\,\pi^{g\ast}_{\sigma}\r)^{\ast}\r].
\end{eqnarray}
\end{subequations}

It is convenient to define new variables from the sum and difference of 
$v_\s$ and $\bar{v}_{\s}$ as~\cite{Prokopec:2006fc,Battarra:2013cha}
\begin{equation}
v_S= \frac{v_\s+\bar{v}_\s}{2},\quad v_D= v_\s-\bar{v}_\s.
\end{equation} 
Then, the reduced density matrix~\eqref{eq:rdm} takes the form
\begin{equation}
\rho_{r}\l(v_{S}+\f{v_{D}}{2},v_{S}-\f{v_{D}}{2}\r)
= \bar{\cN}\, \mathrm{exp}\l[-\f{1}{2}\, 
\l(C_{SS}\,v_{S}^2+C_{DD}\,v_{D}^2+2\,i\, C_{SD}\,v_{S}\, v_{D}\r)\r].
\end{equation}
The phenomenon of decoherence occurs when the following condition is 
satisfied~\cite{schlosshauer2007decoherence,Prokopec:2006fc,Battarra:2013cha}:
\begin{equation}
\f{C_{SS}}{C_{DD}} \ll 1.\label{cond:decoh}
\end{equation} 
We shall see later that this condition corresponds to having a large entanglement 
entropy.

On making use of the expression~\eqref{eq:rdm} for the density matrix in
the definition~\eqref{eq:wf-tfm-dm} of the Wigner function and carrying 
out the resulting integral, we arrive at
\begin{equation}
W(v_{\s},p_{\s},\eta)
= \f{\sqrt{C_{SS}}}{2\, \pi\, \sqrt{C_{DD}}}\;
\mathrm{exp}\l[-\f{C_{SS}}{2}\, v_{\sigma}^2
- \f{1}{2\,C_{DD}}\,\l(p_{\s}+C_{SD}\, v_{\sigma}\r)^2\r].\label{eq:wf-r-tfm}
\end{equation}
Note that this Wigner function has a structure that is quite similar to that 
of the Wigner function in the single field case [cf. Eq.~\eqref{eq:wf-sfm}].
The behavior of the Wigner function in phase space can be tracked by following
the evolution of Wigner ellipse defined by the relation
\begin{equation}
\f{C_{SS}}{2}\,v_{\sigma}^2+\f{1}{2\, C_{DD}}\,
\l(p_{\s}+C_{SD}\, v_{\sigma}^{\rm R}\r)^2=1.\label{eq:we-tfm}
\end{equation}

The reduced Wigner function~\eqref{eq:wf-r-tfm} can again be written in terms 
of the covariance matrix as we had done in the case of single field models [see
Eq.~\eqref{eq:wf-sfm-cm}]. 
However, one finds that, in contrast to the single field case, the area of the 
Wigner ellipse does not remain constant in two field models (in this 
context, see App.~\ref{app:gwe}).
As a result, one requires an extra parameter to describe the covariance matrix. 
Therefore, instead of Eq.~\eqref{eq:cm-sfm}, we 
have~\cite{cariolaro2015quantum,Martin:2021znx}
\begin{eqnarray}\label{eq:cm-tfm}
V \!&=&\! \frac{(2\,N_{\s}+1)}{2}\nn\\
& &\!\!\!\!\times\,\!\!\begin{bmatrix}
\cosh\, (2\,r_\s) +  \sinh\, (2\,r_{\s})\,\cos\, (2\, \varphi_{\s})  
& \sinh\, (2\,r_{\s})\,\sin\, (2\, \varphi_{\s}) \\
\sinh\, (2\,r_{\s})\,\sin (2\, \phi_{\s})   
& \cosh\, (2\,r_{\s}) - \sinh\, (2\,r_{\s}) \,\cos\, (2\, \varphi_{\s}) 
\end{bmatrix}.\nn\\
\end{eqnarray}
In other words, when compared to the single field case, the variances are 
multiplied by an extra factor, leading to 
\begin{subequations}
\begin{eqnarray}
\langle\hat{v}_\s^2\rangle
&=&\vert f_{\sigma}\vert^2+\vert g_{\sigma}\vert^2\nn\\
&=& \f{(2\,N_{\s}+1)}{2 \, k}\,\l[\cosh\,(2\, r_\s)
+\sinh\,(2 \, r_\s)\, \cos\,(2\, \varphi_\s)\r],\label{eq:v2-tfm}\\
\langle\hat{p}_\s^2\rangle 
&=& \vert \pi_{\sigma}^f\vert^2+\vert \pi_{\sigma}^g\vert^2\nn\\
&=&\f{(2\,N_{\s}+1)\,k}{2}\,\l[\cosh\,(2\,r_\s)
-\sinh\,(2\, r_\s)\, \cos\,(2\,\varphi_\s)\r],\qquad\quad\label{eq:p2-tfm}\\
\f{1}{2}\, \langle \hat{v}_\s\, \hat{p}_\s+ \hat{p}_\s\, \hat{v}_\s\rangle 
&=&\f{1}{2}\,\l[\l(f_{\sigma}\,\pi^{f\ast}_{\sigma}+g_{\sigma}\,\pi^{g\ast}_{\sigma}\r)
+ \l(f_{\sigma}\,\pi^{f\ast}_{\sigma}+g_{\sigma}\,\pi^{g\ast}_{\sigma}\r)^{\ast}\r]\nn\\
&=&\f{(2\,N_{\s}+1)}{2}\,\sinh\,(2\, r_\s)\, \sin\,(2\,\varphi_{\s}).\label{vp_mixed}
\end{eqnarray}
\end{subequations}
We should clarify that, again, we have omitted the delta functions for 
simplicity [in this regard, see the comment following Eq.~\eqref{eq:corr}].
We can invert the above relations to express the squeezing parameters $r_\s$
and $\varphi_\s$ as
\begin{subequations}
\begin{eqnarray}
\cosh\,(2\, r_\s)
&=&\f{1}{(2\,N_{\s}+1)}\,
\l[k\, \l(\vert f_{\sigma}\vert^2+\vert g_{\sigma}\vert^2\r)
+\f{1}{k}\,\l(\vert \pi_{\sigma}^f\vert^2+\vert \pi_{\sigma}^g\vert^2\r)\r],\\
\cos\, (2\, \varphi_\s) 
&=&\f{1}{(2\,N_{\s}+1)}\, \f{1}{\sinh\,(2 \, r_\s)}\,
\l[k\, \l(\vert f_{\sigma}\vert^2+\vert g_{\sigma}\vert^2\r)
-\f{1}{k}\,\l(\vert \pi_{\sigma}^f\vert^2+\vert \pi_{\sigma}^g\vert^2\r)\r].\nn\\
\label{eq:r-sigma-phi-sigam}
\end{eqnarray}
\end{subequations}
Also, note that the quantity $N_{\s}$ is determined through the relation
[see Eq.~\eqref{eq:c's}] 
\begin{equation}
(2\,N_{\s}+1)^2
= 4\, \langle\hat{v}_\s^2\rangle\, \langle\hat{p}_\s^2\rangle 
-  \langle \hat{v}_\s\, \hat{p}_\s
+ \hat{p}_\s\, \hat{v}_\s\rangle^2 
= \f{4\,C_{DD}}{C_{SS}}.\label{eq:N-si}
\end{equation}

In a subsection that follows, we shall discuss the behavior of the Wigner function
and the squeezing parameters in specific two field inflationary models, one that 
leads to slow roll inflation and another that leads to enhanced power on small 
scales. 

%%%%%%%%%%%%%%%%%%%%%%%%%%%%%%%%%%%%%%%%%%%%%%%%%%%%%%%%%%%%%%%%%%%%%%%%%%%%%%%

\subsection{Entanglement entropy and quantum discord}

In contrast to the single field inflationary models, in the two field models, 
the presence of the additional degree of freedom leads to an entanglement 
between the two degrees of freedom.
As is well known, a good measure of the degree of quantum entanglement in such 
a bipartite system is the so-called von Neumann entropy of either of the two 
subsystems, since they are equal~\cite{schlosshauer2007decoherence}. 
Given the density matrix $\rho$, the von Neumann entropy $\mathcal{S}(\rho)$ 
is defined as~\cite{Adesso:2007tx,horodecki2009quantum}
\begin{equation}
\mathcal{S}(\rho)= -\mathrm{Tr.}\l(\rho\, \mathrm{ln}\, \rho\r).
\end{equation}
We shall use the von Neumann entropy associated with the reduced density matrix
in \eqref{eq:rdm} to quantify the entanglement~\cite{Prokopec:1992ia,Prokopec:2006fc,
Battarra:2013cha}.
But, before we do so, let us understand the relation between the entanglement
entropy and the quantum discord in the system of our interest.

Over the past two decades or so, various other measures of correlation have been 
discovered that are decidedly quantum in nature (in the sense that they should be zero 
in our classical notion of the world but can be non-zero for systems that have zero 
entanglement). 
The most popular of these ideas is the measure of quantum discord (see 
Refs.~\cite{zurek2002einselection,ollivier2001quantum,henderson2001classical}; for 
reviews in this context, see Refs.~\cite{modi2014pedagogical,bera2017quantum}). 
As we mentioned, quantum discord can be non-zero even when entanglement is zero. 
On the other hand, if the quantum discord is zero, then so is the entanglement 
(for a discussion on how the zero-discord states form a much smaller subset of
the separable---zero entanglement---states, see Refs.~\cite{modi2014pedagogical,
bera2017quantum}).
Thus, quantum discord seems to be a better tool than quantum entanglement to look 
for non-classical correlations in a system.

There are various ways of defining quantum discord \cite{bera2017quantum} and we 
shall be using the original measurement-based version, which is what has been used
previously in the context of cosmology (see, for instance, Refs.~\cite{Lim:2014uea,
Martin:2015qta,Hollowood:2017bil,Martin:2021znx}).
(Note that the quantum discord between two systems depends on which of the two 
systems is chosen for measurement and hence it is not symmetric. 
For two systems, say, $\mathbbm{1}$ and $\mathbbm{2}$, choosing $\mathbbm{2}$ for 
measurements and choosing $\mathbbm{1}$ for measurements will give two different 
values for quantum discord. 
But, as we shall see, quantum discord coincides with entanglement entropy when the 
total system is in a pure state wherein it is symmetric.) 
Moreover, since we will be dealing with Gaussian states, we shall specialize to 
Gaussian quantum discord~\cite{bera2017quantum}. 
In such cases, we have a closed form expression that allows us to calculate quantum 
discord directly from the covariance matrix \cite{adesso2010quantum}.
In fact, this has been used earlier to calculate quantum discord of cosmological
perturbations~\cite{Lim:2014uea}. 
We shall utilize this expression to determine the quantum discord in the two field 
inflationary models of our interest.

Even though we have made it a point to stress that quantum discord is separate from 
entanglement, it turns out that the quantum discord that arises when a system in a 
pure state is divided into two subsystems is identical to the entanglement entropy 
(i.e. the von Neumann entropy)~\cite{bera2017quantum,datta2008quantum}. 
We shall be evaluating the quantum discord when the pure quantum state of perturbations 
in the two field inflationary model is divided into curvature and isocurvature
perturbations. 
Thus, it is enough to calculate the entanglement entropy to obtain the discord. 
Nevertheless, in this subsection, we shall evaluate the complete expression for 
quantum discord and show that it is indeed equivalent to the entanglement entropy.

We shall now describe how Gaussian quantum discord can be evaluated from the 
covariance matrix~\cite{adesso2010quantum}. 
For two degrees of freedom, let $\hat{x}_1, \hat{p}_1,\hat{x}_2,\hat{p}_2$
be the two pairs of canonical operators. 
We arrange them into the vector $\hat{\Z}= (\hat{x}_1, \hat{p}_1,\hat{x}_2,\hat{p}_2)$.
To connect with the earlier notation (in Ref.~\cite{adesso2010quantum}), we define 
a scaled covariance matrix from the covariance matrix in Eq.~\eqref{eq:cm-sfm-def} as
\begin{equation}
\sigma_{ij}= 2\, V_{ij}
= \langle \hat{\Z}_i\,  \hat{\Z}_j+\hat{\Z}_j\, \hat{\Z}_i \rangle.\label{sigma_def}
\end{equation} 
Decomposing this $(4\times4)$ matrix in terms of $(2\times2)$ sub-blocks, we write 
\begin{equation}
\bm{\sigma}= \begin{bmatrix}
\bm{\alpha} &  \bm{\gamma}\\
\bm{\gamma}^T & \bm{\beta}
\end{bmatrix}, \label{sig_form}
\end{equation}
where $\bm{\alpha}$, $\bm{\beta}$ and $\bm{\gamma}$ are $(2\times 2)$ matrices.
Local symplectic operations [symplectic operations that act on the subspaces 
$(\hat{x}_1, \hat{p}_1)$ and $(\hat{x}_2, \hat{p}_2)$ without mixing them] will 
have the four independent invariants~\cite{Serafini:2003ke,Simon:1999lfr}, viz.
\begin{equation}
A=\mathrm{det.}~\bm{\alpha},\quad 
B=\mathrm{det.}~\bm{\beta},\quad
C=\mathrm{det.}~\bm{\gamma},\quad 
D=\mathrm{det.}~\sigma.\label{sub_det} 	
\end{equation}
The entanglement entropy for the $(\hat{x}_2, \hat{p}_2)$ subsystem,
say, $\mathcal{S}_2(\sigma_{\mathbbm{12}})$, may then be directly calculated 
to be~\cite{Serafini:2003ke}
\begin{equation}
\mathcal{S}_2(\sigma_{\mathbbm{12}}) = F(\sqrt{B}),\label{EE:cov}
\end{equation}
with the function~$F(x)$ being given by
\begin{equation}
F(x) = \l(\f{x+1}{2}\r)\, \ln\l(\f{x+1}{2}\r) 
-\l(\f{x-1}{2}\r) \ln\l(\f{x-1}{2}\r).\label{def_f}
\end{equation}	
As is well-known, the entanglement entropy is the same for both subsystems 
of a bipartite pure state~\cite{weedbrook2012gaussian}. 
So, this is also the entanglement entropy of the  $(\hat{x}_1, \hat{p}_1)$ system.

For the wave function~\eqref{eq:ga-tfm} in the two field models of our interest, 
we can evaluate the elements of the covariance matrix using, say, Mathematica. 
We obtain the sub-determinants~\eqref{sub_det} to be
\begin{subequations}\label{sub_det_val}
\begin{eqnarray}
A &=& B=\f{\l(\Omegai_{\s s}\r)^2
+\Omegar_{\s \s}\,\Omegar_{ss}}{\Omegar_{\s\s}\,\Omegar_{ss}
-\l(\Omegar_{\s s}\r)^2}, \label{sub_det_val_A_B} \\\
C &=& -\f{\l(\Omegai_{\s s}\r)^2+\l(\Omegar_{\s s}\r)^2}{\Omegar_{\s \sigma}\,
\Omegar_{ss}-\l(\Omegar_{\s s} \r)^2},\label{sub_det_val_C}\\
D &=& 1. 	
\end{eqnarray}
\end{subequations}
We find that the sub-determinants~$A$ and~$B$ have a simple expression in 
terms of the coefficients~\eqref{the_Cs} that describe the reduced density 
matrix~\eqref{eq:rdm} associated with the curvature perturbation.
It can be established that the sub-determinants have the following simple 
form:
\begin{align}
A=B=\f{4\, C_{DD}}{C_{SS}}.
\end{align}
Therefore, the entanglement entropy can be expressed as
\begin{eqnarray}
\mathcal{S}_2(\sigma_{\mathbbm{12}}) = F\l(\sqrt{\f{4\, C_{DD}}{C_{SS}}}\r), 
\label{Gdiscord-pure-C}
\end{eqnarray}
with the function $F(x)$ defined in Eq.~\eqref{def_f}. 

The function $F(x)$ is monotonic when $x \geq 1$ (which is when the function 
is real).
From Eq.~\eqref{sub_det_val_A_B}, it should be clear that $B$ is greater than or 
equal to one. 
Thus, $B$, or equivalently $4\, C_{DD}/C_{SS}$, is by itself a good measure 
of the entanglement entropy. 
In fact, it is nothing but the quantity $(2\,N_{\s}+1)^2$ [see Eq.~\eqref{eq:N-si}],
and hence is related to the area of the Wigner ellipse [see Eq.~\eqref{Area_in_N}]. 
Therefore, the area of the Wigner ellipse increases as the entropy increases. 
As we had pointed out earlier following Eq.~\eqref{cond:decoh}, the growth in 
entanglement entropy signifies the process of decoherence. 
Consequently, the growth in the area of the Wigner ellipse beyond the value for 
a pure state (wherein $N_{\s}=0$) is a measure of the decoherence of the system
(for related discussions, see Refs.~\cite{Koksma:2010zi,Martin:2021znx}).
	
The symplectic eigenvalues of the covariance matrix $\bm{\sigma}$ can be written 
in terms of the sub-determinants~\cite{adesso2010quantum}. 
Defining $\Delta$ as
\begin{equation}
\Delta = A+B+2\,C,
\end{equation}
we have the following expression for the symplectic eigen values $\nu_{+}$ and 
$\nu_{-}$:
\begin{equation}
\nu_{\pm}^2 = \frac{1}{2}\left(\Delta \pm \sqrt{\Delta^2-4D}\right)~. \label{sym_val}
\end{equation}
The relevant point for us is that quantum discord can be written in terms of the
above invariants~\cite{adesso2010quantum}. 
The quantum discord when the measurements are made on the subsystem represented 
by $(\hat{x}_2, \hat{p}_2)$ can be expressed in a closed form  as follows:
\begin{equation}
\mathcal{D}(\sigma_{\mathbbm{12}}) 
= F(\sqrt{B}) 
+  F(\sqrt{E_{\min}})-\l[F(\nu_-) + F(\nu_+)\r], \label{Gdiscord}
\end{equation}
with the function $F$ defined in Eq.~\eqref{def_f} and
\begin{eqnarray}\label{infdet}
E_{\min}
= \l\{\begin{array}{c}
\f{2\, C^2+ (-1+B)\, (-A+D)+2\, \vert C\vert\,\sqrt{C^2+(-1+B)\, 
(-A+D)}}{(-1+B)^2}\\
\hbox{when}\;\; (D-A\,B)^2 \le (1+B)\, (A + D)\, C^2,\\
\f{{A\, B-C^2+D-\sqrt{C^4+(-A\, B+D)^2-2\, C^2\, (A B+D)}}}{{2\, B}}\\
\hbox{otherwise}.\end{array} \r.
\end{eqnarray}
When the full system is in a pure state, quantum discord is supposed to coincide
with the entanglement entropy (i.e. the von Neumann entropy in our case), and we
should obtain~\cite{datta2008quantum}
\begin{equation}
\mathcal{D}(\sigma_{\mathbbm{12}})= F(\sqrt{B}), \label{Gdiscord-pure}
\end{equation}
i.e. the entanglement entropy for the $(\hat{x}_2, \hat{p}_2)$ subsystem, the 
system that is measured. 
We have explicitly checked the result in Eq.~\eqref{Gdiscord-pure} for the general
two field wave function in Eq.~\eqref{eq:ga-tfm}.
Upon using Eq.~\eqref{sym_val}, we obtain the symplectic eigen values to be
$\nu_{+}=\nu_{-}=1$.
Moreover, $E_{\min}$ from Eq.~\eqref{infdet} also turns out to be unity.
We should mention that, in evaluating the above, we have set $\vert C\vert $ 
to be $-C$, since $C$ is negative [see Eq.~\eqref{sub_det_val_C}]. 
(The numerator is obviously negative. The denominator needs to be positive for 
the normalizability of the wave function.)
On substituting these values into Eq.~\eqref{Gdiscord}, and using $F(1)=0$ as 
can be proved using a limiting process, we obtain the equality in
Eq.~\eqref{Gdiscord-pure}, which implies that the quantum discord coincides with 
the entanglement entropy. 

%%%%%%%%%%%%%%%%%%%%%%%%%%%%%%%%%%%%%%%%%%%%%%%%%%%%%%%%%%%%%%%%%%%%%%%%%%%%%%%

\subsection{Behavior in specific two field inflationary models}

In this section, we shall evaluate the Wigner function, squeezing 
parameters and entanglement entropy in specific two field models
of inflation described by the action~\eqref{eq:a-tfm-f}.
We shall consider a potential $V(\phi,\chi)$ that is separable and is 
given by
\begin{equation}\label{eq:double-v-PBH}
V(\phi,\chi) =\f{m^2}{2}\, (\phi^2+\chi^2).
\end{equation}
Therefore, the interaction between the two fields~$\phi$ and~$\chi$
arises only due to the function $b(\phi)$.
We shall work with the following form for the function $b(\phi)$:
\begin{equation}
b(\phi)=\f{b_1}{2}\, \biggl\{1+\tanh\,\l[\alpha\,(\phi-\phi_0)\r]\biggr\}.
\label{eq:b-phi}
\end{equation}

We shall first consider a slow roll scenario wherein $b_1=0$. 
Since $b_1=0$, $b(\phi)$ vanishes, and the field $\chi$ reduces to that 
of a canonical scalar field.
Also, there arises no interaction between the two fields. 
To ensure COBE normalization on large scales, we shall set $m=9\times 
10^{-6}\, \Mpl$.
We choose the initial values of the fields and their velocities to be 
$\phi_\mathrm{i} = \chi_\mathrm{i}=11.5\,\Mpl$ and $\dot{\phi}_\mathrm{i} 
= \dot{\chi}_\mathrm{i}=-3.68\times 10^{-6}\,\Mpl^2$.
We find that, under these conditions, inflation lasts for about $66$ e-folds.
As the potential $V(\phi,\chi)$ has the same form along the two field directions, 
for the above choice of initial conditions, both the fields evolve in the same 
manner. 
In other words, the evolution of the background is radial in field space and, 
hence, the behavior of the perturbations effectively reduces to the behavior 
in a single field model.

\begin{figure}[!t]
\centering
\includegraphics[width=0.475\linewidth]{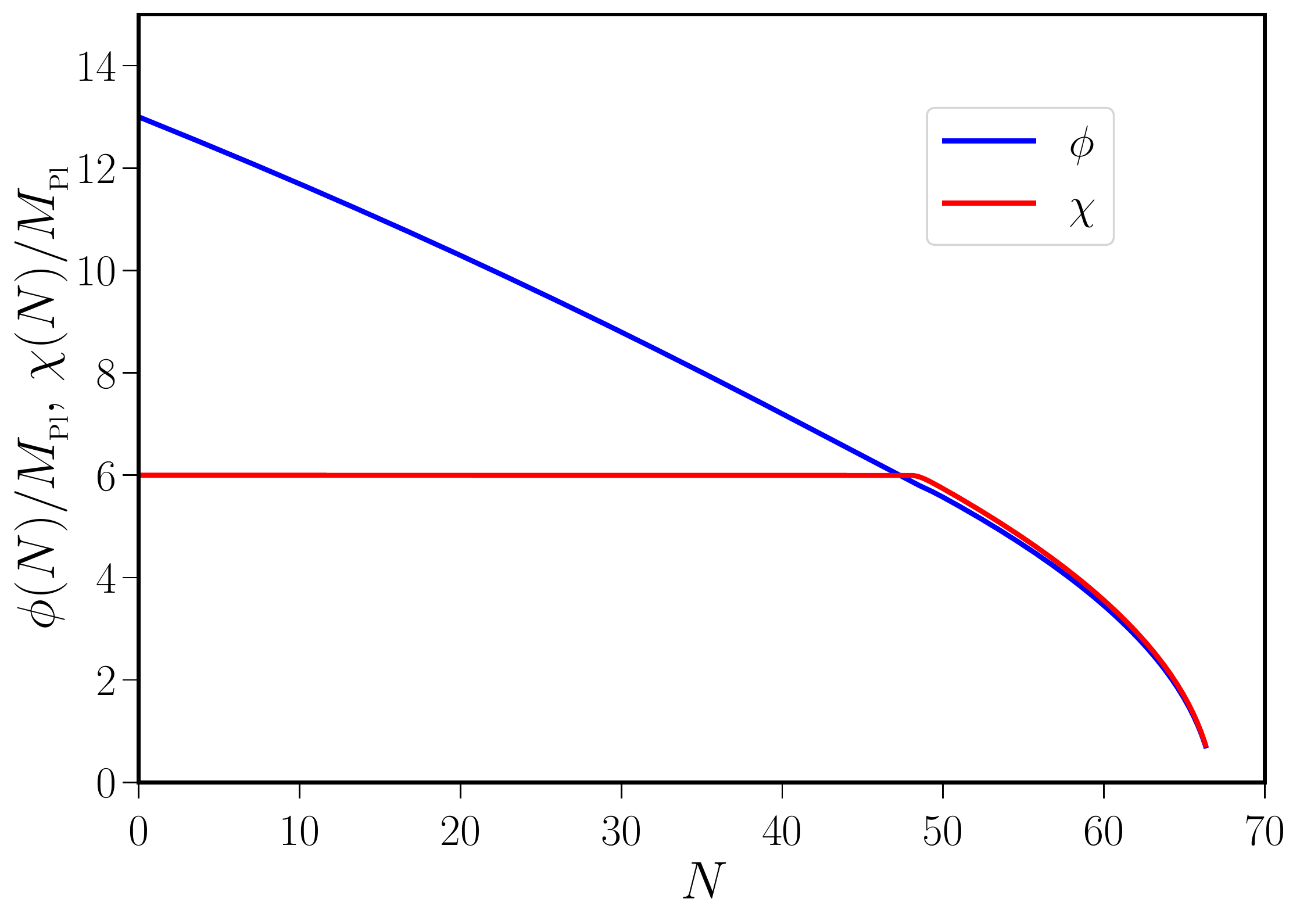}
\hskip 5pt
\includegraphics[width=0.475\linewidth]{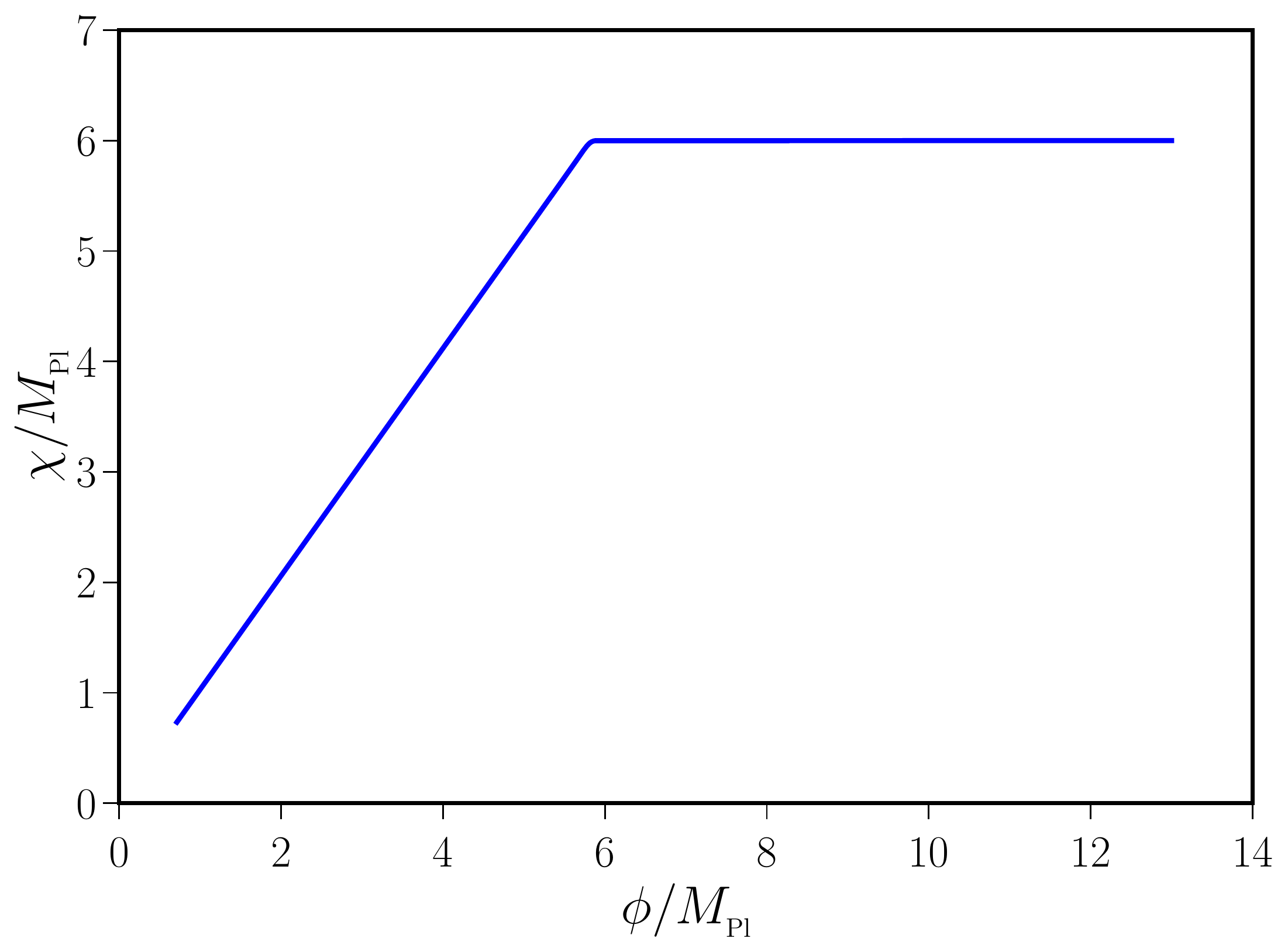}
\vskip 5pt
\includegraphics[width=0.475\linewidth]{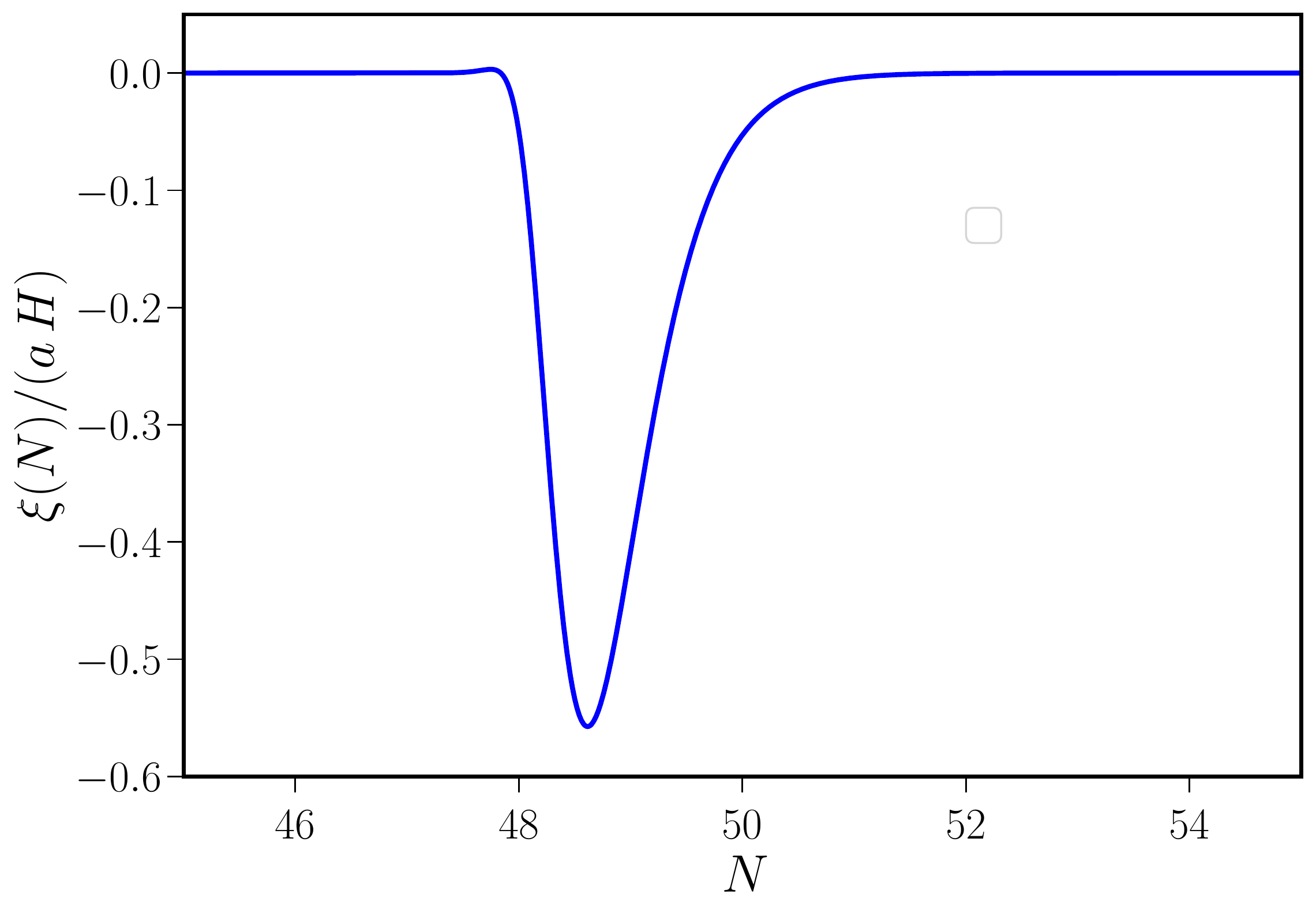}
\hskip 5pt
\includegraphics[width=0.475\linewidth]{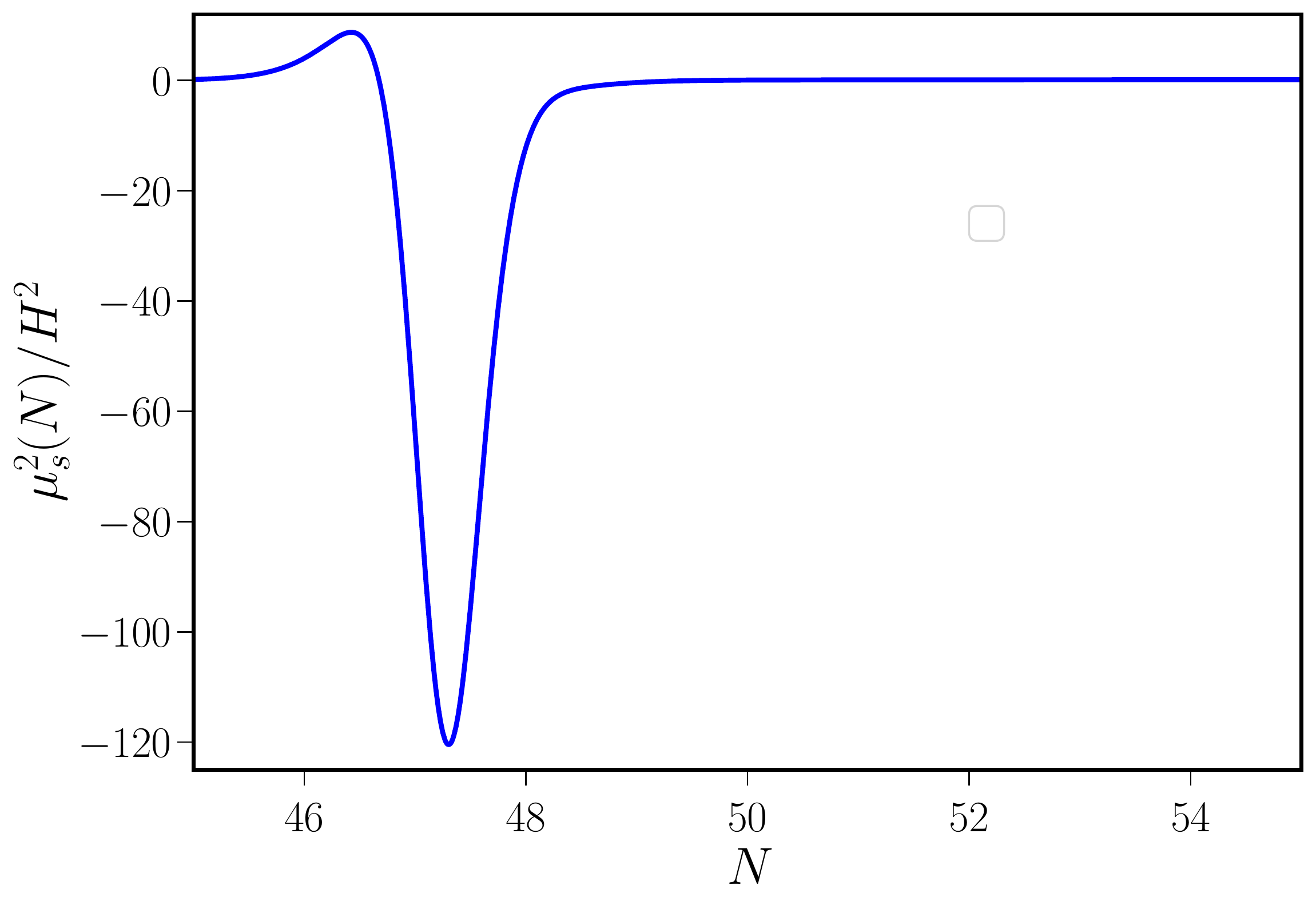}
\caption{The evolution of the fields $\phi$ (in blue) and $\chi$ (in red) 
in the two field inflationary model of our interest with a non-zero $b_1$ 
has been plotted against the number of e-folds (on top left) as well as in 
field space (on top right).
We have also plotted interaction strength $\xi$ between the curvature and 
the isocurvature perturbations (on bottom left) and the effective mass 
squared $\mu_s^2$ of the isocurvature perturbations (on bottom right).
It should be clear from these figures that $\xi$ increases (during 
$47\lesssim N\lesssim 50$) and $\mu_s^2$ becomes negative (during 
$46\lesssim N \lesssim 48$) as the field turns in space.
These two effects combine to result in enhanced power over wave numbers
which leave the Hubble radius during this period.}\label{fig:be-tfm}
\end{figure}

For achieving the scenario which leads to enhanced power on small scales, we 
shall set $m=1.03\times 10^{-5}\,\Mpl$, $b_1=15$, $\alpha=10\, \Mpl^{-1}$ and 
$\phi_0=6\,\Mpl$.
We shall choose the initial conditions to be $(\phi_\mathrm{i},\chi_\mathrm{i})
= (13\,\Mpl,6\,\Mpl)$ and $(\dot{\phi}_\mathrm{i},\dot{\chi}_\mathrm{i})
= (-5.44\times 10^{-6}\,\Mpl^2,0)$.
For such values and conditions, we find that inflation proceeds 
for about $66$ e-folds before it is terminated.
The non-zero value for $b_1$ induces an interaction between the two fields.
In Fig.~\ref{fig:be-tfm}, we have plotted the evolution of the two fields 
$\phi$ and $\chi$ as a function of e-folds.
%%%%%%%%%%%%%%%%%%%%%%%%%%%%%%%%%%%%%%%%%%%%%%%%%%%%%%%%%%%%%%%%%%%%%%%%%%%%%%%
%%%%%%%%%%%%%%%%%%%%%%%%%%%%%%%%%%%%%%%%%%%%%%%%%%%%%%%%%%%%%%%%%%%%%%%%%%%%%%%
We have also illustrated their evolution in field space.
Moreover, in the figure, we have plotted the quantity $\xi$ which characterizes 
the strength of the interaction between the curvature and the isocurvature 
perturbations [see Eqs.~\eqref{eq:de-vsi-vs}] as well as the effective mass 
squared of the isocurvature perturbations $\mu_{s}^2$ [see Eq.~\eqref{eq:mus2}].
We find that the initial stage of inflation is driven by the field~$\phi$, 
whereas the latter stage is driven by both the fields~$\phi$ and~$\chi$.
The form of the function $b(\phi)$ [see Eq.~\eqref{eq:b-phi}] induces a 
turning in the field space as the field~$\phi$ approaches~$\phi_0$.
The turn in the field space briefly increases the interaction strength $\xi$ 
(during the e-folds $47 < N < 50$) between the curvature and the isocurvature 
perturbations.
In the course of this turning (in fact, during $46 < N < 48$), the quantity
$\mu _{s}^2$ becomes negative creating a temporary tachyonic instability. 
As we shall soon illustrate, the interaction between the curvature and the 
isocurvature perturbations, coupled with the tachyonic instability, leads 
to enhanced power on small scales.

Before we proceed, it is important that we clarify a couple of points concerning
our choice~\eqref{eq:double-v-PBH} for the potential $V(\phi,\chi)$ and the 
functional form~\eqref{eq:b-phi} of $b(\phi)$.
Recall that, in the single field case, the potentials have to be suitably designed 
to achieve the desired background dynamics and enhanced power on small scales.
In contrast, in two field models, we have worked with simple potentials and it
is the form of the function $b(\phi)$ that is responsible for the non-trivial 
background dynamics.
While the specific form of $b(\phi)$ that we have worked with [as given by 
Eq.~\eqref{eq:b-phi}] seems fine tuned, one can work with simpler forms of $b(\phi)$
and potentials with an additional parameter to achieve similar scalar power 
spectra.
It is the richer dynamics of the two field models, specifically the presence of the
non-canonical kinetic term involving $b(\phi)$ and its non-trivial effects on the 
evolution of the perturbations, that permits such a possibility (in this context, 
see, for instance, Refs.~\cite{Braglia:2020eai,Braglia:2020fms}). 

\begin{figure}[!t]
\centering
\includegraphics[width=0.75\linewidth]{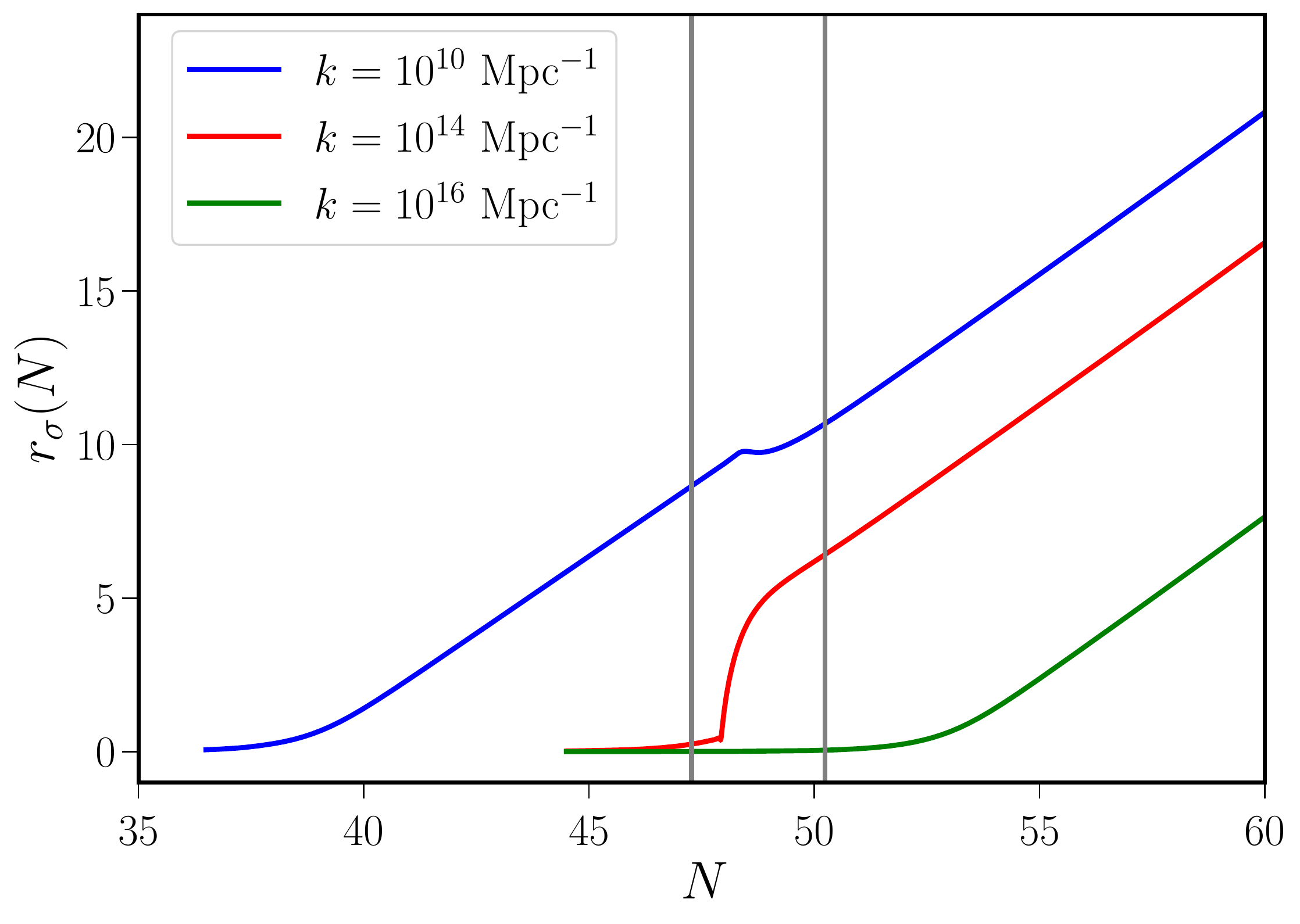}
\vskip 5pt
\includegraphics[width=0.75\linewidth]{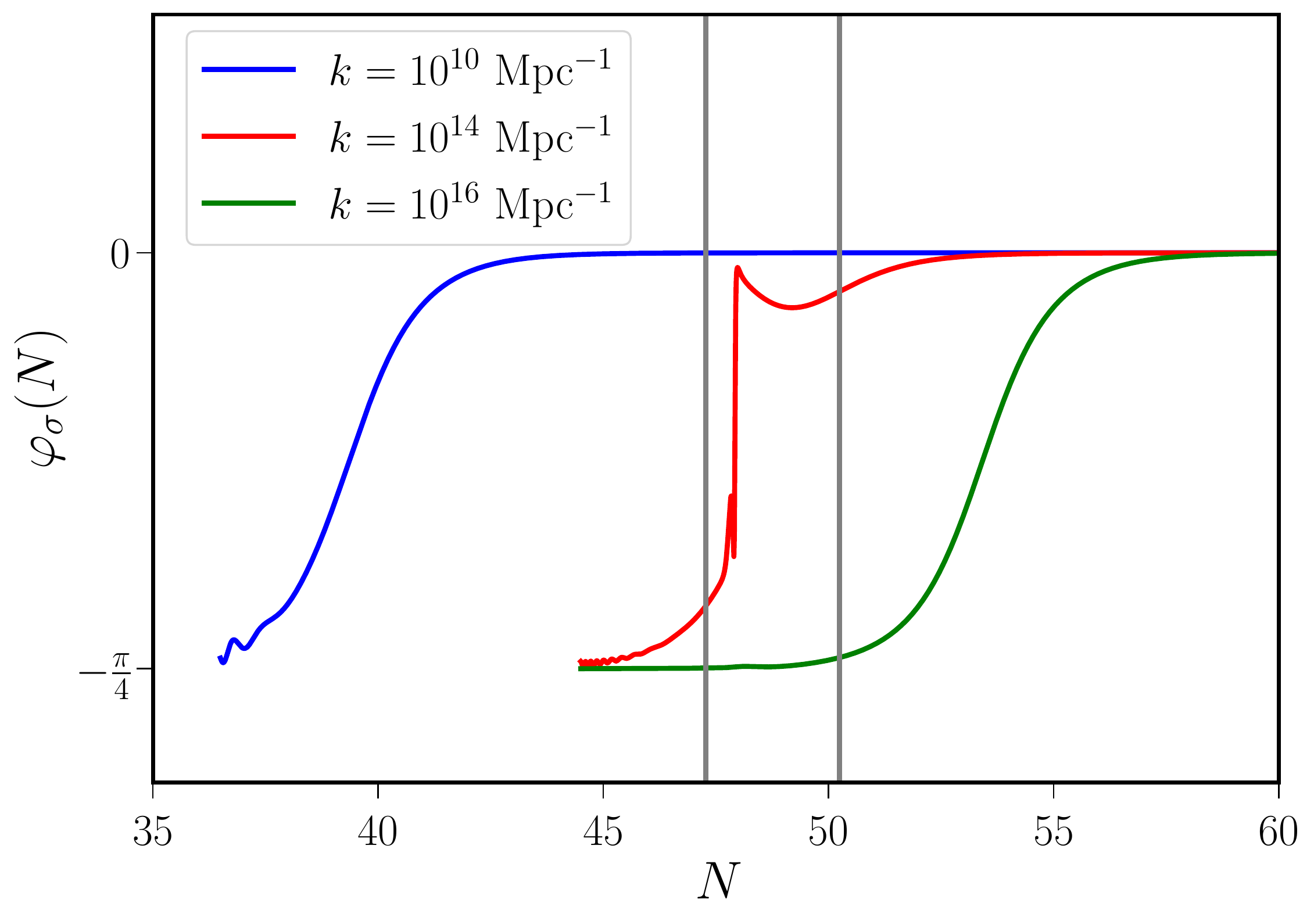}
\caption{The evolution of the squeezing amplitude~$r_{\s}$ associated with 
the curvature perturbation and the corresponding squeezing angle~$\varphi_{\s}$ 
has been plotted as a function of e-folds (on top and at the bottom, 
respectively) in the two field inflationary model which leads to enhanced 
power on small scales.
We have plotted the evolution of the squeezing parameters for modes with the 
wave numbers $k=(10^{10}, 10^{14}, 10^{16})\,\mathrm{Mpc}^{-1}$ (in blue, red
and green).
We have also demarcated the period when the turning in field space occurs (by
gray vertical lines).
It should be clear that, during this period, both the squeezing parameters 
exhibit non-trivial behavior.
We find that, as in the single field case, at late times, while $r_{\s}(N)
\propto N$, $\varphi_{\s}$ approaches zero.}\label{fig:sp-tfm}
\end{figure}

In Fig.~\ref{fig:sp-tfm}, we have plotted the evolution of the squeezing 
amplitude~$r_{\s}$ associated with the curvature perturbation and the 
associated squeezing angle~$\varphi_{\s}$, as a function of e-folds in
the scenario which leads to enhanced power on small scales.
%%%%%%%%%%%%%%%%%%%%%%%%%%%%%%%%%%%%%%%%%%%%%%%%%%%%%%%%%%%%%%%%%%%%%%%%%%%%%%%
%%%%%%%%%%%%%%%%%%%%%%%%%%%%%%%%%%%%%%%%%%%%%%%%%%%%%%%%%%%%%%%%%%%%%%%%%%%%%%%
We have plotted the squeezing parameters for three different wave numbers which
leave the Hubble radius around the time when the turning in the field space occurs.
It is clear from the figure that the turning in field space induces non-trivial
evolution of the squeezing parameters.
However, at adequately late times, after the turn is complete and when the modes 
are on super-Hubble scales, while $r_{\s}\propto N$, $\varphi_{\s}$ tends to zero
as in the single field case.

\begin{figure}[!t]
\centering
\includegraphics[width=0.75\linewidth]{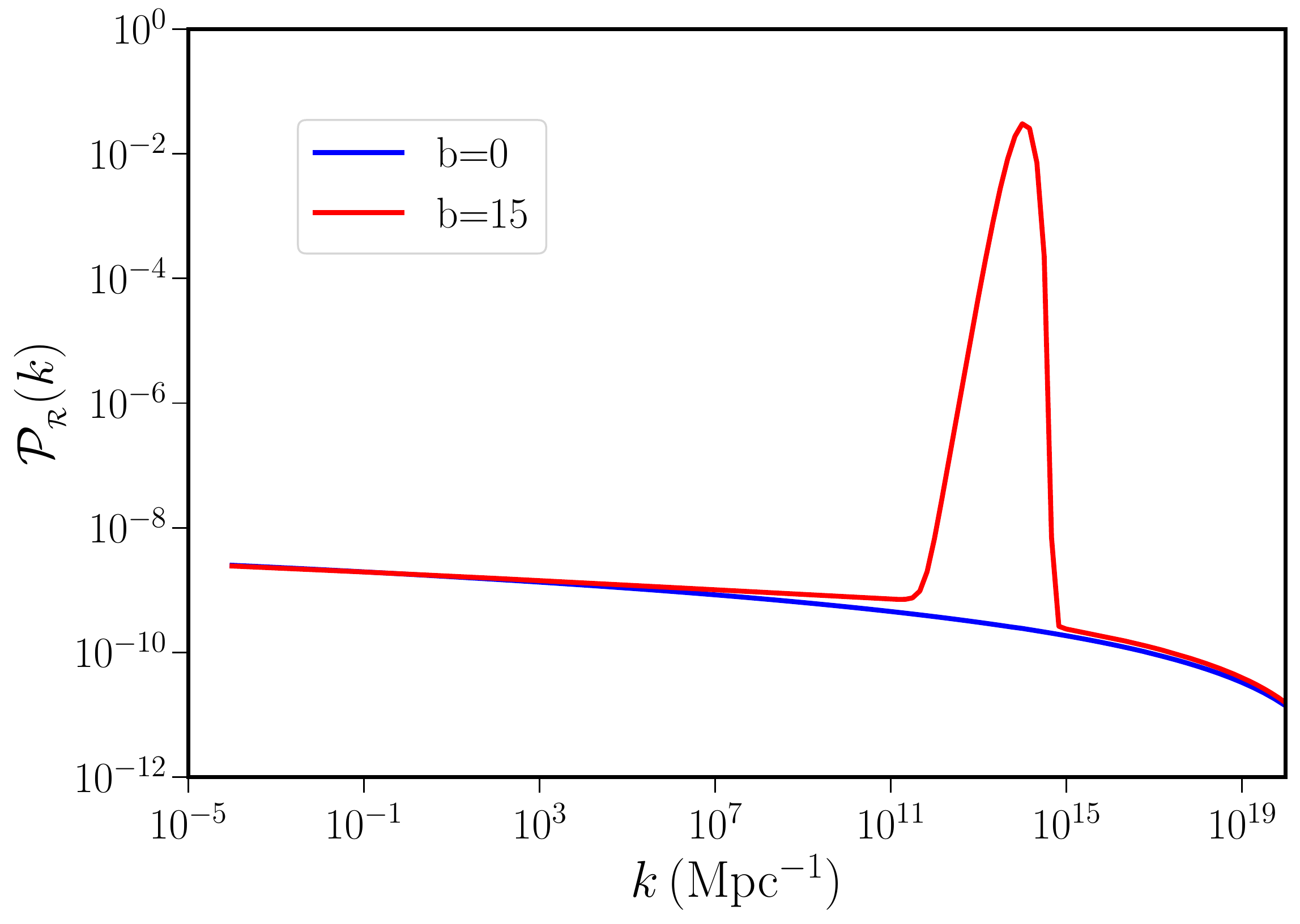}
\vskip 5pt
\includegraphics[width=0.75\linewidth]{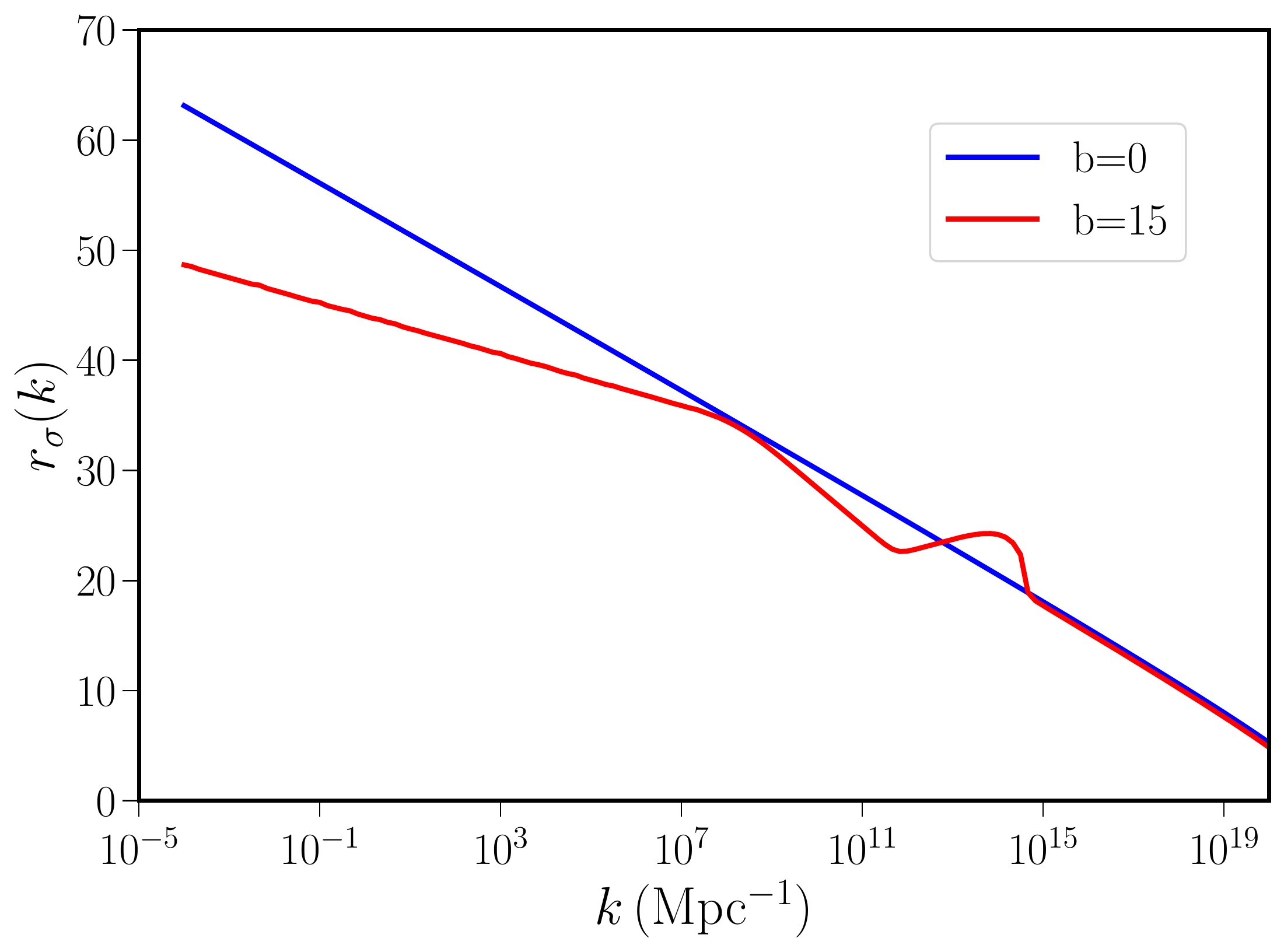}
\caption{The power spectrum of the curvature perturbations $\ps(k)$ (on top) 
and the squeezing amplitude $r_{\s}(k)$ (at the bottom), evaluated towards 
the end of inflation, have been plotted as a function of wave numbers in  
both the two field inflationary models we have considered, viz. the one that 
permits only slow roll inflation and another that leads to a turn in the 
field space.
While the spectrum of curvature perturbations is nearly scale invariant in 
the slow roll model, the scenario involving the turn in field space leads 
to a sharp peak in the spectrum.
The non-trivial evolution that is responsible for the peak in the power spectrum
also modifies the shape of $r_{\s}(k)$, which too exhibits a small peak around
the wave number at which the scalar power spectrum $\ps(k)$ contains the peak.}
\label{fig:ps-sp-tfm}
\end{figure}

In two field inflationary models, the power spectrum of the curvature perturbations 
is defined as in Eq.~\eqref{eq:ps-d}, with~$\hat{v}$ replaced by~$\hat{v}_{\s}$.
Upon using the expression~\eqref{eq:v2-tfm} for $\langle \hat{v}_{\s}^2\rangle$,
we obtain that 
\begin{eqnarray}
\ps(k) &=&\f{k^2}{4\,\pi^2\,z^2}\,(2\,N_{\s}+1)\,
\l[\cosh\,(2\, r_\s)
+\sinh\,(2 \, r_\s)\, \cos\,(2\, \varphi_\s)\r]\nn\\
&=&\f{k^3}{2\,\pi^2\,z^2}\,
\l(\vert f_{\sigma}\vert^2+\vert g_{\sigma}\vert^2\r),
\end{eqnarray}
where $N_{\s}$ is determined by the relation~\eqref{eq:N-si}.
In Fig.~\ref{fig:ps-sp-tfm}, we have plotted the power spectrum of the curvature
perturbation and the squeezing amplitude in the slow roll scenario as well as the 
scenario involving a turn in field space.
%%%%%%%%%%%%%%%%%%%%%%%%%%%%%%%%%%%%%%%%%%%%%%%%%%%%%%%%%%%%%%%%%%%%%%%%%%%%%%%
%%%%%%%%%%%%%%%%%%%%%%%%%%%%%%%%%%%%%%%%%%%%%%%%%%%%%%%%%%%%%%%%%%%%%%%%%%%%%%%	
We have plotted these quantities for a wide range of wave numbers.
Clearly, while we obtain a nearly scale invariant spectrum in the slow roll
case, the scenario involving the turn in the field space leads to enhanced 
amplitude on smaller scales with a peak in the power spectrum. 
Also, as in the single field models, at the end of inflation, $r_{\s}$ is 
roughly determined by the amount of time the modes spend on super-Hubble 
scales.
We find that the scenario involving the turn in field space leads to slightly
lower or higher values of $r_{\s}$ for modes that leave the Hubble radius
during the turn.
Interestingly, the quantity $r_{\s}(k)$ exhibits a peak around the same 
location as the spectrum of curvature perturbations~$\ps(k)$.

\begin{figure}[!t]
\centering
\includegraphics[width=0.85\linewidth]{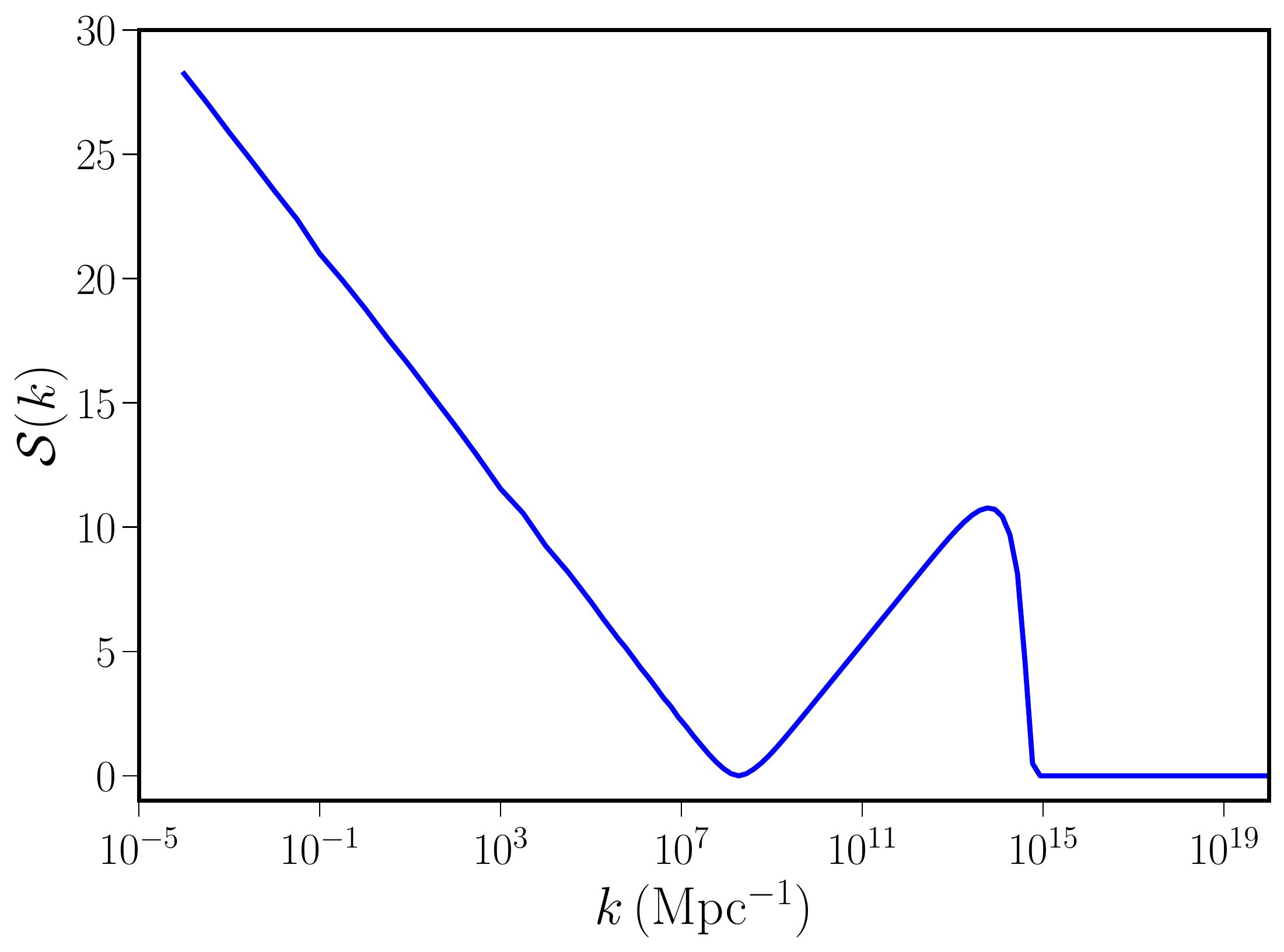}
\caption{The entanglement entropy---or, equivalently, the quantum 
discord---$\mathcal{S}(k)$ associated with the curvature perturbations 
in the two field model has been plotted for the scenario involving a 
turn in the field space.
Note that the entanglement entropy is higher over large scales and is 
lower over small scales.
We find that the modes that leave the Hubble radius during the turning 
in field space exhibit a higher level of entanglement entropy, with the 
peak corresponding to the peak in the spectrum of curvature 
perturbations.}\label{fig:ee-tfm}
\end{figure}

Let us now turn to understand the behavior of the entanglement entropy, 
which is an additional measure describing the evolution of the curvature
perturbations in two field models.
It is easy to establish that, in the slow roll scenario, the entanglement 
entropy $\mathcal{S}(k)$ associated with the curvature perturbations vanishes 
identically.
This is because of the fact that, since the dynamics of the two fields
are the same (due to the choice of the potential and the initial conditions),
the system effectively behaves as in the case of a single scalar field.
In Fig.~\ref{fig:ee-tfm}, we have plotted the entanglement entropy~$\mathcal{S}(k)$ 
in the non-trivial scenario involving the turn in field space. 
%%%%%%%%%%%%%%%%%%%%%%%%%%%%%%%%%%%%%%%%%%%%%%%%%%%%%%%%%%%%%%%%%%%%%%%%%%%%%%%
%%%%%%%%%%%%%%%%%%%%%%%%%%%%%%%%%%%%%%%%%%%%%%%%%%%%%%%%%%%%%%%%%%%%%%%%%%%%%%%	
We find that the entanglement entropy is higher over large scales when 
compared to smaller scales.
Interestingly, in a manner similar to the behavior of the squeezing
amplitude~$r_{\s}(k)$, we find that the entanglement entropy~$\mathcal{S}(k)$ 
also exhibits a peak around the same location as the spectrum of curvature
perturbations~$\ps(k)$.

%%%%%%%%%%%%%%%%%%%%%%%%%%%%%%%%%%%%%%%%%%%%%%%%%%%%%%%%%%%%%%%%%%%%%%%%%%%%%%%

\section{Discussion}\label{sec:d}

In this work, we have examined the evolution of the quantum state of the 
curvature perturbations in single and two field models of inflation.
In the case of single field models, we have used mathematical measures 
such as the behavior of the Wigner function and the squeezing parameters 
to understand the manner in which the quantum state of the system evolves.
In the case of two field models, apart from the measures mentioned above,
we have also examined the behavior of the entanglement entropy (arrived
at by integrating the degrees of freedom associated with the isocurvature
perturbations) or, equivalently, quantum discord.
We have compared the behavior of these measures in slow roll inflation 
with the corresponding behavior in models that lead to departures from
slow roll and enhanced power on small scales.
We find that the departures from slow roll inflation can modify or boost
these measures to only to a limited extent.

In the case of the single field models, we had partitioned the system 
into real and imaginary parts of the Fourier modes of the Mukhanov-Sasaki 
variable.
Since these quantities were decoupled, no entanglement entropy or quantum 
discord arose in this case. 
On the other hand, if we work in terms of the Fourier modes themselves, 
entanglement and discord can arise if we partition the system into $\bmk$ 
and $-\bmk$ sectors~\cite{Martin:2015qta}. 
(We have, in fact, checked that, upon using the wave function provided in 
Ref.~\cite{Martin:2019wta} in Eq.~\eqref{Gdiscord-pure}, we can arrive at
expression for quantum discord obtained in Ref.~\cite{Martin:2015qta}.) 
Even in the case of two field models, one may partition the system into 
$\bmk$ and $-\bmk$ sectors and study the resulting entanglement entropy 
and quantum discord. 
We shall leave these avenues for future work. 

Another topic to be explored is decoherence due to interactions with an 
environment. 
There are very many ways in which such an environment may be constructed 
(in this context, see, for example, Ref.~\cite{Kiefer:2008ku}). 
One may consider interactions with other fields. 
Even otherwise, the full non-linear theory will have interactions between
different modes and between the scalar, vector and tensor sectors of the 
cosmological perturbations. 
One could also consider interaction between perturbations in different 
spatial regions, for example the regions inside and outside the Hubble 
radius. 
Even the case of two scalar fields, as we have considered, has been claimed 
to be sufficient to provide decoherence in certain 
cases~\cite{Prokopec:2006fc,Battarra:2013cha}. 
We observe a similar dynamics in our model, and we plan to pursue the topic 
of decoherence in more detail in a future publication.

%%%%%%%%%%%%%%%%%%%%%%%%%%%%%%%%%%%%%%%%%%%%%%%%%%%%%%%%%%%%%%%%%%%%%%%%%%%%%%%

\section*{Data availability}

Data sharing is not applicable to this article as no data sets were generated 
or analyzed during the current study.

%%%%%%%%%%%%%%%%%%%%%%%%%%%%%%%%%%%%%%%%%%%%%%%%%%%%%%%%%%%%%%%%%%%%%%%%%%%%%%%

\section*{Acknowledgements}

The authors wish to thank Sagarika Tripathy for independently checking our
numerical results in the two field models.
RNR is supported by the Research Associateship of the Indian Association 
for the Cultivation of Science, Kolkata, India.
KP would like to thank the Department of Physics, Cochin University of Science 
and Technology, Kochi, India, for kind hospitality during a stay.
LS wishes to acknowledge support from the Science and Engineering Research 
Board, Department of Science and Technology, Government of India, through 
the Core Research Grant~CRG/2018/002200.	

%%%%%%%%%%%%%%%%%%%%%%%%%%%%%%%%%%%%%%%%%%%%%%%%%%%%%%%%%%%%%%%%%%%%%%%%%%%%%%%

\appendix

%%%%%%%%%%%%%%%%%%%%%%%%%%%%%%%%%%%%%%%%%%%%%%%%%%%%%%%%%%%%%%%%%%%%%%%%%%%%%%%

\section*{Appendices}

\section{Geometry of the Wigner ellipse}\label{app:gwe}

For a Gaussian Wigner function of the form
\begin{equation}
W(x,p) \propto \mathrm{exp}\l[-\l(\cA\,x^2+\cB\,x\,y+\cC\,y^2\r)\r],\label{eq:wf-ge}
\end{equation}
the Wigner ellipse can be defined as the contour described by the quadratic 
equation
\begin{equation}
\cA\,x^2+\cB\,x\,y+\cC\,y^2=1.\label{eq:ellipse}
\end{equation}
The geometry of an ellipse can be described by specifying the semi-major 
axis~$\bar{a}$, the semi-minor axis~$\bar{b}$ and the inclination angle~$\delta$, 
which we shall take to be the angle between the $x$-axis and the major axis, 
measured anticlockwise from the~$x$-axis. 
While the semi-major and semi-minor axes of the ellipse~\eqref{eq:ellipse}
are given by~\cite{kendig2005conics}
\begin{equation}
\bar{a}^2 =\f{2}{\cA+\cC- \sqrt{(\cA-\cC)^2+\cB^2}},\quad  
\bar{b}^2 =\f{2}{\cA+\cC+ \sqrt{(\cA-\cC)^2+\cB^2}},\label{eq:mma}
\end{equation}		
the angle $\delta$ is determined by the equations
\begin{equation}
\sin\, (2\,\delta)
=\f{-\cB}{\sqrt{(\cA-\cC)^2+\cB^2}},\quad 
\cos\,(2\,\delta) 
= \f{C-A}{\sqrt{(\cA-\cC)^2+\cB^2}}.\label{eq:ep} 
\end{equation}
We should stress that {\it both}\/ these equations are needed to specify
the angle~$\delta$ in the right quadrant. 
We take the angle $\delta$ to lie in the range $(-\pi/2,\pi/2]$.

Recall that the Wigner ellipse can be expressed in terms of the covariance
matrix~$V$ as follows [see Eqs.~\eqref{eq:dcv}, \eqref{eq:wf-sfm-cm} 
and~\eqref{eq:cm-sfm-def}] 
\begin{equation}
\f{\Z^T\, V^{-1}\,\Z}{2}=\f{V_{22}\,\tilde{v}^2
-2\, V_{12}\,\tilde{v}\,\tilde{p}
+V_{11}\,\tilde{p}^2}{2~\textrm{det}~V} = 1.\label{eq:ellipse-V}
\end{equation}
We can apply the formulae~\eqref{eq:mma} and~\eqref{eq:ep} to such a Wigner 
ellipse, with the general covariance matrix for a mixed Gaussian state given
in Eq.~\eqref{eq:cm-tfm}.
When we do so, we see that the squeezing amplitude $r_\s$ is related to the 
ratio between the minor and major axes and the squeezing angle $\varphi_\s$ 
is just the inclination angle $\delta$ between the $\tilde{v}$-axis and
the major axis of the ellipse.
In other words, we obtain that
\begin{equation}
\f{\bar{b}^2}{\bar{a}^2}
= \mathrm{e}^{-4\,r_{\s}},\quad  \delta=\varphi_\s.
\end{equation}  
We find that the additional parameter $N_{\s}$ we had encountered in the
two field case [see Eq.~\eqref{eq:cm-tfm}] is related to the area of the
ellipse by the formula
\begin{equation}
\mathrm{Area}= \pi\, \bar{a}\, \bar{b}
=2\, \pi\, \l(2\, N_{\s}+1\r).\label{Area_in_N}
\end{equation}
Note that, in the case of single field models, since $N_{\s}=0$, the 
area of the Wigner ellipse does not change.

%%%%%%%%%%%%%%%%%%%%%%%%%%%%%%%%%%%%%%%%%%%%%%%%%%%%%%%%%%%%%%%%%%%%%%%%%%%%%%%

\section{Conventions of the squeezing parameters}\label{app:conventions-sq}

We should point out that, for the squeezing parameters, we have followed the 
convention adopted earlier in some of the literature (for instance, in 
Ref.~\cite{Polarski:1995jg}). 
Let us consider the case of single field inflationary models.
The convention is most transparent in the following parametrization of the 
Bogoliubov coefficients:
\begin{equation}
{\tilde \alpha} =\mathrm{e}^{-i\, \vartheta}\,\cosh r,\quad
{\tilde \beta}=\mathrm{e}^{i\,(2\, \varphi +\vartheta)}\,\sinh r,\label{eq:sp-2}
\end{equation}
with the quantity $r$ assumed to be positive.
However, since we have not introduced the Bogoliubov coefficients in this work,
we shall explain our convention in terms of variances. 

From the discussion in App.~\ref{app:gwe}, it should be clear that the quantity
$\mathrm{e}^{-4\,r}$ corresponds to the ratio between the minor and the major 
axes of the Wigner ellipse, while $\varphi$ corresponds to the angle of the major 
axis of the ellipse with respect to the $\tilde{v}$-axis. 
Note that the parameter~$\vartheta$ does not appear in the covariance 
matrix~\eqref{eq:cm-sfm}. 
One requires only one angle to specify the orientation of the Wigner ellipse,
and this orientation may be reached by many different combinations of squeezing 
and rotation (in this context, see Ref.~\cite{cariolaro2015quantum}). 
In the literature on cosmology, the evolution from the vacuum state is usually 
parametrized as a rotation followed by a squeezing~\cite{Martin:2015qta}. 
In this case, the rotation operator acting on the initial vacuum state leaves 
it invariant and only the squeezing parameters appear in the final state. 

As we have discussed earlier, in most inflationary scenarios, at late times, 
$\varphi \to 0$ for a wide range of modes.
This implies that, towards the end of inflation, the major axis of the Wigner 
ellipse will lie along the $\tilde{v}$-axis. 
In such a limit, the variances are given by
\begin{equation}
\langle \hat{v}^2\rangle 
= \f{\mathrm{e}^{2\,r}}{2\,k}, \quad 
\langle \hat{p}^2\rangle = \f{k}{2}\,\mathrm{e}^{-2r}.
\end{equation}
As the squeezing amplitude~$r$ increases on super-Hubble scales, the variance in 
momentum becomes very small, whereas the variance in position becomes very large.
In other words, at late times, the Wigner ellipse is highly squeezed along the 
$\tilde{v}$-direction and lies on the $\tilde{v}$-axis.

We should mention that our convention matches with that of 
Ref.~\cite{Grishchuk:1993ds}
(except for a change in sign of $\vartheta$, which anyway does 
not appear in our covariance matrix), but differs from the 
convention used in Refs.~\cite{Martin:2015qta,Martin:2021znx} 
where the squeezing angle $\varphi$ is measured from 
the~$\tilde{p}$-axis.

%%%%%%%%%%%%%%%%%%%%%%%%%%%%%%%%%%%%%%%%%%%%%%%%%%%%%%%%%%%%%%%%%%%%%%%%%%%%%%%

%%%%%%%%%%%%%%%%%%%%%%%%%%%%%%%%%%%%%%%%%%%%%%%%%%%%%%%%%%%%%%%%%%%%%%%%%%%%%%%
%\bibliography{sn-bibliography}
%\bibliographystyle{JHEP}
\bibliography{sps-mb-august-2017,mybibliography-gravity,mybibliography-cosmology,
mybibliography-quantum,mybibliography-maths}


\begin{thebibliography}{86}
\ifx \bisbn   \undefined \def \bisbn  #1{ISBN #1}\fi
\ifx \binits  \undefined \def \binits#1{#1}\fi
\ifx \bauthor  \undefined \def \bauthor#1{#1}\fi
\ifx \batitle  \undefined \def \batitle#1{#1}\fi
\ifx \bjtitle  \undefined \def \bjtitle#1{#1}\fi
\ifx \bvolume  \undefined \def \bvolume#1{\textbf{#1}}\fi
\ifx \byear  \undefined \def \byear#1{#1}\fi
\ifx \bissue  \undefined \def \bissue#1{#1}\fi
\ifx \bfpage  \undefined \def \bfpage#1{#1}\fi
\ifx \blpage  \undefined \def \blpage #1{#1}\fi
\ifx \burl  \undefined \def \burl#1{\textsf{#1}}\fi
\ifx \doiurl  \undefined \def \doiurl#1{\url{https://doi.org/#1}}\fi
\ifx \betal  \undefined \def \betal{\textit{et al.}}\fi
\ifx \binstitute  \undefined \def \binstitute#1{#1}\fi
\ifx \binstitutionaled  \undefined \def \binstitutionaled#1{#1}\fi
\ifx \bctitle  \undefined \def \bctitle#1{#1}\fi
\ifx \beditor  \undefined \def \beditor#1{#1}\fi
\ifx \bpublisher  \undefined \def \bpublisher#1{#1}\fi
\ifx \bbtitle  \undefined \def \bbtitle#1{#1}\fi
\ifx \bedition  \undefined \def \bedition#1{#1}\fi
\ifx \bseriesno  \undefined \def \bseriesno#1{#1}\fi
\ifx \blocation  \undefined \def \blocation#1{#1}\fi
\ifx \bsertitle  \undefined \def \bsertitle#1{#1}\fi
\ifx \bsnm \undefined \def \bsnm#1{#1}\fi
\ifx \bsuffix \undefined \def \bsuffix#1{#1}\fi
\ifx \bparticle \undefined \def \bparticle#1{#1}\fi
\ifx \barticle \undefined \def \barticle#1{#1}\fi
\bibcommenthead
\ifx \bconfdate \undefined \def \bconfdate #1{#1}\fi
\ifx \botherref \undefined \def \botherref #1{#1}\fi
\ifx \url \undefined \def \url#1{\textsf{#1}}\fi
\ifx \bchapter \undefined \def \bchapter#1{#1}\fi
\ifx \bbook \undefined \def \bbook#1{#1}\fi
\ifx \bcomment \undefined \def \bcomment#1{#1}\fi
\ifx \oauthor \undefined \def \oauthor#1{#1}\fi
\ifx \citeauthoryear \undefined \def \citeauthoryear#1{#1}\fi
\ifx \endbibitem  \undefined \def \endbibitem {}\fi
\ifx \bconflocation  \undefined \def \bconflocation#1{#1}\fi
\ifx \arxivurl  \undefined \def \arxivurl#1{\textsf{#1}}\fi
\csname PreBibitemsHook\endcsname

%%% 1
\bibitem{Planck:2015mrs}
\begin{barticle}
\bauthor{\bsnm{Adam}, \binits{R.}}, \betal:
\batitle{{Planck 2015 results. I. Overview of products and scientific
  results}}.
\bjtitle{Astron. Astrophys.}
\bvolume{594},
\bfpage{1}
(\byear{2016})
{\href{https://arxiv.org/abs/1502.01582}{{arXiv:1502.01582}}}
{[astro-ph.CO]}.
\doiurl{10.1051/0004-6361/201527101}
\end{barticle}
\endbibitem

%%% 2
\bibitem{Planck:2018nkj}
\begin{barticle}
\bauthor{\bsnm{Aghanim}, \binits{N.}}, \betal:
\batitle{{Planck 2018 results. I. Overview and the cosmological legacy of
  Planck}}.
\bjtitle{Astron. Astrophys.}
\bvolume{641},
\bfpage{1}
(\byear{2020})
{\href{https://arxiv.org/abs/1807.06205}{{arXiv:1807.06205}}}
{[astro-ph.CO]}.
\doiurl{10.1051/0004-6361/201833880}
\end{barticle}
\endbibitem

%%% 3
\bibitem{BICEPKeck:2021bsr}
\begin{barticle}
\bauthor{\bsnm{Ade}, \binits{P.A.R.}}, \betal:
\batitle{{Bicep/KeckXV: The Bicep3 Cosmic Microwave Background Polarimeter and
  the First Three-year Data Set}}.
\bjtitle{Astrophys. J.}
\bvolume{927}(\bissue{1}),
\bfpage{77}
(\byear{2022})
{\href{https://arxiv.org/abs/2110.00482}{{arXiv:2110.00482}}}
{[astro-ph.IM]}.
\doiurl{10.3847/1538-4357/ac4886}
\end{barticle}
\endbibitem

%%% 4
\bibitem{Mukhanov:1990me}
\begin{barticle}
\bauthor{\bsnm{Mukhanov}, \binits{V.F.}},
\bauthor{\bsnm{Feldman}, \binits{H.A.}},
\bauthor{\bsnm{Brandenberger}, \binits{R.H.}}:
\batitle{{Theory of cosmological perturbations. Part 1. Classical
  perturbations. Part 2. Quantum theory of perturbations. Part 3. Extensions}}.
\bjtitle{Phys. Rept.}
\bvolume{215},
\bfpage{203}--\blpage{333}
(\byear{1992}).
\doiurl{10.1016/0370-1573(92)90044-Z}
\end{barticle}
\endbibitem

%%% 5
\bibitem{Martin:2003bt}
\begin{barticle}
\bauthor{\bsnm{Martin}, \binits{J.}}:
\batitle{{Inflation and precision cosmology}}.
\bjtitle{Braz. J. Phys.}
\bvolume{34},
\bfpage{1307}--\blpage{1321}
(\byear{2004})
{\href{https://arxiv.org/abs/astro-ph/0312492}{{arXiv:astro-ph/0312492}}}.
\doiurl{10.1590/S0103-97332004000700005}
\end{barticle}
\endbibitem

%%% 6
\bibitem{Martin:2004um}
\begin{barticle}
\bauthor{\bsnm{Martin}, \binits{J.}}:
\batitle{{Inflationary cosmological perturbations of quantum-mechanical
  origin}}.
\bjtitle{Lect. Notes Phys.}
\bvolume{669},
\bfpage{199}--\blpage{244}
(\byear{2005})
{\href{https://arxiv.org/abs/hep-th/0406011}{{arXiv:hep-th/0406011}}}.
\doiurl{10.1007/11377306_7}
\end{barticle}
\endbibitem

%%% 7
\bibitem{Bassett:2005xm}
\begin{barticle}
\bauthor{\bsnm{Bassett}, \binits{B.A.}},
\bauthor{\bsnm{Tsujikawa}, \binits{S.}},
\bauthor{\bsnm{Wands}, \binits{D.}}:
\batitle{{Inflation dynamics and reheating}}.
\bjtitle{Rev. Mod. Phys.}
\bvolume{78},
\bfpage{537}--\blpage{589}
(\byear{2006})
{\href{https://arxiv.org/abs/astro-ph/0507632}{{arXiv:astro-ph/0507632}}}.
\doiurl{10.1103/RevModPhys.78.537}
\end{barticle}
\endbibitem

%%% 8
\bibitem{Sriramkumar:2009kg}
\begin{botherref}
\oauthor{\bsnm{Sriramkumar}, \binits{L.}}:
{An introduction to inflation and cosmological perturbation theory}
(2009)
{\href{https://arxiv.org/abs/0904.4584}{{arXiv:0904.4584}}}
{[astro-ph.CO]}
\end{botherref}
\endbibitem

%%% 9
\bibitem{Baumann:2008bn}
\begin{barticle}
\bauthor{\bsnm{Baumann}, \binits{D.}},
\bauthor{\bsnm{Peiris}, \binits{H.V.}}:
\batitle{{Cosmological Inflation: Theory and Observations}}.
\bjtitle{Adv. Sci. Lett.}
\bvolume{2},
\bfpage{105}--\blpage{120}
(\byear{2009})
{\href{https://arxiv.org/abs/0810.3022}{{arXiv:0810.3022}}}
{[astro-ph]}.
\doiurl{10.1166/asl.2009.1019}
\end{barticle}
\endbibitem

%%% 10
\bibitem{Baumann:2009ds}
\begin{bchapter}
\bauthor{\bsnm{Baumann}, \binits{D.}}:
\bctitle{Inflation}.
In: \bbtitle{Theoretical Advanced Study Institute in Elementary Particle
  Physics: Physics of the Large and the Small},
pp. \bfpage{523}--\blpage{686}
(\byear{2011}).
\doiurl{10.1142/9789814327183_0010}
\end{bchapter}
\endbibitem

%%% 11
\bibitem{Sriramkumar:2012mik}
\begin{bbook}
\bauthor{\bsnm{Sriramkumar}, \binits{L.}}:
In: \beditor{\bsnm{Sriramkumar}, \binits{L.}},
\beditor{\bsnm{Seshadri}, \binits{T.R.}} (eds.)
\bbtitle{{On the generation and evolution of perturbations during inflation and
  reheating}},
pp. \bfpage{207}--\blpage{249}
(\byear{2012}).
\doiurl{10.1142/9789814322072_0008}
\end{bbook}
\endbibitem

%%% 12
\bibitem{Linde:2014nna}
\begin{bchapter}
\bauthor{\bsnm{Linde}, \binits{A.}}:
\bctitle{Inflationary cosmology after planck 2013}.
In: \bbtitle{100e Ecole d'Ete de Physique: Post-Planck Cosmology},
pp. \bfpage{231}--\blpage{316}
(\byear{2015}).
\doiurl{10.1093/acprof:oso/9780198728856.003.0006}
\end{bchapter}
\endbibitem

%%% 13
\bibitem{Martin:2015dha}
\begin{barticle}
\bauthor{\bsnm{Martin}, \binits{J.}}:
\batitle{{The Observational Status of Cosmic Inflation after Planck}}.
\bjtitle{Astrophys. Space Sci. Proc.}
\bvolume{45},
\bfpage{41}--\blpage{134}
(\byear{2016})
{\href{https://arxiv.org/abs/1502.05733}{{arXiv:1502.05733}}}
{[astro-ph.CO]}.
\doiurl{10.1007/978-3-319-44769-8_2}
\end{barticle}
\endbibitem

%%% 14
\bibitem{Albrecht:1992kf}
\begin{barticle}
\bauthor{\bsnm{Albrecht}, \binits{A.}},
\bauthor{\bsnm{Ferreira}, \binits{P.}},
\bauthor{\bsnm{Joyce}, \binits{M.}},
\bauthor{\bsnm{Prokopec}, \binits{T.}}:
\batitle{{Inflation and squeezed quantum states}}.
\bjtitle{Phys. Rev.}
\bvolume{D50},
\bfpage{4807}--\blpage{4820}
(\byear{1994})
{\href{https://arxiv.org/abs/astro-ph/9303001}{{arXiv:astro-ph/9303001}}}
{[astro-ph]}.
\doiurl{10.1103/PhysRevD.50.4807}
\end{barticle}
\endbibitem

%%% 15
\bibitem{Polarski:1995jg}
\begin{barticle}
\bauthor{\bsnm{Polarski}, \binits{D.}},
\bauthor{\bsnm{Starobinsky}, \binits{A.A.}}:
\batitle{{Semiclassicality and decoherence of cosmological perturbations}}.
\bjtitle{Class. Quant. Grav.}
\bvolume{13},
\bfpage{377}--\blpage{392}
(\byear{1996})
{\href{https://arxiv.org/abs/gr-qc/9504030}{{arXiv:gr-qc/9504030}}}
{[gr-qc]}.
\doiurl{10.1088/0264-9381/13/3/006}
\end{barticle}
\endbibitem

%%% 16
\bibitem{Kiefer:1998qe}
\begin{barticle}
\bauthor{\bsnm{Kiefer}, \binits{C.}},
\bauthor{\bsnm{Polarski}, \binits{D.}},
\bauthor{\bsnm{Starobinsky}, \binits{A.A.}}:
\batitle{{Quantum to classical transition for fluctuations in the early
  universe}}.
\bjtitle{Int. J. Mod. Phys.}
\bvolume{D7},
\bfpage{455}--\blpage{462}
(\byear{1998})
{\href{https://arxiv.org/abs/gr-qc/9802003}{{arXiv:gr-qc/9802003}}}
{[gr-qc]}.
\doiurl{10.1142/S0218271898000292}
\end{barticle}
\endbibitem

%%% 17
\bibitem{Kiefer:1998jk}
\begin{barticle}
\bauthor{\bsnm{Kiefer}, \binits{C.}},
\bauthor{\bsnm{Polarski}, \binits{D.}}:
\batitle{{Emergence of classicality for primordial fluctuations: Concepts and
  analogies}}.
\bjtitle{Annalen Phys.}
\bvolume{7},
\bfpage{137}--\blpage{158}
(\byear{1998})
{\href{https://arxiv.org/abs/gr-qc/9805014}{{arXiv:gr-qc/9805014}}}.
\doiurl{10.1002/andp.2090070302}
\end{barticle}
\endbibitem

%%% 18
\bibitem{Kiefer:2006je}
\begin{barticle}
\bauthor{\bsnm{Kiefer}, \binits{C.}},
\bauthor{\bsnm{Lohmar}, \binits{I.}},
\bauthor{\bsnm{Polarski}, \binits{D.}},
\bauthor{\bsnm{Starobinsky}, \binits{A.A.}}:
\batitle{{Pointer states for primordial fluctuations in inflationary
  cosmology}}.
\bjtitle{Class. Quant. Grav.}
\bvolume{24},
\bfpage{1699}--\blpage{1718}
(\byear{2007})
{\href{https://arxiv.org/abs/astro-ph/0610700}{{arXiv:astro-ph/0610700}}}
{[astro-ph]}.
\doiurl{10.1088/0264-9381/24/7/002}
\end{barticle}
\endbibitem

%%% 19
\bibitem{Martin:2007bw}
\begin{barticle}
\bauthor{\bsnm{Martin}, \binits{J.}}:
\batitle{{Inflationary perturbations: The Cosmological Schwinger effect}}.
\bjtitle{Lect. Notes Phys.}
\bvolume{738},
\bfpage{193}--\blpage{241}
(\byear{2008})
{\href{https://arxiv.org/abs/0704.3540}{{arXiv:0704.3540}}}
{[hep-th]}.
\doiurl{10.1007/978-3-540-74353-8_6}
\end{barticle}
\endbibitem

%%% 20
\bibitem{Kiefer:2008ku}
\begin{barticle}
\bauthor{\bsnm{Kiefer}, \binits{C.}},
\bauthor{\bsnm{Polarski}, \binits{D.}}:
\batitle{{Why do cosmological perturbations look classical to us?}}
\bjtitle{Adv. Sci. Lett.}
\bvolume{2},
\bfpage{164}--\blpage{173}
(\byear{2009})
{\href{https://arxiv.org/abs/0810.0087}{{arXiv:0810.0087}}}
{[astro-ph]}.
\doiurl{10.1166/asl.2009.1023}
\end{barticle}
\endbibitem

%%% 21
\bibitem{Martin:2012pea}
\begin{barticle}
\bauthor{\bsnm{Martin}, \binits{J.}},
\bauthor{\bsnm{Vennin}, \binits{V.}},
\bauthor{\bsnm{Peter}, \binits{P.}}:
\batitle{{Cosmological Inflation and the Quantum Measurement Problem}}.
\bjtitle{Phys. Rev.}
\bvolume{D86},
\bfpage{103524}
(\byear{2012})
{\href{https://arxiv.org/abs/1207.2086}{{arXiv:1207.2086}}}
{[hep-th]}.
\doiurl{10.1103/PhysRevD.86.103524}
\end{barticle}
\endbibitem

%%% 22
\bibitem{Martin:2012ua}
\begin{barticle}
\bauthor{\bsnm{Martin}, \binits{J.}}:
\batitle{{The Quantum State of Inflationary Perturbations}}.
\bjtitle{J. Phys. Conf. Ser.}
\bvolume{405},
\bfpage{012004}
(\byear{2012})
{\href{https://arxiv.org/abs/1209.3092}{{arXiv:1209.3092}}}
{[hep-th]}.
\doiurl{10.1088/1742-6596/405/1/012004}
\end{barticle}
\endbibitem

%%% 23
\bibitem{Martin:2015qta}
\begin{barticle}
\bauthor{\bsnm{Martin}, \binits{J.}},
\bauthor{\bsnm{Vennin}, \binits{V.}}:
\batitle{{Quantum Discord of Cosmic Inflation: Can we Show that CMB
  Anisotropies are of Quantum-Mechanical Origin?}}
\bjtitle{Phys. Rev.}
\bvolume{D93}(\bissue{2}),
\bfpage{023505}
(\byear{2016})
{\href{https://arxiv.org/abs/1510.04038}{{arXiv:1510.04038}}}
{[astro-ph.CO]}.
\doiurl{10.1103/PhysRevD.93.023505}
\end{barticle}
\endbibitem

%%% 24
\bibitem{Planck:2015sxf}
\begin{barticle}
\bauthor{\bsnm{Ade}, \binits{P.A.R.}}, \betal:
\batitle{{Planck 2015 results. XX. Constraints on inflation}}.
\bjtitle{Astron. Astrophys.}
\bvolume{594},
\bfpage{20}
(\byear{2016})
{\href{https://arxiv.org/abs/1502.02114}{{arXiv:1502.02114}}}
{[astro-ph.CO]}.
\doiurl{10.1051/0004-6361/201525898}
\end{barticle}
\endbibitem

%%% 25
\bibitem{Planck:2018jri}
\begin{barticle}
\bauthor{\bsnm{Akrami}, \binits{Y.}}, \betal:
\batitle{{Planck 2018 results. X. Constraints on inflation}}.
\bjtitle{Astron. Astrophys.}
\bvolume{641},
\bfpage{10}
(\byear{2020})
{\href{https://arxiv.org/abs/1807.06211}{{arXiv:1807.06211}}}
{[astro-ph.CO]}.
\doiurl{10.1051/0004-6361/201833887}
\end{barticle}
\endbibitem

%%% 26
\bibitem{Martin:2013tda}
\begin{barticle}
\bauthor{\bsnm{Martin}, \binits{J.}},
\bauthor{\bsnm{Ringeval}, \binits{C.}},
\bauthor{\bsnm{Vennin}, \binits{V.}}:
\batitle{{Encyclop\ae{}dia Inflationaris}}.
\bjtitle{Phys. Dark Univ.}
\bvolume{5-6},
\bfpage{75}--\blpage{235}
(\byear{2014})
{\href{https://arxiv.org/abs/1303.3787}{{arXiv:1303.3787}}}
{[astro-ph.CO]}.
\doiurl{10.1016/j.dark.2014.01.003}
\end{barticle}
\endbibitem

%%% 27
\bibitem{Martin:2013nzq}
\begin{barticle}
\bauthor{\bsnm{Martin}, \binits{J.}},
\bauthor{\bsnm{Ringeval}, \binits{C.}},
\bauthor{\bsnm{Trotta}, \binits{R.}},
\bauthor{\bsnm{Vennin}, \binits{V.}}:
\batitle{{The Best Inflationary Models After Planck}}.
\bjtitle{JCAP}
\bvolume{03},
\bfpage{039}
(\byear{2014})
{\href{https://arxiv.org/abs/1312.3529}{{arXiv:1312.3529}}}
{[astro-ph.CO]}.
\doiurl{10.1088/1475-7516/2014/03/039}
\end{barticle}
\endbibitem

%%% 28
\bibitem{Bird:2016dcv}
\begin{barticle}
\bauthor{\bsnm{Bird}, \binits{S.}},
\bauthor{\bsnm{Cholis}, \binits{I.}},
\bauthor{\bsnm{Mu\~noz}, \binits{J.B.}},
\bauthor{\bsnm{Ali-Ha\"\i{}moud}, \binits{Y.}},
\bauthor{\bsnm{Kamionkowski}, \binits{M.}},
\bauthor{\bsnm{Kovetz}, \binits{E.D.}},
\bauthor{\bsnm{Raccanelli}, \binits{A.}},
\bauthor{\bsnm{Riess}, \binits{A.G.}}:
\batitle{{Did LIGO detect dark matter?}}
\bjtitle{Phys. Rev. Lett.}
\bvolume{116}(\bissue{20}),
\bfpage{201301}
(\byear{2016})
{\href{https://arxiv.org/abs/1603.00464}{{arXiv:1603.00464}}}
{[astro-ph.CO]}.
\doiurl{10.1103/PhysRevLett.116.201301}
\end{barticle}
\endbibitem

%%% 29
\bibitem{DeLuca:2020qqa}
\begin{barticle}
\bauthor{\bsnm{De~Luca}, \binits{V.}},
\bauthor{\bsnm{Franciolini}, \binits{G.}},
\bauthor{\bsnm{Pani}, \binits{P.}},
\bauthor{\bsnm{Riotto}, \binits{A.}}:
\batitle{{Primordial Black Holes Confront LIGO/Virgo data: Current situation}}.
\bjtitle{JCAP}
\bvolume{06},
\bfpage{044}
(\byear{2020})
{\href{https://arxiv.org/abs/2005.05641}{{arXiv:2005.05641}}}
{[astro-ph.CO]}.
\doiurl{10.1088/1475-7516/2020/06/044}
\end{barticle}
\endbibitem

%%% 30
\bibitem{Jedamzik:2020ypm}
\begin{barticle}
\bauthor{\bsnm{Jedamzik}, \binits{K.}}:
\batitle{{Primordial Black Hole Dark Matter and the LIGO/Virgo observations}}.
\bjtitle{JCAP}
\bvolume{09},
\bfpage{022}
(\byear{2020})
{\href{https://arxiv.org/abs/2006.11172}{{arXiv:2006.11172}}}
{[astro-ph.CO]}.
\doiurl{10.1088/1475-7516/2020/09/022}
\end{barticle}
\endbibitem

%%% 31
\bibitem{Jedamzik:2020omx}
\begin{barticle}
\bauthor{\bsnm{Jedamzik}, \binits{K.}}:
\batitle{{Consistency of Primordial Black Hole Dark Matter with LIGO/Virgo
  Merger Rates}}.
\bjtitle{Phys. Rev. Lett.}
\bvolume{126}(\bissue{5}),
\bfpage{051302}
(\byear{2021})
{\href{https://arxiv.org/abs/2007.03565}{{arXiv:2007.03565}}}
{[astro-ph.CO]}.
\doiurl{10.1103/PhysRevLett.126.051302}
\end{barticle}
\endbibitem

%%% 32
\bibitem{Chongchitnan:2006wx}
\begin{barticle}
\bauthor{\bsnm{Chongchitnan}, \binits{S.}},
\bauthor{\bsnm{Efstathiou}, \binits{G.}}:
\batitle{{Accuracy of slow-roll formulae for inflationary perturbations:
  implications for primordial black hole formation}}.
\bjtitle{JCAP}
\bvolume{01},
\bfpage{011}
(\byear{2007})
{\href{https://arxiv.org/abs/astro-ph/0611818}{{arXiv:astro-ph/0611818}}}.
\doiurl{10.1088/1475-7516/2007/01/011}
\end{barticle}
\endbibitem

%%% 33
\bibitem{Garcia-Bellido:2017mdw}
\begin{barticle}
\bauthor{\bsnm{Garcia-Bellido}, \binits{J.}},
\bauthor{\bsnm{Ruiz~Morales}, \binits{E.}}:
\batitle{{Primordial black holes from single field models of inflation}}.
\bjtitle{Phys. Dark Univ.}
\bvolume{18},
\bfpage{47}--\blpage{54}
(\byear{2017})
{\href{https://arxiv.org/abs/1702.03901}{{arXiv:1702.03901}}}
{[astro-ph.CO]}.
\doiurl{10.1016/j.dark.2017.09.007}
\end{barticle}
\endbibitem

%%% 34
\bibitem{Ballesteros:2017fsr}
\begin{barticle}
\bauthor{\bsnm{Ballesteros}, \binits{G.}},
\bauthor{\bsnm{Taoso}, \binits{M.}}:
\batitle{{Primordial black hole dark matter from single field inflation}}.
\bjtitle{Phys. Rev. D}
\bvolume{97}(\bissue{2}),
\bfpage{023501}
(\byear{2018})
{\href{https://arxiv.org/abs/1709.05565}{{arXiv:1709.05565}}}
{[hep-ph]}.
\doiurl{10.1103/PhysRevD.97.023501}
\end{barticle}
\endbibitem

%%% 35
\bibitem{Germani:2017bcs}
\begin{barticle}
\bauthor{\bsnm{Germani}, \binits{C.}},
\bauthor{\bsnm{Prokopec}, \binits{T.}}:
\batitle{{On primordial black holes from an inflection point}}.
\bjtitle{Phys. Dark Univ.}
\bvolume{18},
\bfpage{6}--\blpage{10}
(\byear{2017})
{\href{https://arxiv.org/abs/1706.04226}{{arXiv:1706.04226}}}
{[astro-ph.CO]}.
\doiurl{10.1016/j.dark.2017.09.001}
\end{barticle}
\endbibitem

%%% 36
\bibitem{Kannike:2017bxn}
\begin{barticle}
\bauthor{\bsnm{Kannike}, \binits{K.}},
\bauthor{\bsnm{Marzola}, \binits{L.}},
\bauthor{\bsnm{Raidal}, \binits{M.}},
\bauthor{\bsnm{Veerm\"ae}, \binits{H.}}:
\batitle{{Single Field Double Inflation and Primordial Black Holes}}.
\bjtitle{JCAP}
\bvolume{09},
\bfpage{020}
(\byear{2017})
{\href{https://arxiv.org/abs/1705.06225}{{arXiv:1705.06225}}}
{[astro-ph.CO]}.
\doiurl{10.1088/1475-7516/2017/09/020}
\end{barticle}
\endbibitem

%%% 37
\bibitem{Dalianis:2018frf}
\begin{barticle}
\bauthor{\bsnm{Dalianis}, \binits{I.}},
\bauthor{\bsnm{Kehagias}, \binits{A.}},
\bauthor{\bsnm{Tringas}, \binits{G.}}:
\batitle{{Primordial black holes from \ensuremath{\alpha}-attractors}}.
\bjtitle{JCAP}
\bvolume{01},
\bfpage{037}
(\byear{2019})
{\href{https://arxiv.org/abs/1805.09483}{{arXiv:1805.09483}}}
{[astro-ph.CO]}.
\doiurl{10.1088/1475-7516/2019/01/037}
\end{barticle}
\endbibitem

%%% 38
\bibitem{Ragavendra:2020sop}
\begin{barticle}
\bauthor{\bsnm{Ragavendra}, \binits{H.V.}},
\bauthor{\bsnm{Saha}, \binits{P.}},
\bauthor{\bsnm{Sriramkumar}, \binits{L.}},
\bauthor{\bsnm{Silk}, \binits{J.}}:
\batitle{{Primordial black holes and secondary gravitational waves from
  ultraslow roll and punctuated inflation}}.
\bjtitle{Phys. Rev. D}
\bvolume{103}(\bissue{8}),
\bfpage{083510}
(\byear{2021})
{\href{https://arxiv.org/abs/2008.12202}{{arXiv:2008.12202}}}
{[astro-ph.CO]}.
\doiurl{10.1103/PhysRevD.103.083510}
\end{barticle}
\endbibitem

%%% 39
\bibitem{Palma:2020ejf}
\begin{barticle}
\bauthor{\bsnm{Palma}, \binits{G.A.}},
\bauthor{\bsnm{Sypsas}, \binits{S.}},
\bauthor{\bsnm{Zenteno}, \binits{C.}}:
\batitle{{Seeding primordial black holes in multifield inflation}}.
\bjtitle{Phys. Rev. Lett.}
\bvolume{125}(\bissue{12}),
\bfpage{121301}
(\byear{2020})
{\href{https://arxiv.org/abs/2004.06106}{{arXiv:2004.06106}}}
{[astro-ph.CO]}.
\doiurl{10.1103/PhysRevLett.125.121301}
\end{barticle}
\endbibitem

%%% 40
\bibitem{Fumagalli:2020adf}
\begin{botherref}
\oauthor{\bsnm{Fumagalli}, \binits{J.}},
\oauthor{\bsnm{Renaux-Petel}, \binits{S.}},
\oauthor{\bsnm{Ronayne}, \binits{J.W.}},
\oauthor{\bsnm{Witkowski}, \binits{L.T.}}:
{Turning in the landscape: a new mechanism for generating Primordial Black
  Holes}
(2020)
{\href{https://arxiv.org/abs/2004.08369}{{arXiv:2004.08369}}}
{[hep-th]}
\end{botherref}
\endbibitem

%%% 41
\bibitem{Braglia:2020eai}
\begin{barticle}
\bauthor{\bsnm{Braglia}, \binits{M.}},
\bauthor{\bsnm{Hazra}, \binits{D.K.}},
\bauthor{\bsnm{Finelli}, \binits{F.}},
\bauthor{\bsnm{Smoot}, \binits{G.F.}},
\bauthor{\bsnm{Sriramkumar}, \binits{L.}},
\bauthor{\bsnm{Starobinsky}, \binits{A.A.}}:
\batitle{{Generating PBHs and small-scale GWs in two-field models of
  inflation}}.
\bjtitle{JCAP}
\bvolume{08},
\bfpage{001}
(\byear{2020})
{\href{https://arxiv.org/abs/2005.02895}{{arXiv:2005.02895}}}
{[astro-ph.CO]}.
\doiurl{10.1088/1475-7516/2020/08/001}
\end{barticle}
\endbibitem

%%% 42
\bibitem{Choi:2021yxz}
\begin{barticle}
\bauthor{\bsnm{Choi}, \binits{K.-Y.}},
\bauthor{\bsnm{Kang}, \binits{S.-b.}},
\bauthor{\bsnm{Raveendran}, \binits{R.N.}}:
\batitle{{Reconstruction of potentials of hybrid inflation in the light of
  primordial black hole formation}}.
\bjtitle{JCAP}
\bvolume{06},
\bfpage{054}
(\byear{2021})
{\href{https://arxiv.org/abs/2102.02461}{{arXiv:2102.02461}}}
{[astro-ph.CO]}.
\doiurl{10.1088/1475-7516/2021/06/054}
\end{barticle}
\endbibitem

%%% 43
\bibitem{dePutter:2019xxv}
\begin{barticle}
\bauthor{\bparticle{de} \bsnm{Putter}, \binits{R.}},
\bauthor{\bsnm{Dor\'e}, \binits{O.}}:
\batitle{{In search of an observational quantum signature of the primordial
  perturbations in slow-roll and ultraslow-roll inflation}}.
\bjtitle{Phys. Rev. D}
\bvolume{101}(\bissue{4}),
\bfpage{043511}
(\byear{2020})
{\href{https://arxiv.org/abs/1905.01394}{{arXiv:1905.01394}}}
{[gr-qc]}.
\doiurl{10.1103/PhysRevD.101.043511}
\end{barticle}
\endbibitem

%%% 44
\bibitem{Figueroa:2021zah}
\begin{barticle}
\bauthor{\bsnm{Figueroa}, \binits{D.G.}},
\bauthor{\bsnm{Raatikainen}, \binits{S.}},
\bauthor{\bsnm{Rasanen}, \binits{S.}},
\bauthor{\bsnm{Tomberg}, \binits{E.}}:
\batitle{{Implications of stochastic effects for primordial black hole
  production in ultra-slow-roll inflation}}.
\bjtitle{JCAP}
\bvolume{05}(\bissue{05}),
\bfpage{027}
(\byear{2022})
{\href{https://arxiv.org/abs/2111.07437}{{arXiv:2111.07437}}}
{[astro-ph.CO]}.
\doiurl{10.1088/1475-7516/2022/05/027}
\end{barticle}
\endbibitem

%%% 45
\bibitem{Mukhanov-Winitzky}
\begin{bbook}
\bauthor{\bsnm{Mukhanov}, \binits{V.}},
\bauthor{\bsnm{Winitzki}, \binits{S.}}:
\bbtitle{Introduction to Quantum Effects in Gravity},
\bedition{1st} edn.
\bpublisher{Cambridge University Press},
\blocation{Cambridge, UK}
(\byear{2007}).
\doiurl{10.1017/CBO9780511809149}
\end{bbook}
\endbibitem

%%% 46
\bibitem{Battarra:2013cha}
\begin{barticle}
\bauthor{\bsnm{Battarra}, \binits{L.}},
\bauthor{\bsnm{Lehners}, \binits{J.-L.}}:
\batitle{{Quantum-to-classical transition for ekpyrotic perturbations}}.
\bjtitle{Phys. Rev.}
\bvolume{D89}(\bissue{6}),
\bfpage{063516}
(\byear{2014})
{\href{https://arxiv.org/abs/1309.2281}{{arXiv:1309.2281}}}
{[hep-th]}.
\doiurl{10.1103/PhysRevD.89.063516}
\end{barticle}
\endbibitem

%%% 47
\bibitem{Kiefer:1991xy}
\begin{barticle}
\bauthor{\bsnm{Kiefer}, \binits{C.}}:
\batitle{{Functional Schrodinger equation for scalar QED}}.
\bjtitle{Phys. Rev.}
\bvolume{D45},
\bfpage{2044}--\blpage{2056}
(\byear{1992}).
\doiurl{10.1103/PhysRevD.45.2044}
\end{barticle}
\endbibitem

%%% 48
\bibitem{Hollowood:2017bil}
\begin{barticle}
\bauthor{\bsnm{Hollowood}, \binits{T.J.}},
\bauthor{\bsnm{McDonald}, \binits{J.I.}}:
\batitle{{Decoherence, discord and the quantum master equation for cosmological
  perturbations}}.
\bjtitle{Phys. Rev.}
\bvolume{D95}(\bissue{10}),
\bfpage{103521}
(\byear{2017})
{\href{https://arxiv.org/abs/1701.02235}{{arXiv:1701.02235}}}
{[gr-qc]}.
\doiurl{10.1103/PhysRevD.95.103521}
\end{barticle}
\endbibitem

%%% 49
\bibitem{Grishchuk:1993ds}
\begin{barticle}
\bauthor{\bsnm{Grishchuk}, \binits{L.P.}}:
\batitle{{Quantum effects in cosmology}}.
\bjtitle{Class. Quant. Grav.}
\bvolume{10},
\bfpage{2449}--\blpage{2478}
(\byear{1993})
{\href{https://arxiv.org/abs/gr-qc/9302036}{{arXiv:gr-qc/9302036}}}.
\doiurl{10.1088/0264-9381/10/12/006}
\end{barticle}
\endbibitem

%%% 50
\bibitem{Hillery:1983ms}
\begin{barticle}
\bauthor{\bsnm{Hillery}, \binits{M.}},
\bauthor{\bsnm{O'Connell}, \binits{R.F.}},
\bauthor{\bsnm{Scully}, \binits{M.O.}},
\bauthor{\bsnm{Wigner}, \binits{E.P.}}:
\batitle{{Distribution functions in physics: Fundamentals}}.
\bjtitle{Phys. Rept.}
\bvolume{106},
\bfpage{121}--\blpage{167}
(\byear{1984}).
\doiurl{10.1016/0370-1573(84)90160-1}
\end{barticle}
\endbibitem

%%% 51
\bibitem{case2008wigner}
\begin{barticle}
\bauthor{\bsnm{Case}, \binits{W.B.}}:
\batitle{Wigner functions and weyl transforms for pedestrians}.
\bjtitle{American Journal of Physics}
\bvolume{76}(\bissue{10}),
\bfpage{937}--\blpage{946}
(\byear{2008})
\end{barticle}
\endbibitem

%%% 52
\bibitem{weedbrook2012gaussian}
\begin{barticle}
\bauthor{\bsnm{Weedbrook}, \binits{C.}},
\bauthor{\bsnm{Pirandola}, \binits{S.}},
\bauthor{\bsnm{Garc{\'\i}a-Patr{\'o}n}, \binits{R.}},
\bauthor{\bsnm{Cerf}, \binits{N.J.}},
\bauthor{\bsnm{Ralph}, \binits{T.C.}},
\bauthor{\bsnm{Shapiro}, \binits{J.H.}},
\bauthor{\bsnm{Lloyd}, \binits{S.}}:
\batitle{Gaussian quantum information}.
\bjtitle{Reviews of Modern Physics}
\bvolume{84}(\bissue{2}),
\bfpage{621}
(\byear{2012})
\end{barticle}
\endbibitem

%%% 53
\bibitem{narcowich1990geometry}
\begin{barticle}
\bauthor{\bsnm{Narcowich}, \binits{F.J.}}:
\batitle{Geometry and uncertainty}.
\bjtitle{Journal of mathematical physics}
\bvolume{31}(\bissue{2}),
\bfpage{354}--\blpage{364}
(\byear{1990})
\end{barticle}
\endbibitem

%%% 54
\bibitem{Koksma:2010zi}
\begin{barticle}
\bauthor{\bsnm{Koksma}, \binits{J.F.}},
\bauthor{\bsnm{Prokopec}, \binits{T.}},
\bauthor{\bsnm{Schmidt}, \binits{M.G.}}:
\batitle{{Entropy and Correlators in Quantum Field Theory}}.
\bjtitle{Annals Phys.}
\bvolume{325},
\bfpage{1277}--\blpage{1303}
(\byear{2010})
{\href{https://arxiv.org/abs/1002.0749}{{arXiv:1002.0749}}}
{[hep-th]}.
\doiurl{10.1016/j.aop.2010.02.016}
\end{barticle}
\endbibitem

%%% 55
\bibitem{Martin:2021znx}
\begin{barticle}
\bauthor{\bsnm{Martin}, \binits{J.}},
\bauthor{\bsnm{Micheli}, \binits{A.}},
\bauthor{\bsnm{Vennin}, \binits{V.}}:
\batitle{{Discord and decoherence}}.
\bjtitle{JCAP}
\bvolume{04}(\bissue{04}),
\bfpage{051}
(\byear{2022})
{\href{https://arxiv.org/abs/2112.05037}{{arXiv:2112.05037}}}
{[quant-ph]}.
\doiurl{10.1088/1475-7516/2022/04/051}
\end{barticle}
\endbibitem

%%% 56
\bibitem{cariolaro2015quantum}
\begin{bbook}
\bauthor{\bsnm{Cariolaro}, \binits{G.}}:
\bbtitle{Quantum Communications}.
\bpublisher{Springer},
\blocation{Switzerland}
(\byear{2015})
\end{bbook}
\endbibitem

%%% 57
\bibitem{Starobinsky:1979ty}
\begin{barticle}
\bauthor{\bsnm{Starobinsky}, \binits{A.A.}}:
\batitle{{Spectrum of relict gravitational radiation and the early state of the
  universe}}.
\bjtitle{JETP Lett.}
\bvolume{30},
\bfpage{682}--\blpage{685}
(\byear{1979})
\end{barticle}
\endbibitem

%%% 58
\bibitem{Starobinsky:1980te}
\begin{barticle}
\bauthor{\bsnm{Starobinsky}, \binits{A.A.}}:
\batitle{{A New Type of Isotropic Cosmological Models Without Singularity}}.
\bjtitle{Phys. Lett. B}
\bvolume{91},
\bfpage{99}--\blpage{102}
(\byear{1980}).
\doiurl{10.1016/0370-2693(80)90670-X}
\end{barticle}
\endbibitem

%%% 59
\bibitem{Hazra:2012yn}
\begin{barticle}
\bauthor{\bsnm{Hazra}, \binits{D.K.}},
\bauthor{\bsnm{Sriramkumar}, \binits{L.}},
\bauthor{\bsnm{Martin}, \binits{J.}}:
\batitle{{BINGO: A code for the efficient computation of the scalar
  bi-spectrum}}.
\bjtitle{JCAP}
\bvolume{05},
\bfpage{026}
(\byear{2013})
{\href{https://arxiv.org/abs/1201.0926}{{arXiv:1201.0926}}}
{[astro-ph.CO]}.
\doiurl{10.1088/1475-7516/2013/05/026}
\end{barticle}
\endbibitem

%%% 60
\bibitem{Ragavendra:2020old}
\begin{botherref}
\oauthor{\bsnm{Ragavendra}, \binits{H.V.}},
\oauthor{\bsnm{Chowdhury}, \binits{D.}},
\oauthor{\bsnm{Sriramkumar}, \binits{L.}}:
{Suppression of scalar power on large scales and associated bispectra}
(2020)
{\href{https://arxiv.org/abs/2003.01099}{{arXiv:2003.01099}}}
{[astro-ph.CO]}
\end{botherref}
\endbibitem

%%% 61
\bibitem{Balaji:2022zur}
\begin{botherref}
\oauthor{\bsnm{Balaji}, \binits{S.}},
\oauthor{\bsnm{Ragavendra}, \binits{H.V.}},
\oauthor{\bsnm{Sethi}, \binits{S.K.}},
\oauthor{\bsnm{Silk}, \binits{J.}},
\oauthor{\bsnm{Sriramkumar}, \binits{L.}}:
{Observing nulling of primordial correlations via the 21 cm signal}
(2022)
{\href{https://arxiv.org/abs/2206.06386}{{arXiv:2206.06386}}}
{[astro-ph.CO]}
\end{botherref}
\endbibitem

%%% 62
\bibitem{Franciolini:2022pav}
\begin{botherref}
\oauthor{\bsnm{Franciolini}, \binits{G.}},
\oauthor{\bsnm{Urbano}, \binits{A.}}:
{Primordial black hole dark matter from inflation: the reverse engineering
  approach}
(2022)
{\href{https://arxiv.org/abs/2207.10056}{{arXiv:2207.10056}}}
{[astro-ph.CO]}
\end{botherref}
\endbibitem

%%% 63
\bibitem{Iacconi:2021ltm}
\begin{barticle}
\bauthor{\bsnm{Iacconi}, \binits{L.}},
\bauthor{\bsnm{Assadullahi}, \binits{H.}},
\bauthor{\bsnm{Fasiello}, \binits{M.}},
\bauthor{\bsnm{Wands}, \binits{D.}}:
\batitle{{Revisiting small-scale fluctuations in \ensuremath{\alpha}-attractor
  models of inflation}}.
\bjtitle{JCAP}
\bvolume{06}(\bissue{06}),
\bfpage{007}
(\byear{2022})
{\href{https://arxiv.org/abs/2112.05092}{{arXiv:2112.05092}}}
{[astro-ph.CO]}.
\doiurl{10.1088/1475-7516/2022/06/007}
\end{barticle}
\endbibitem

%%% 64
\bibitem{DiMarco:2002eb}
\begin{barticle}
\bauthor{\bsnm{Di~Marco}, \binits{F.}},
\bauthor{\bsnm{Finelli}, \binits{F.}},
\bauthor{\bsnm{Brandenberger}, \binits{R.}}:
\batitle{{Adiabatic and isocurvature perturbations for multifield generalized
  Einstein models}}.
\bjtitle{Phys. Rev. D}
\bvolume{67},
\bfpage{063512}
(\byear{2003})
{\href{https://arxiv.org/abs/astro-ph/0211276}{{arXiv:astro-ph/0211276}}}.
\doiurl{10.1103/PhysRevD.67.063512}
\end{barticle}
\endbibitem

%%% 65
\bibitem{Lalak:2007vi}
\begin{barticle}
\bauthor{\bsnm{Lalak}, \binits{Z.}},
\bauthor{\bsnm{Langlois}, \binits{D.}},
\bauthor{\bsnm{Pokorski}, \binits{S.}},
\bauthor{\bsnm{Turzynski}, \binits{K.}}:
\batitle{{Curvature and isocurvature perturbations in two-field inflation}}.
\bjtitle{JCAP}
\bvolume{0707},
\bfpage{014}
(\byear{2007})
{\href{https://arxiv.org/abs/0704.0212}{{arXiv:0704.0212}}}
{[hep-th]}.
\doiurl{10.1088/1475-7516/2007/07/014}
\end{barticle}
\endbibitem

%%% 66
\bibitem{Raveendran:2018yyh}
\begin{barticle}
\bauthor{\bsnm{Raveendran}, \binits{R.N.}},
\bauthor{\bsnm{Sriramkumar}, \binits{L.}}:
\batitle{{Primordial features from ekpyrotic bounces}}.
\bjtitle{Phys. Rev. D}
\bvolume{99}(\bissue{4}),
\bfpage{043527}
(\byear{2019})
{\href{https://arxiv.org/abs/1809.03229}{{arXiv:1809.03229}}}
{[astro-ph.CO]}.
\doiurl{10.1103/PhysRevD.99.043527}
\end{barticle}
\endbibitem

%%% 67
\bibitem{Braglia:2020fms}
\begin{barticle}
\bauthor{\bsnm{Braglia}, \binits{M.}},
\bauthor{\bsnm{Hazra}, \binits{D.K.}},
\bauthor{\bsnm{Sriramkumar}, \binits{L.}},
\bauthor{\bsnm{Finelli}, \binits{F.}}:
\batitle{{Generating primordial features at large scales in two field models of
  inflation}}.
\bjtitle{JCAP}
\bvolume{08},
\bfpage{025}
(\byear{2020})
{\href{https://arxiv.org/abs/2004.00672}{{arXiv:2004.00672}}}
{[astro-ph.CO]}.
\doiurl{10.1088/1475-7516/2020/08/025}
\end{barticle}
\endbibitem

%%% 68
\bibitem{Malik:2008im}
\begin{barticle}
\bauthor{\bsnm{Malik}, \binits{K.A.}},
\bauthor{\bsnm{Wands}, \binits{D.}}:
\batitle{{Cosmological perturbations}}.
\bjtitle{Phys. Rept.}
\bvolume{475},
\bfpage{1}--\blpage{51}
(\byear{2009})
{\href{https://arxiv.org/abs/0809.4944}{{arXiv:0809.4944}}}
{[astro-ph]}.
\doiurl{10.1016/j.physrep.2009.03.001}
\end{barticle}
\endbibitem

%%% 69
\bibitem{Vachaspati:2018hcu}
\begin{barticle}
\bauthor{\bsnm{Vachaspati}, \binits{T.}},
\bauthor{\bsnm{Zahariade}, \binits{G.}}:
\batitle{{Classical-Quantum Correspondence for Fields}}.
\bjtitle{JCAP}
\bvolume{1909},
\bfpage{015}
(\byear{2019})
{\href{https://arxiv.org/abs/1807.10282}{{arXiv:1807.10282}}}
{[hep-th]}.
\doiurl{10.1088/1475-7516/2019/09/015}
\end{barticle}
\endbibitem

%%% 70
\bibitem{Prokopec:2006fc}
\begin{barticle}
\bauthor{\bsnm{Prokopec}, \binits{T.}},
\bauthor{\bsnm{Rigopoulos}, \binits{G.I.}}:
\batitle{{Decoherence from Isocurvature perturbations in Inflation}}.
\bjtitle{JCAP}
\bvolume{0711},
\bfpage{029}
(\byear{2007})
{\href{https://arxiv.org/abs/astro-ph/0612067}{{arXiv:astro-ph/0612067}}}
{[astro-ph]}.
\doiurl{10.1088/1475-7516/2007/11/029}
\end{barticle}
\endbibitem

%%% 71
\bibitem{schlosshauer2007decoherence}
\begin{bbook}
\bauthor{\bsnm{Schlosshauer}, \binits{M.A.}}:
\bbtitle{Decoherence: and the Quantum-to-classical Transition}.
\bpublisher{Springer},
\blocation{Berlin Heidelberg}
(\byear{2007})
\end{bbook}
\endbibitem

%%% 72
\bibitem{Adesso:2007tx}
\begin{barticle}
\bauthor{\bsnm{Adesso}, \binits{G.}},
\bauthor{\bsnm{Illuminati}, \binits{F.}}:
\batitle{{Entanglement in continuous variable systems: Recent advances and
  current perspectives}}.
\bjtitle{J. Phys. A}
\bvolume{40},
\bfpage{7821}--\blpage{7880}
(\byear{2007})
{\href{https://arxiv.org/abs/quant-ph/0701221}{{arXiv:quant-ph/0701221}}}.
\doiurl{10.1088/1751-8113/40/28/S01}
\end{barticle}
\endbibitem

%%% 73
\bibitem{horodecki2009quantum}
\begin{barticle}
\bauthor{\bsnm{Horodecki}, \binits{R.}},
\bauthor{\bsnm{Horodecki}, \binits{P.}},
\bauthor{\bsnm{Horodecki}, \binits{M.}},
\bauthor{\bsnm{Horodecki}, \binits{K.}}:
\batitle{Quantum entanglement}.
\bjtitle{Reviews of modern physics}
\bvolume{81}(\bissue{2}),
\bfpage{865}
(\byear{2009})
\end{barticle}
\endbibitem

%%% 74
\bibitem{Prokopec:1992ia}
\begin{barticle}
\bauthor{\bsnm{Prokopec}, \binits{T.}}:
\batitle{{Entropy of the squeezed vacuum}}.
\bjtitle{Class. Quant. Grav.}
\bvolume{10},
\bfpage{2295}--\blpage{2306}
(\byear{1993}).
\doiurl{10.1088/0264-9381/10/11/012}
\end{barticle}
\endbibitem

%%% 75
\bibitem{zurek2002einselection}
\begin{bchapter}
\bauthor{\bsnm{Zurek}, \binits{W.H.}}:
\bctitle{Einselection and decoherence from an information theory perspective}.
In: \bbtitle{Quantum Communication, Computing, and Measurement 3},
pp. \bfpage{115}--\blpage{125}.
\bpublisher{Kluwer Academic},
\blocation{New York}
(\byear{2002})
\end{bchapter}
\endbibitem

%%% 76
\bibitem{ollivier2001quantum}
\begin{barticle}
\bauthor{\bsnm{Ollivier}, \binits{H.}},
\bauthor{\bsnm{Zurek}, \binits{W.H.}}:
\batitle{Quantum discord: a measure of the quantumness of correlations}.
\bjtitle{Physical review letters}
\bvolume{88}(\bissue{1}),
\bfpage{017901}
(\byear{2001})
{\href{https://arxiv.org/abs/quant-ph/0105072}{{arXiv:quant-ph/0105072}}}
\end{barticle}
\endbibitem

%%% 77
\bibitem{henderson2001classical}
\begin{barticle}
\bauthor{\bsnm{Henderson}, \binits{L.}},
\bauthor{\bsnm{Vedral}, \binits{V.}}:
\batitle{Classical, quantum and total correlations}.
\bjtitle{Journal of physics A: mathematical and general}
\bvolume{34}(\bissue{35}),
\bfpage{6899}
(\byear{2001})
{\href{https://arxiv.org/abs/quant-ph/0105028}{{arXiv:quant-ph/0105028}}}
\end{barticle}
\endbibitem

%%% 78
\bibitem{modi2014pedagogical}
\begin{barticle}
\bauthor{\bsnm{Modi}, \binits{K.}}:
\batitle{A pedagogical overview of quantum discord}.
\bjtitle{Open Systems \& Information Dynamics}
\bvolume{21}(\bissue{01n02}),
\bfpage{1440006}
(\byear{2014})
{\href{https://arxiv.org/abs/1312.7676}{{arXiv:1312.7676}}}
{[quant-ph]}
\end{barticle}
\endbibitem

%%% 79
\bibitem{bera2017quantum}
\begin{barticle}
\bauthor{\bsnm{Bera}, \binits{A.}},
\bauthor{\bsnm{Das}, \binits{T.}},
\bauthor{\bsnm{Sadhukhan}, \binits{D.}},
\bauthor{\bsnm{Roy}, \binits{S.S.}},
\bauthor{\bsnm{De}, \binits{A.S.}},
\bauthor{\bsnm{Sen}, \binits{U.}}:
\batitle{Quantum discord and its allies: a review of recent progress}.
\bjtitle{Reports on Progress in Physics}
\bvolume{81}(\bissue{2}),
\bfpage{024001}
(\byear{2017})
{\href{https://arxiv.org/abs/1703.10542}{{arXiv:1703.10542}}}
{[quant-ph]}
\end{barticle}
\endbibitem

%%% 80
\bibitem{Lim:2014uea}
\begin{barticle}
\bauthor{\bsnm{Lim}, \binits{E.A.}}:
\batitle{{Quantum information of cosmological correlations}}.
\bjtitle{Phys. Rev.}
\bvolume{D91}(\bissue{8}),
\bfpage{083522}
(\byear{2015})
{\href{https://arxiv.org/abs/1410.5508}{{arXiv:1410.5508}}}
{[hep-th]}.
\doiurl{10.1103/PhysRevD.91.083522}
\end{barticle}
\endbibitem

%%% 81
\bibitem{adesso2010quantum}
\begin{barticle}
\bauthor{\bsnm{Adesso}, \binits{G.}},
\bauthor{\bsnm{Datta}, \binits{A.}}:
\batitle{Quantum versus classical correlations in gaussian states}.
\bjtitle{Physical review letters}
\bvolume{105}(\bissue{3}),
\bfpage{030501}
(\byear{2010})
{\href{https://arxiv.org/abs/1003.4979}{{arXiv:1003.4979}}}
{[quant-ph]}
\end{barticle}
\endbibitem

%%% 82
\bibitem{datta2008quantum}
\begin{barticle}
\bauthor{\bsnm{Datta}, \binits{A.}},
\bauthor{\bsnm{Shaji}, \binits{A.}},
\bauthor{\bsnm{Caves}, \binits{C.M.}}:
\batitle{Quantum discord and the power of one qubit}.
\bjtitle{Physical review letters}
\bvolume{100}(\bissue{5}),
\bfpage{050502}
(\byear{2008})
{\href{https://arxiv.org/abs/0709.0548}{{arXiv:0709.0548}}}
{[quant-ph]}
\end{barticle}
\endbibitem

%%% 83
\bibitem{Serafini:2003ke}
\begin{barticle}
\bauthor{\bsnm{Serafini}, \binits{A.}},
\bauthor{\bsnm{Illuminati}, \binits{F.}},
\bauthor{\bsnm{De~Siena}, \binits{S.}}:
\batitle{{Symplectic invariants, entropic measures and correlations of Gaussian
  states}}.
\bjtitle{J. Phys. B}
\bvolume{37},
\bfpage{21}
(\byear{2004})
{\href{https://arxiv.org/abs/quant-ph/0307073}{{arXiv:quant-ph/0307073}}}.
\doiurl{10.1088/0953-4075/37/2/L02}
\end{barticle}
\endbibitem

%%% 84
\bibitem{Simon:1999lfr}
\begin{barticle}
\bauthor{\bsnm{Simon}, \binits{R.}}:
\batitle{{Peres-Horodecki Separability Criterion for Continuous Variable
  Systems}}.
\bjtitle{Phys. Rev. Lett.}
\bvolume{84},
\bfpage{2726}--\blpage{2729}
(\byear{2000})
{\href{https://arxiv.org/abs/quant-ph/9909044}{{arXiv:quant-ph/9909044}}}.
\doiurl{10.1103/PhysRevLett.84.2726}
\end{barticle}
\endbibitem

%%% 85
\bibitem{Martin:2019wta}
\begin{barticle}
\bauthor{\bsnm{Martin}, \binits{J.}}:
\batitle{{Cosmic Inflation, Quantum Information and the Pioneering Role of John
  S Bell in Cosmology}}.
\bjtitle{Universe}
\bvolume{5}(\bissue{4}),
\bfpage{92}
(\byear{2019})
{\href{https://arxiv.org/abs/1904.00083}{{arXiv:1904.00083}}}
{[quant-ph]}.
\doiurl{10.3390/universe5040092}
\end{barticle}
\endbibitem

%%% 86
\bibitem{kendig2005conics}
\begin{bbook}
\bauthor{\bsnm{Kendig}, \binits{K.}}:
\bbtitle{Conics}.
\bpublisher{The Mathematical Association of America},
\blocation{Washington, USA}
(\byear{2005})
\end{bbook}
\endbibitem

\end{thebibliography}
%%%%%%%%%%%%%%%%%%%%%%%%%%%%%%%%%%%%%%%%%%%%%%%%%%%%%%%%%%%%%%%%%%%%%%%%%%%%%%%
\end{document}